\documentclass[aps,prd,amsmath,amsfonts,amssymb,eqsecnum,nofootinbib,
  floatfix,secnumarabic,preprintnumbers,superscriptaddress,nobalancelastpage,
  twocolumn]{revtex4}
\usepackage{color,multirow,graphicx}
\usepackage{slashed}

\newcommand{\ie} {{\it i.e.}}

\newcommand {\beq} {\begin{equation}}
\newcommand {\eeq} {\end{equation}}
\newcommand {\bea} {\begin{eqnarray}}
\newcommand {\eea} {\end{eqnarray}}

\definecolor{red1}{cmyk}{0,1,1,0.1}
\definecolor{blue1}{cmyk}{1,0,0,0}

\newcommand{\ignore}[1]{}
 \renewcommand{\bf}{\textbf}

\newcommand{\GeV}{{\rm\ GeV}}

\newcommand{\TeV}{{\rm\ TeV}}

\newcommand{\met}{\slashed{E}_T}

\newcommand{\LM}{\mathcal{L}}

\newcommand{\amc}{{\sc MadGraph5\textunderscore}a{\sc MC@NLO}}

\begin{document}

\hspace*{125mm}{\large \tt FERMILAB-PUB-16-125-T}

\date{\today}

\title{Probing TeV scale Top-Philic Resonances with Boosted Top-Tagging at the High Luminosity LHC}

\author{Jeong Han Kim} \email{jeonghan.kim@kaist.ac.kr} 
\affiliation{Department of Physics, Korea Advanced Institute of Science and Technology, 335 Gwahak-ro, Yuseong-gu, Daejeon 305-701, Korea, and Center for Theoretical Physics of the Universe, Institute for Basic Science (IBS), Daejeon 34051, Republic of Korea}
\author{Kyoungchul Kong} \email{kckong@ku.edu} 
\affiliation{Department of Physics and Astronomy, University of Kansas, Lawrence, KS, 66045, USA} 
\author{Seung J. Lee} \email{sjjlee@korea.edu} 
\affiliation{Department of Physics, Korea University, Seoul 136-713, Korea and \\
 School of Physics, Korea Institute for Advanced Study, Seoul 130-722, Korea} 
\author{Gopolang Mohlabeng} \email{gopolang.mohlabeng@ku.edu}
\affiliation{Department of Physics and Astronomy, University of Kansas, Lawrence, KS, 66045, USA} 
\affiliation{Theoretical Physics Department, Fermilab, Batavia, IL 60510, USA}

\begin{abstract}

We investigate the discovery potential of singly produced top-philic resonances at the high luminosity (HL) LHC in the four-top final state. Our analysis spans over the fully-hadronic, semi-leptonic, and same-sign dilepton channels where we present concrete search strategies adequate to a boosted kinematic regime and high jet-multiplicity environments. We utilize the Template Overlap Method (TOM) with newly developed template observables for tagging boosted top quarks, a large-radius jet variable $M_J$ and customized b-tagging tactics for background discrimination. Our results show that the same-sign dilepton channel gives the best sensitivity among the considered channels, with an improvement of significance up to 10\%-20\% when combined with boosted-top tagging. Both the fully-hadronic and semi-leptonic channels yield comparable discovery potential and contribute to further enhancements in the sensitivity by combining all channels. Finally, we show the sensitivity of a top-philic resonance at the LHC and HL-LHC by showing the $2\sigma$ exclusion limit and $5\sigma$ discovery reach, including a combination of all three channels. 
%
\end{abstract}
\maketitle

\section{Introduction}\label{sec:intro}

The discovery of the Higgs boson at the LHC has completed the particle content of the Standard Model (SM).
Precision measurements of the Higgs interaction to the SM particles provides an excellent opportunity in understanding
the nature of electroweak symmetry breaking (EWSB) and in the search for new physics (NP) beyond the SM.
Among myriad possibilities, the large Yukawa coupling of the top quark to the Higgs boson makes the top quark 
one of the most interesting candles in searching for NP underlying the EWSB.
By the same token, for the class of models addressing the naturalness of the EWSB scale, it provides the most important window for NP.

Some of these models often introduce new particles which interact strongly to the top sector.
Examples include two Higgs doublet models \cite{Branco:2011iw,Gori:2016zto,Dev:2014yca}, 
left-right extensions of the SM \cite{Mohapatra:1974gc},
models with a color-sextet or color-octet \cite{Dobrescu:2007yp,Chen:2008hh,Bai:2010dj,Berger:2010fy,Zhang:2010kr,Gerbush:2007fe}, 
models with composite particles \cite{ Matsedonskyi:2012ym,Lillie:2007hd, Gripaios:2014pqa,Kumar:2009vs,Liu:2015hxi,Cacciapaglia:2015eqa,     Kaplan:1983fs, Kaplan:1983sm, Georgi:1984ef, Banks:1984gj, Georgi:1984af, Dugan:1984hq, ArkaniHamed:2002qy, Contino:2003ve, Giudice:2007fh, Barbieri:2007bh, Panico:2011pw, DeCurtis:2011yx, Marzocca:2012zn, Bellazzini:2012tv, Panico:2012uw, Bellazzini:2014yua}, etc. 
NP that couples strongly to top quarks might manifest itself as heavy resonant states that can be produced at the LHC. 
Therefore, searching for $t \bar{t}$ resonances at hadron colliders is of particular importance.
Experimental collaborations have been searching for them and null results have provided stringent bounds on the production of the $t \bar t$ resonances. Current limits on the resonance mass lies at the TeV scale, depending on models.
However, in most
cases 
the $t\bar t$ resonances are produced via $q\bar q$ annihilations with sizable couplings to the light quarks, which
indicate that the $t\bar t$ resonance may couple to each generation differently. 
A good example would be the Kaluza-Klein gluon in RS models \cite{Agashe:2006hk, Lillie:2007yh} and Kaluza-Klein gauge bosons in flat extra dimensions with boundary terms and bulk masses \cite{Flacke:2013pla}.
Only recently there has been an attempt to perform a model-independent study on collider phenomenology and dark matter extension of a $t \bar t$ resonance without involving its couplings to the light quarks \cite{Greiner:2014qna,Liu:2015hxi,Cox:2015afa,Jackson:2013rqp,Jackson:2009kg,Servant:2010zza,Brooijmans:2010tn}. 

In this paper we take the same philosophy and approach in the context of a simplified model, 
where a $t \bar t$ resonance couples primarily to the top quark and weakly to the light quarks (top-philic), such that we can ignore all the other couplings except for the one with tops.
We investigate the discovery potential of such top-philic resonances at the high luminosity (HL) LHC. 
There are several possible production modes, among which we focus on the four-top final state in our study, with two tops originating from the decay of the top-philic resonance and the other two are spectators.
A top-philic color singlet vector resonance is a good example that fits into such criteria, and we study it in this analysis. 
%
Depending on the mass of this resonance, the top-pair from its decay may be boosted and appear in the detector as two collimated fat jets, in which case, boosted techniques will be useful in the reconstruction of the resonance mass and backgrounds reduction.
Furthermore when such a resonance is heavy (of order TeV), the resulting top from its decay is highly boosted, and one needs to use jet substructure methods to tag the boosted top.
We use the TemplateTagger implementation of the Template Overlap Method (TOM) \cite{Almeida:2010pa, Almeida:2011aa,  Backovic:2012jj, Backovic:2013bga} with newly developed template observables in our analysis. In particular, to cope with high-multiplicity final states, we combine the TOM and jet-trimming methods to reduce soft radiation and achieve better mass resolution\footnote{A similar study has been performed in the associated production of a heavy higgs with $b\bar b$ and $t\bar{t}$ in Ref. \cite{Chen:2015fca}}.

This paper is organized as follows. 
In Section \ref{sec:section2}, we introduce a simplified model of a top-philic resonance and discuss its production and decay with current bounds. 
Detailed information on the MC simulation and the top-tagging is presented in Section \ref{sec:simulation}. 
We show our results in Section \ref{sec:results} in three different channels as well as their combination.
Section \ref{sec:conclusions} is reserved for summary.

\section{A Top-Philic Resonance: Simplified Model}\label{sec:section2}
\subsection{Setup}\label{sec:model}

We consider a color singlet vector particle ($V_1$) which dominantly couples to top and anti-top.
Assuming that all other interactions are weak, 
the relevant interaction is given by the following renormalizable Lagrangian
  \begin{eqnarray}
  \LM_{int} &=& \bar{t} \, \gamma_{\mu} (c_{L} P_{L} + c_{R} P_{R}) \, t \, V_{1}^{\mu} \nonumber \\
  &=& c_t \, \bar{t} \, \gamma_{\mu} (\cos \theta P_{L} + \sin \theta P_{R}) \, t \, V_{1}^{\mu},
 \label{v0inttop} 
  \end{eqnarray}
  where  $P_{R/L} \, = \, (1 \pm \gamma_{5})/2$, $c_t \, = \, \sqrt{(c_{L})^{2} + (c_{R})^{2}}$ and $\tan \theta \, = \, \frac{c_{R}}{c_{L}}$ are the projection operators, coupling of the vector singlet with the top quarks and tangent of the chirality angle respectively. 
The decay width is given by
  \begin{eqnarray}
  \Gamma(V_1 \rightarrow t \bar{t}) &=& \frac{c_t^{2} M_{V_1}}{8\pi} \sqrt{1 - \frac{4m_{t}^{2}}{M_{V_1}^{2}}}   \nonumber \\
&\times&  \Big  [1 - \frac{m_{t}^{2}}{M_{V_1}^{2}}\Big (1 - 3\sin(2 \theta) \Big )   \Big ] .
   \label{width} 
  \end{eqnarray}
For $m_{t} \ll M_{V_{1}}$, $\frac{\Gamma}{M_{V_{1}}} \approx \frac{c_{t}^{2}}{8\pi}$ and $V_{1}$ must be a narrow resonance, 
if it weakly couples to a top pair. 

\subsection{Production and Decay}\label{sec:Production}

\begin{figure*}[t]
\begin{center}
\begin{tabular}{ccc}
\includegraphics[width=.32\textwidth]{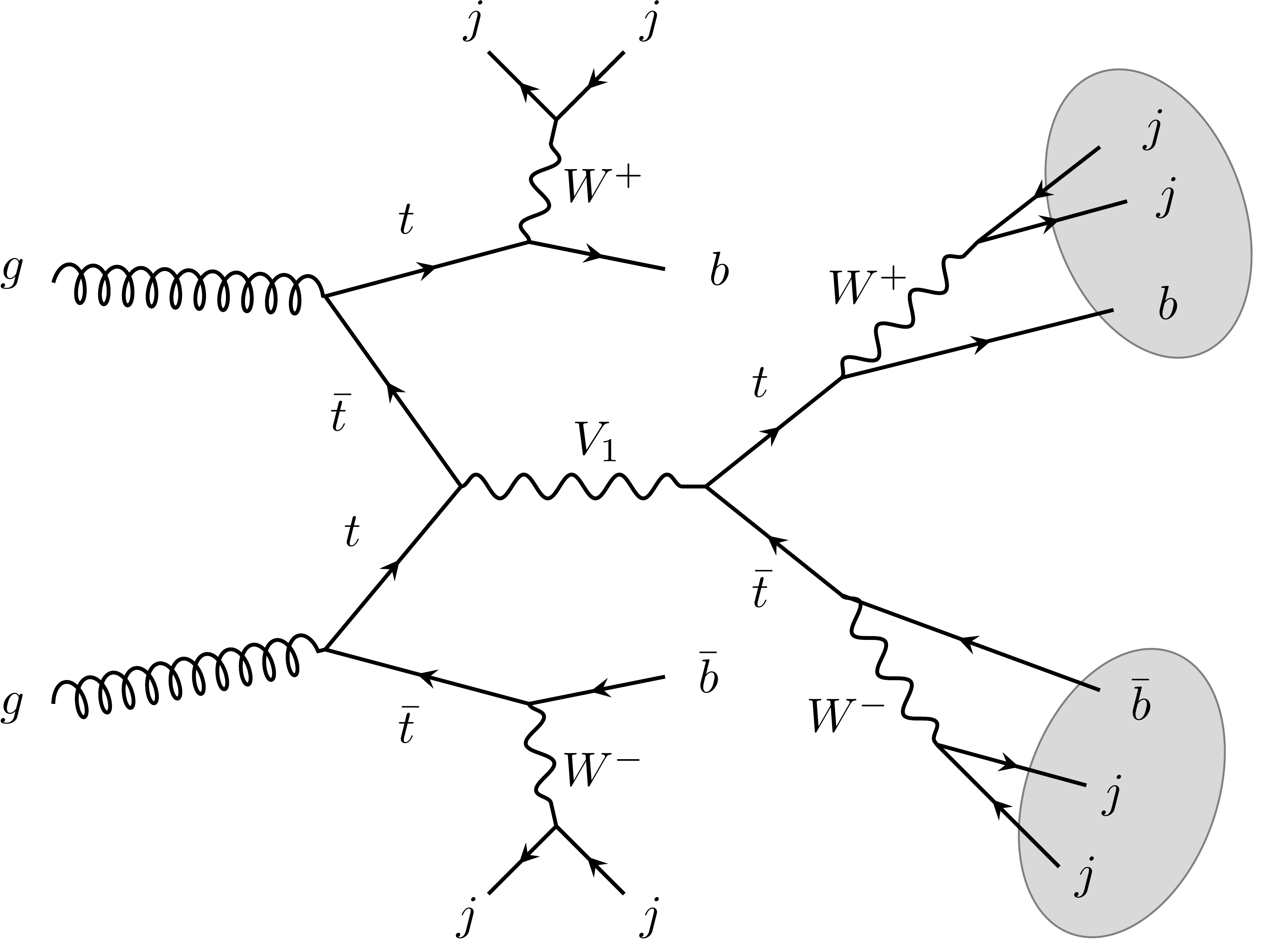} &
\hspace{0.15cm} \includegraphics[width=.32\textwidth]{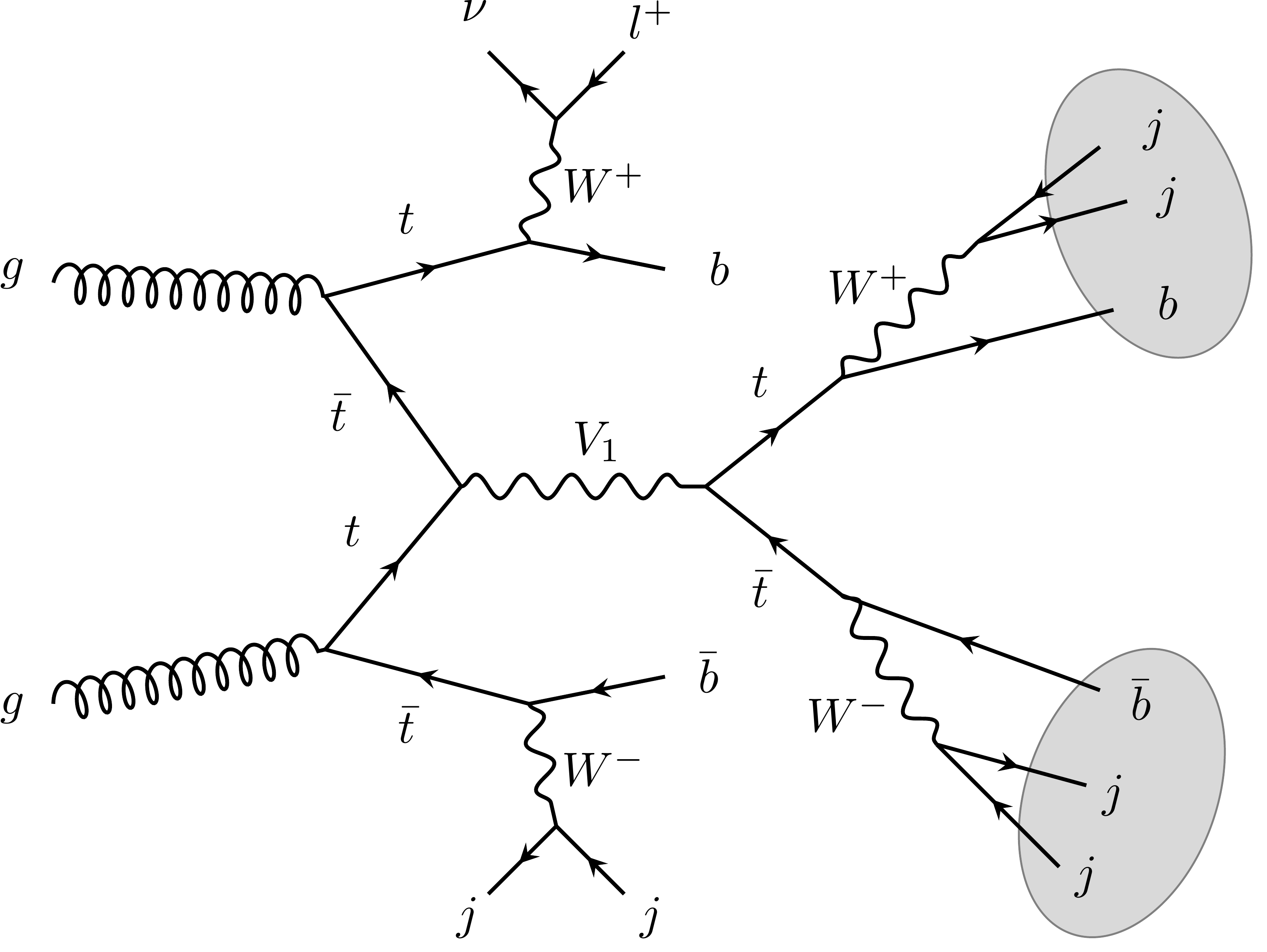}  &
\hspace{0.15cm}\includegraphics[width=.32\textwidth]{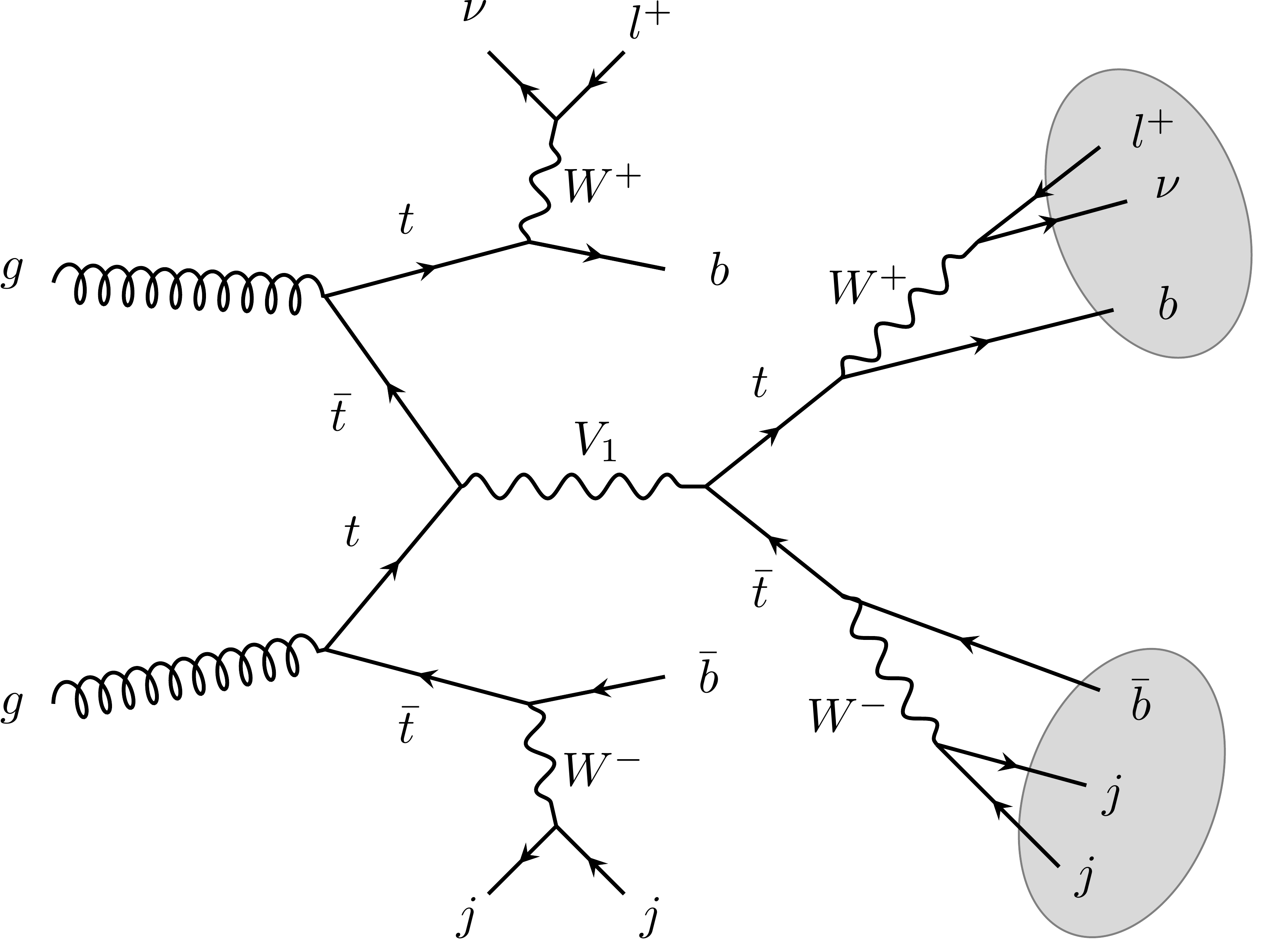}
\end{tabular}
\end{center}
\caption{The tree-level single production of the $V_1$ decaying into the fully-hadronic (left), semi-leptonic (middle) and same-sign-dileptonic (right) channels. As all three channels exhibit disjoint final states, we set a different search strategy on each channel and estimate its sensitivity at the HL-LHC. The top and anti-top pair from the resonance decay are highly-boosted and they are characteristically distinguished by the other two non-boosted spectator tops.}
\label{fig:Diagram}
\end{figure*}

In our study, we choose a model independent approach and do not consider any underlying or fundamental theory which might generate Eq. (\ref{v0inttop}). We focus on the two body decay of $V_{1}$ into $ t \bar{t}$ with $M_{V_{1}}$ in the TeV range
(For possible decay modes below $M_{V_1} < 2 m_t$, see Ref. \cite{Cox:2015afa}.).
There are three free parameters, the vector resonance mass ($M_{V_{1}}$), 
the overall coupling strength ($c_{t}$) and the chirality ($\theta$).

There are two ways to produce a top-philic resonance at the LHC: at one-loop and at tree-level \cite{Greiner:2014qna}. 
 %
%
\begin{enumerate}
\item At one loop, on-shell $V_1$ is produced in association with a jet, i.e. $ p \, p  \rightarrow V_1 + j$ (figure 8 in Ref. \cite{Greiner:2014qna}) and 
its cross section is dependent on all three parameters ($M_{V_1}$, $c_t$, $\theta$).
It is one-loop suppressed but has an advantage over the tree-level production ($t \bar t V_1$) in term of phase space. 
It turns out that the loop-production exhibits an additional enhancement in the case of the axial coupling and its cross section can be much larger than the tree-level production cross section near $\theta = \frac{3}{4}\pi$ \cite{vanderBij:1988ac,Greiner:2014qna}.
In addition, $V_1$ may be produced off-shell in the $ g g \to V_1 \to t\bar t$ process (figure 10 in Ref. \cite{Greiner:2014qna}) and contribute to the cross section measurement of $t\bar t$ production, which provides the most stringent bounds on the model. 
Similar to the on-shell case, the off-shell production becomes the largest for the axial coupling.
\item Tree-level production is essentially top production with $V_1$-strahlung: $t \bar{t} + V_1$, $t W + V_1$ and $t j + V_1$, with $V_1 \to t \bar{t}$.
The largest contribution at the LHC comes from the four top-quark final state as shown in figure \ref{fig:Diagram}. 
For a wide range of $M_{V_1}$, $tjV_1$ production is smaller than $t\bar t V_1$ roughly by a factor of 2 while $tW V_1$ production is smaller by a factor of 4.
Unlike the loop-production ($t\bar t$ and $t\bar t j$) or electroweak production ($t W V_1$ and $t j V_1$) channels, 
strong production ($t\bar{t} t \bar{t}$) is independent of $\theta$ and depends only on ($M_{V_1}$, $c_t$). 
We set $\theta=\frac{\pi}{2}$ for the rest of discussion in this paper\footnote{In our study, perturbative unitarity should not be an issue due to negligible couplings in the other sectors. See Ref. \cite{Han:2015rys} for details.}.

\end{enumerate}

\begin{figure*}[t]
\centerline{
\includegraphics[scale=0.92]{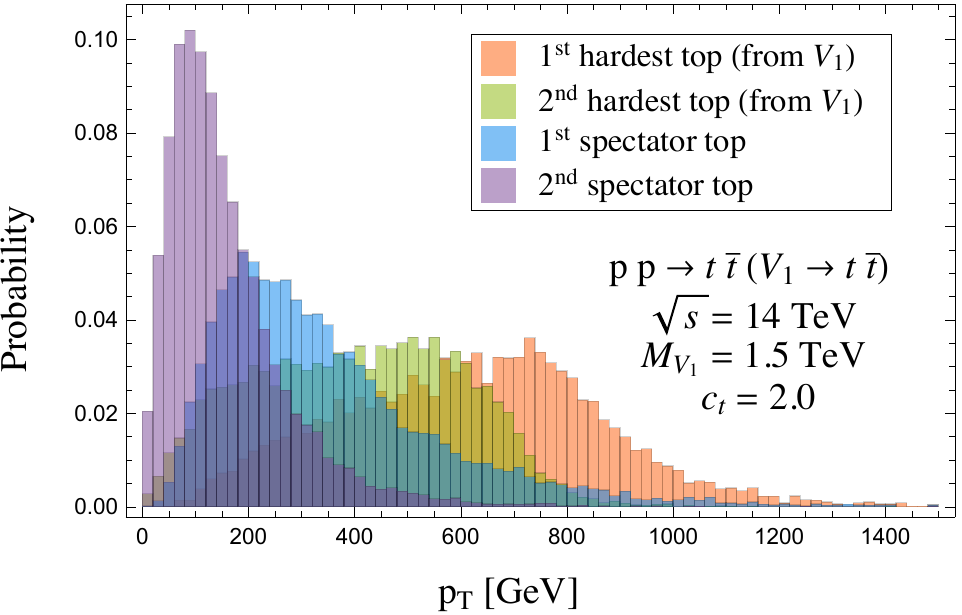}
\includegraphics[scale=0.92]{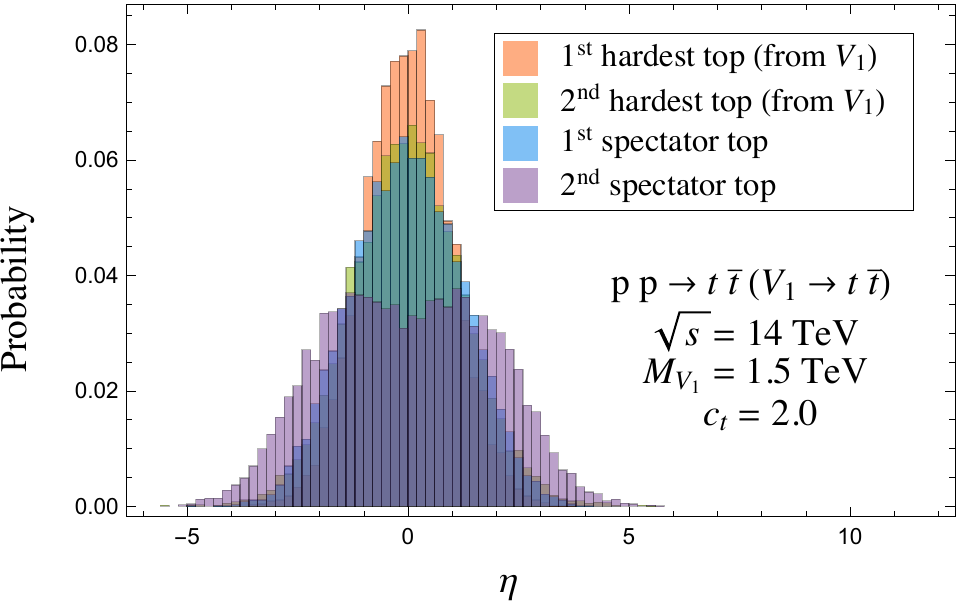} }
\caption{$p_T$ (left) and $\eta$ (right) distributions of the four tops at the parton-level for a benchmark point, $M_{V_1} = 1.5 \TeV$ and $c_t=2.0$.}
\label{fig:parton}
\end{figure*}
In this paper, we investigate the tree-level production of a heavy top-philic resonance in the four-top final state.
In figure \ref{fig:parton}, we plot the $p_{T}$ and $\eta$ distributions of the parton level top-quarks, from the resonance decay and those produced in association with the resonance. For the $p_{T}$ distribution we see that the hardest top, peaks roughly at $p_{T} = M_{V_{1}}/2$, while the second hardest top peaks at a slightly lower value. We also notice that one of the spectator tops is broad in $p_T$ and peaks at a value higher than $m_{t}$, which implies some contamination from the spectator tops to the boosted tops, similar to the case in \cite{Liu:2015hxi}. 
As shown in figure \ref{fig:parton}, the two tops from the resonance are boosted and the use of jet substructure observables for top-tagging would play an important role in maximizing the sensitivity.

Four top final states can be looked for through various experimental searches, which we prioritize into two classes: one with two hadronically-decaying boosted tops and the other with the same-sign dilepton (right, figure \ref{fig:Diagram}). The former derives benefit from the hadronic ditop-tagging to effectively reduce backgrounds. It is further classified in the fully-hadronic (left, figure \ref{fig:Diagram}), semi-leptonic (middle, figure \ref{fig:Diagram}) and dileptonic decay modes of the two spectator tops with the corresponding branching ratios of $\sim20\%$, $\sim13\%$ and $\sim2.2\%$ respectively. The latter has the smallest set of backgrounds with a branching ratio of $\sim 4.4 \%$. Under the given classifications, we discuss the main final states we will focus on for $V_1$ detection.
\begin{enumerate} 
\item Fully-hadronic channel  (section \ref{sec:had}):\\
The fully-hadronic decays (left, figure \ref{fig:Diagram}) of four tops render the largest branching ratio, but suffer from the enormous multi-jet QCD background. Here we will show that the ditop-tagging technique combined with b-tagging is able to suppress the QCD and $t \bar{t}$ backgrounds sufficiently making this channel competitive.

\item Semi-leptonic channel  (section \ref{sec:semi}):\\
Hadronic decays of two boosted tops and semi-leptonic decays of two spectators (middle, figure \ref{fig:Diagram}) have an advantage of evading the QCD background by the requirement of a hard isolated lepton in signal events. On top of that, since the dominant semi-leptonic $t\bar{t}$ background contains a single hadronic top, the ditop-tagging can further suppress it hence bringing it into the regime where $t\bar{t} t\bar{t}$ is effectively the only background left. The resulting sensitivity turns out to be comparable with the fully-hadronic channel.

\item Dileptonic channel:\\
Hadronic decays of two boosted tops and dileptonic decays of two spectators are strongly suppressed due to the small branching ratio. The ditop-tagging further reduces the signal rate making this channel less competitive, and therefore we do not  consider the dileptonic channel in the rest of our paper.

\item Same-sign dilepton channel (section \ref{sec:SameSign}): \\
Unlike the other channels, the same-sign dilepton (SSDL) channel (right, figure \ref{fig:Diagram}) can evade the dominant $t \bar{t}$ background and provide the largest sensitivity even with a small branching ratio. We will show that the boosted techniques can further improve the discovery potential with better background reduction due to the extra capability of resonance reconstruction in the SSDL channel. 

\end{enumerate}

\subsection{Experimental Bounds}\label{sec:constraints}

Experimental bounds on the top-philic resonance are obtained from production at both tree- ($t\bar{t} t\bar{t}$) and loop-level ($t\bar{t}$ and $t\bar{t} + j$) \cite{Greiner:2014qna}.
The leading order SM production cross-section for $t\bar{t} t\bar{t}$ at the 8 TeV LHC is on the order of 1 fb and next-to-leading-order (NLO) corrections can increase this cross-section up to 30$\%$ \cite{Barger:2010uw}. 
However it can be significantly enhanced due to the production of $t\bar{t} t\bar{t}$ from a heavy resonance as in our study. 
The CMS collaboration has placed an upper limit of 32 fb at 95$\%$ confidence level on the SM production of $t\bar{t} t\bar{t}$ at the 8 TeV LHC \cite{Khachatryan:2014sca}. 
For loop production, the $t\bar{t}$ cross section measurement provides the most stringent limit. 
The loop production (e.g., $g g \to V_1 g$) cross section is dominated by the axial vector part of the cross section.
In the axial coupling the top-philic resonance is predominantly longitudinally polarized and the cross section rises like ${\hat s}/{M_{V_1}^2}$, while it is transversely polarized for the vector coupling.
Therefore loop production becomes especially important in the axial coupling limit, $\theta= \frac{3}{4}\pi$ \cite{vanderBij:1988ac,Greiner:2014qna}.

\begin{figure}[t]
\includegraphics[scale=0.55]{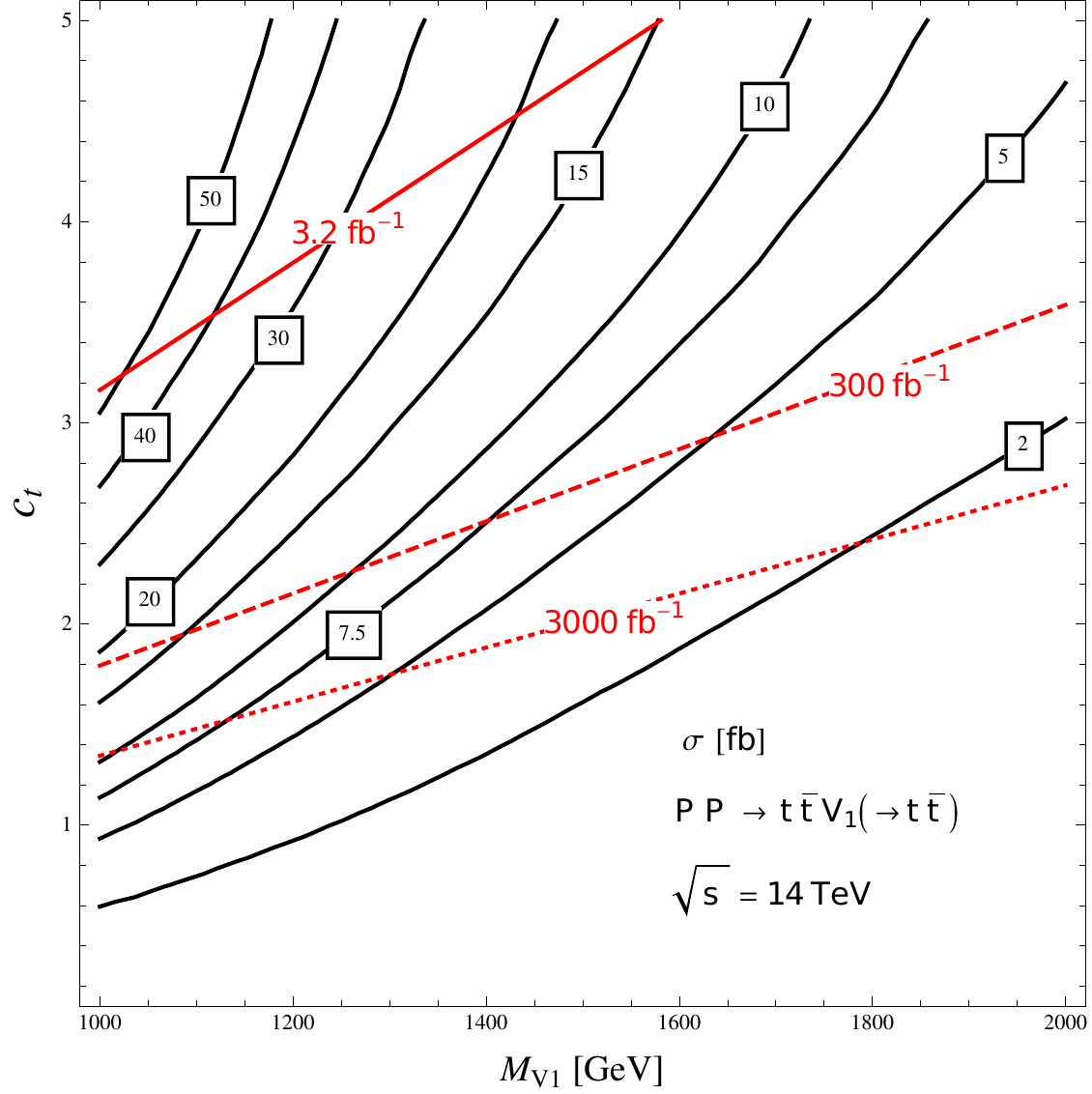} 
\caption{Parton level production cross-section of $p \, p \to t\bar{t} V_1 \to t\bar{t} t\bar{t}$ (in fb, black-soild) at $\sqrt{s} = 14 ~{\rm TeV}$ in the $M_{V_1}$-$c_{t}$ parameter space. Red solid curve (obtained from $ c_t^2 / (2 M_{V_1}^2) = 5.0 ~\text{TeV}^{-2}$) represents current ATLAS bounds from 13 TeV LHC with 3.2 fb$^{-1}$ of data \cite{ATLAS13TeV}. 
Projected bounds for 300 fb$^{-1}$ (3000 fb$^{-1}$) may be obtained via a naive rescaling, and are shown in dashed (dotted) curve.}
\label{fig:4_top14}
\end{figure}

In figure \ref{fig:4_top14} we show the $t\bar{t} t\bar{t}$ production cross-section at the 14 TeV LHC in the ($M_{V_{1}}$, $c_{t}$) parameter space. 
Following Ref. \cite{Greiner:2014qna}, we have checked that the four-top production cross section is small compared to the 8 TeV bound in the high mass region of our interests. 
Since the $t\bar{t}$ cross section measurement places strong bounds only on the parameter space near the axial coupling, one could choose any values of $\theta$ away from the axial limit for the four top production, which is independent of $\theta$.
Therefore in the rest of our discussion, we will not consider any experimental bounds from current LHC analyses and refer readers to Ref. \cite{Greiner:2014qna} for more details.

Recently the ATLAS collaboration has set limits on the four top production cross section in the context of a contact interaction for the right-handed top quark (corresponding to $\theta=\pi/2$ in our case) with 13 TeV data \cite{ATLAS13TeV}\footnote{We note that the four-top production cross section is independent of the choice of $\theta$, whereas the loop-production cross section varies significantly as a function of $\theta$.}. 
For comparison we integrate out the top-philic resonance $V_1$ for $\theta=\pi/2$ in Eq. (\ref{v0inttop}) and obtained the following contact interaction,
\begin{equation}
\frac{1}{2} \frac{c_t^2}{M^2_{V_1}} \, \big ( \bar{t}_R \,  \gamma_\mu \, t_R \big )  \, \big (  \bar{t}_R \,  \gamma^\mu \,  t_R \big )\, .
\end{equation}
The region in the top-left corner in figure \ref{fig:4_top14}, corresponding to $ c_t^2 / (2 M_{V_1}^2) > 5.0 ~\text{TeV}^{-2}$, is excluded at 95\% C.L at the 13 TeV LHC with 3.2 fb$^{-1}$ \cite{ATLAS13TeV}\footnote{8 TeV results are slightly weaker, $c_t^2 / (2 M^2_{V_1}) > 6.6 ~\text{TeV}^{-2}$ \cite{Aad:2015kqa}.}.
The dashed (dotted) curve in red is the projected exclusion with 300 fb$^{-1}$ (3000 fb$^{-1}$) via a naive rescaling with current data.

\section{Monte Carlo simulation and Analysis Method }\label{sec:simulation}

We simulate signal and background events with the \amc\ \cite{Alwall:2011uj,Alwall:2014hca} framework at $\sqrt{s} = 14$ TeV $pp$ center of mass energy, using the \verb|nn23nlo| parton distribution function \cite{Ball:2012cx}. Our model implementation is based on the Lagrangian of Eq. (\ref{v0inttop}) with parameters $M_{V_1}$, $c_t$ and $\theta$.


We generate all event samples at leading order accuracy in QCD, and normalize all background samples by multiplying by a conservative $K$-factor of 2. At the generation level, we require all partons to pass cuts of $p_T > 15$~GeV, ~$| \eta |< 5$, while leptons are required to have $p_T > 10$~GeV and ~$| \eta |< 2.5$. The preselection demands a strong $H_T$ cut for each indivisible channel to improve the statistics in the SM backgrounds and signals, where $H_T$ denotes the scalar sum of the transverse momenta of all final state partons. The numerical values of background cross sections after the $H_T$ cuts are summarized in Tables \ref{tab:TotalBackGroundsHad}, \ref{tab:SemiBackGround} and  \ref{tab:SSDLBackGround}.
Then we shower the events with \verb|PYTHIA 6| \cite{Sjostrand:2006za}  using the modified MLM-matching scheme~\cite{Maltoni:2002qb,Mangano:2006rw}, and cluster all showered events with the \verb|FastJet| \cite{Cacciari:2011ma} implementation of the anti-$k_T$ algorithm \cite{Cacciari:2008gp}. 

When it comes to a search for high-multiplicity and high-$H_T$ final states, non-negligible initial-state-radiation (ISR) sources arise. Since the contamination from the ISR scales like a fat jet radius, the smaller size of a fat jet we choose, the less pollution we have. Therefore, a proper size of a fat jet should be optimized specifically depending on the final states of interest and its characteristic $p_T$ scale. For our purpose, we fix the cone size $R = 0.7$ in the fully hadronic and semi-leptonic channels to reduce the ISR effects as much as possible, while we increase it to $R = 0.8$ in the SSDL channel since the final states become less busy. Finally, for non-forward light jets ($i.e.$ $|\eta| < 2.5$) including the $b$-jets, we use a cone size of $r = 0.4$.

\subsection{Boosted Top Tagging}
\label{sec:TOM}

Tagging heavy boosted objects has become a central topic in probing new physics at the TeV scale. With such high scale masses, their decay products are strongly boosted and collimated into the same directions with characteristic internal structures. It requires, therefore, a detailed inquiry at the sub-jet level to classify and distinguish boosted heavy objects such as Higgs, top and $W/Z$ bosons from each other. In recent years, numerous studies have attempted to develop and design jet substructure observables \cite{Butterworth:2008iy, Almeida:2011aa, Backovic:2012jj, Schlaffer:2014osa, Ellis:2007ib, Abdesselam:2010pt, Salam:2009jx, Nath:2010zj, Almeida:2011ud, Plehn:2011tg, Altheimer:2012mn, Soper:2011cr, Soper:2012pb, Jankowiak:2011qa, Krohn:2009th, Ellis:2009me, Backovic:2012jk, Backovic:2013bga, Hook:2011cq, Thaler:2010tr, Thaler:2011gf, Thaler:2008ju, Almeida:2008yp,Almeida:2008tp,Almeida:2010pa, Rentala:2014bxa, Cogan:2014oua, Larkoski:2014wba, Almeida:2015jua,Larkoski:2014zma}.

In this paper, as illustration for the jet substructure analysis, we use the \verb| TemplateTagger v.1.0 |\cite{Backovic:2012jk} implementation of the Template Overlap Method (TOM) \cite{Almeida:2010pa, Almeida:2011aa,  Backovic:2012jj, Backovic:2013bga} with newly developed template obeservables. TOM continuously attempts to match the energy distribution of jets onto the parton-like configuration of a top decay, until it maximizes an overlap score $Ov$ which measures the probability of a fat jet being a top jet. For the purpose of our analysis, we generate 17 sets of both three body top templates at fixed $p_T$, starting from $p_T = 500 \GeV$ in steps of $50 \GeV$. We use a template resolution parameter $\sigma = 0.4$, and template sub-cone sizes $r_{\rm sub} = \{ 0.2, 0.1, 0.22 \}$ optimized for the fully-hadronic, semi-leptonic and SSDL channels respectively (cf. Ref. \cite{Backovic:2012jj}).

To maximize the performance of TOM and reduce the mis-tag rate, we introduce an extra cut, which is explained as follows. 
As an illustration, we generate two benchmark event samples under the scheme described in section \ref{sec:simulation}: the semi-leptonic production of the $t \bar{t}$ process without additional jets and $j Z(\rightarrow \nu \bar{\nu})$. The samples are chosen such that the former sample contains one hadronically-decaying top (representing a signal of interest) and the latter contains a mono-jet in the event. We shower the events with \verb|PYTHIA 6| \cite{Sjostrand:2006za} and cluster all showered events with the \verb|FastJet| \cite{Cacciari:2011ma} implementation of the anti-$k_T$ algorithm \cite{Cacciari:2008gp}. We fix a cone size of $R = 0.7$ to cluster a fat jet while varying template sub-cone size $r_{\rm sub} = \{ 0.1, 0.15 \}$. For this analysis, we only use the hardest fat jet with $p_T > 500 \GeV$ and $|\eta|< 2.5$.
\begin{figure}
\includegraphics[scale=0.4]{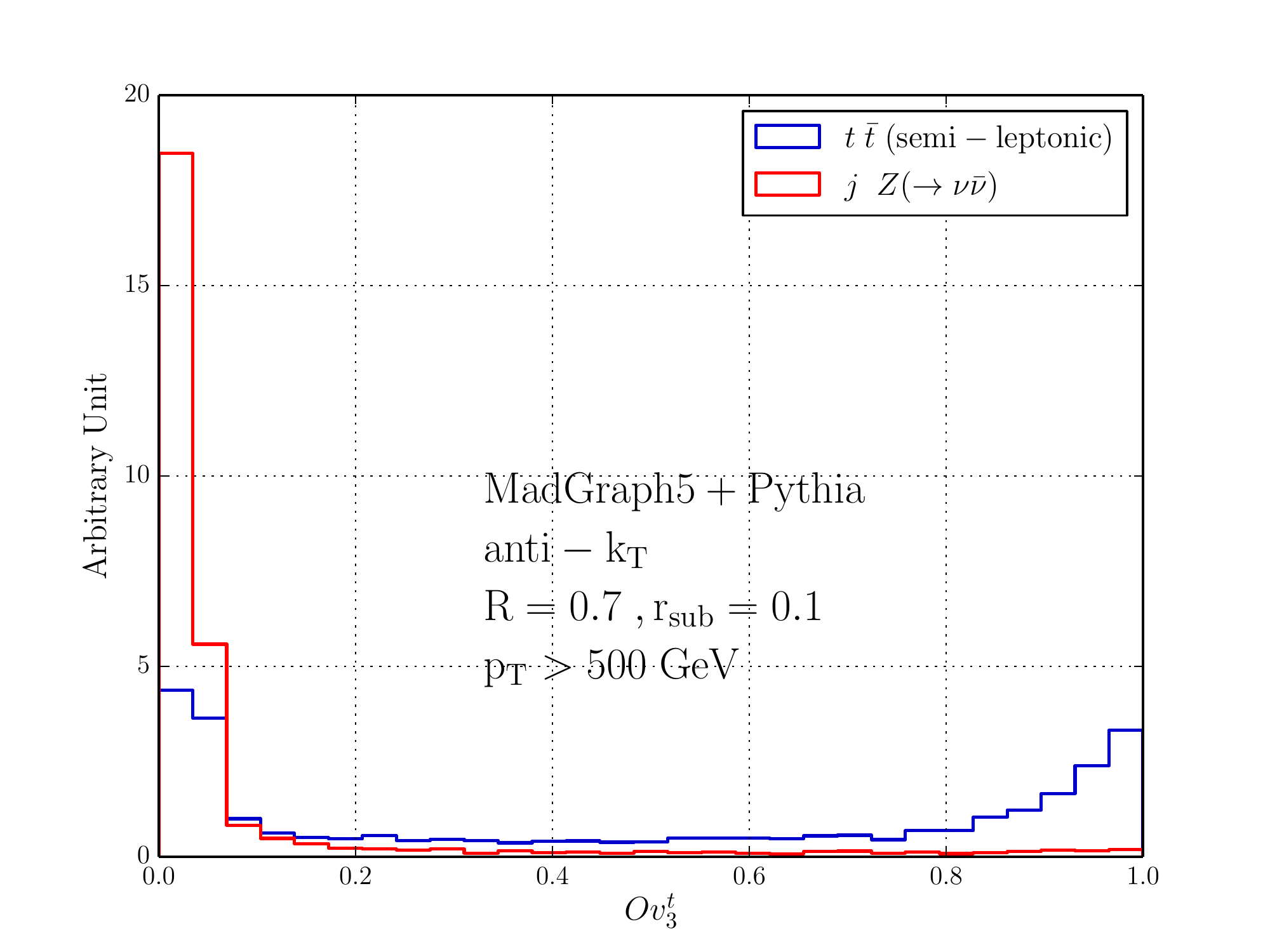} \\[.2cm]
\includegraphics[scale=0.4]{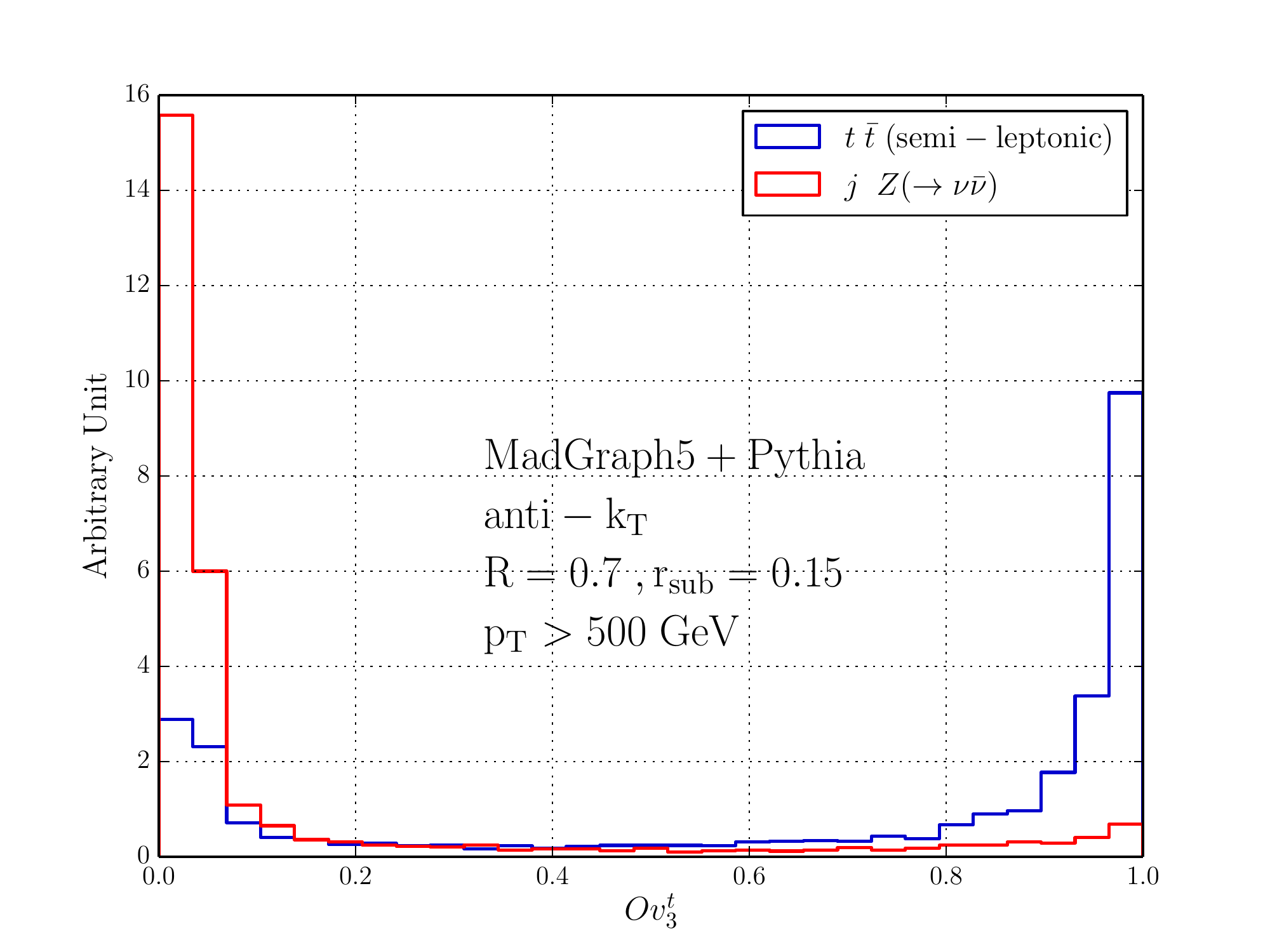}
\caption{$Ov_3^t$ distributions of the hardest $R = 0.7$ fat jet with (top panel) $r_{\rm sub} = 0.1$ and (bottom panel) $r_{\rm sub} = 0.15$ for benchmark event samples: the Standard Model semi-leptonic $t \bar{t}$ process without additional jets and $j Z(\rightarrow \nu \bar{\nu})$. }
\label{fig:Ov}
\end{figure}

\begin{figure*}[t]
\begin{center}
\setlength{\tabcolsep}{0em}
{\renewcommand{\arraystretch}{1}
\begin{tabular}{cc}
\hspace{-20pt} \includegraphics[scale=0.15]{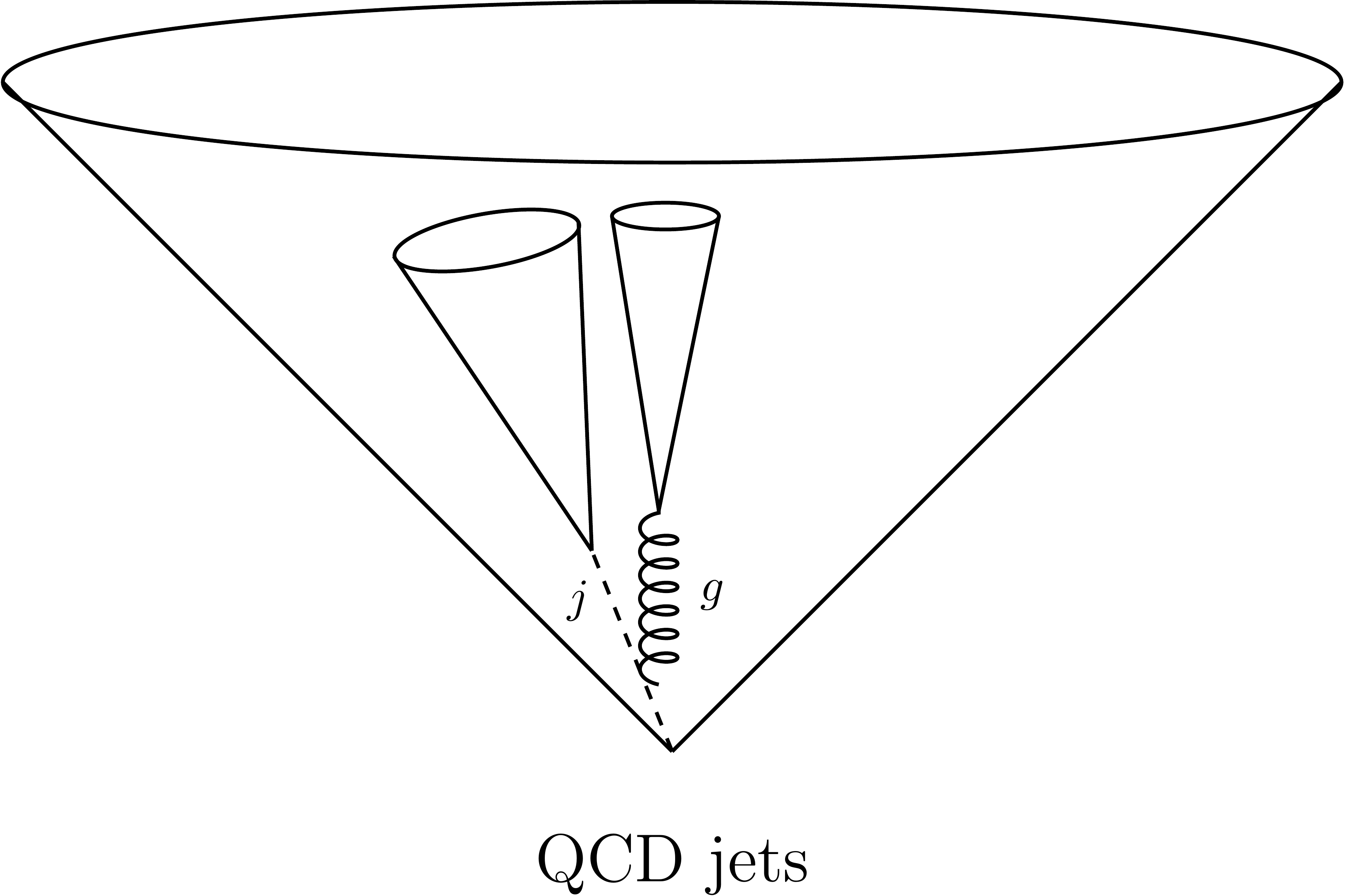} &
\hspace{40pt} \includegraphics[scale=0.15]{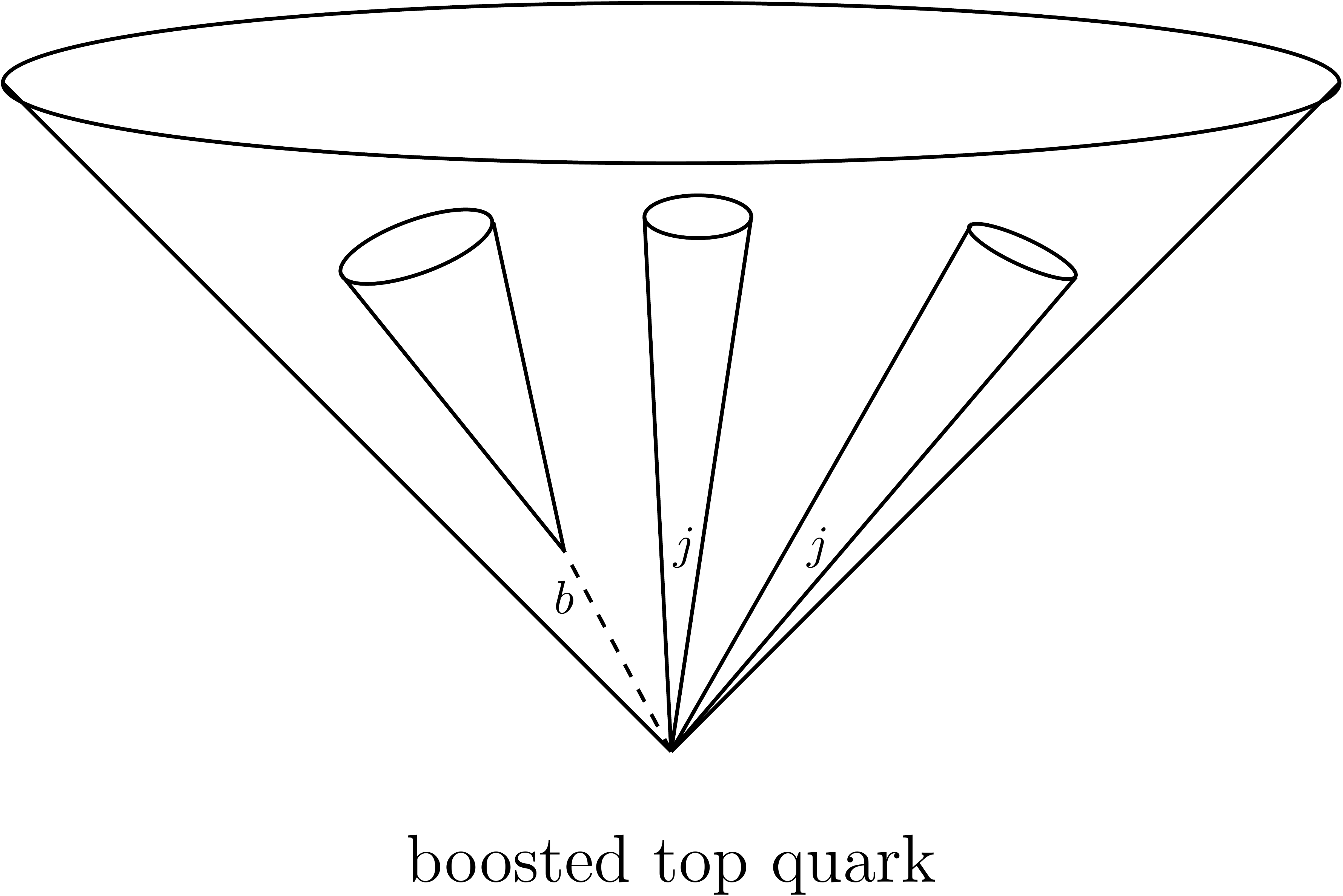}  \\
\end{tabular}
}
\end{center}
\caption{(left) For a typical top-faking QCD jet, there is a $p_T$ hierarchy between the collinear jets, while (right) for a boosted top, its decay products share a symmetric $p_T$ scale with a relatively large angular separation. We can implement the difference of these features into the Template Overlap Method by introducing a new measure $ty$ in Eq.(\ref{ty}).
}
\label{fig:splitting}
\end{figure*}

Figure \ref{fig:Ov} illustrates $Ov_3^t$ distributions of the hardest fat jets with (top panel) $r_{\rm sub} = 0.1$ and (bottom panel) $r_{\rm sub} = 0.15$. First, we see that a significant amount of $j Z(\rightarrow \nu \bar{\nu})$ events is saturated in the region of $Ov_3^t \sim 0$, whereas a sizable portion of semi-leptonic $t \bar{t}$ samples is observed in the region of $Ov_3^t \sim 1$. Such a sharp contrast allows us to disentangle the majority of top-free $j Z(\rightarrow \nu \bar{\nu})$ events by demanding a cut of $Ov_3^t >0.6$. Second, we notice that reducing $r_{\rm sub}$ directly impacts on the signal efficiency in the $t \bar{t}$ samples where approximately half of the population is cut by the top-tagging requirement $Ov_3^t >0.6$. This action does not accompany any extra reduction on the top-faking $j Z(\rightarrow \nu \bar{\nu})$ samples, therefore only harming the signal efficiency.

Although keeping a high signal efficiency is mostly preferred, if we are in the situation where the gigantic QCD background overwhelms the signal rate, then the focus should be directed to reducing the mis-tag rate at the cost of the signal efficiency. In what follows, we propose a new way to reduce the mis-tag rate of QCD jets aiming for an intermediate efficiency by introducing a new measure on the sub-jet level of $r_{\rm sub} \sim 0.1$. TOM has additional degrees of freedom, maximally-matched three-prong top templates with a sub-cone size $r_{\rm sub}$, where one can in principle manipulate them to exploit additional information at the sub-jet level. When a boosted top decays into three jets, they share a symmetric $p_T$ scale with each other. In contrast, for a typical top-faking QCD jet, there is a $p_T$ hierarchy between collinear jets resulting from jet-splitting (see figure \ref{fig:splitting}). We can implement the difference of these features into the Template Overlap Method by introducing a new scale-dependent measure $ty$:
\begin{eqnarray}
	ty &=& \text{min}(p_{Ti}, p_{Tj}) \Delta R_{ij}   
	   = \sqrt{ d_{i j} R^2} \, ,
\label{ty}
\end{eqnarray}
where the template-prong indices, $i$ and $j$, denote
the pair of template-prongs with the smallest angular distance among maximally-matched three template prongs,
and 
$d_{i j} =  \text{min}(p^2_{Ti}, p^2_{Tj}) \Delta R^2_{ij} / R^2$.

\begin{figure}
\includegraphics[scale=0.4]{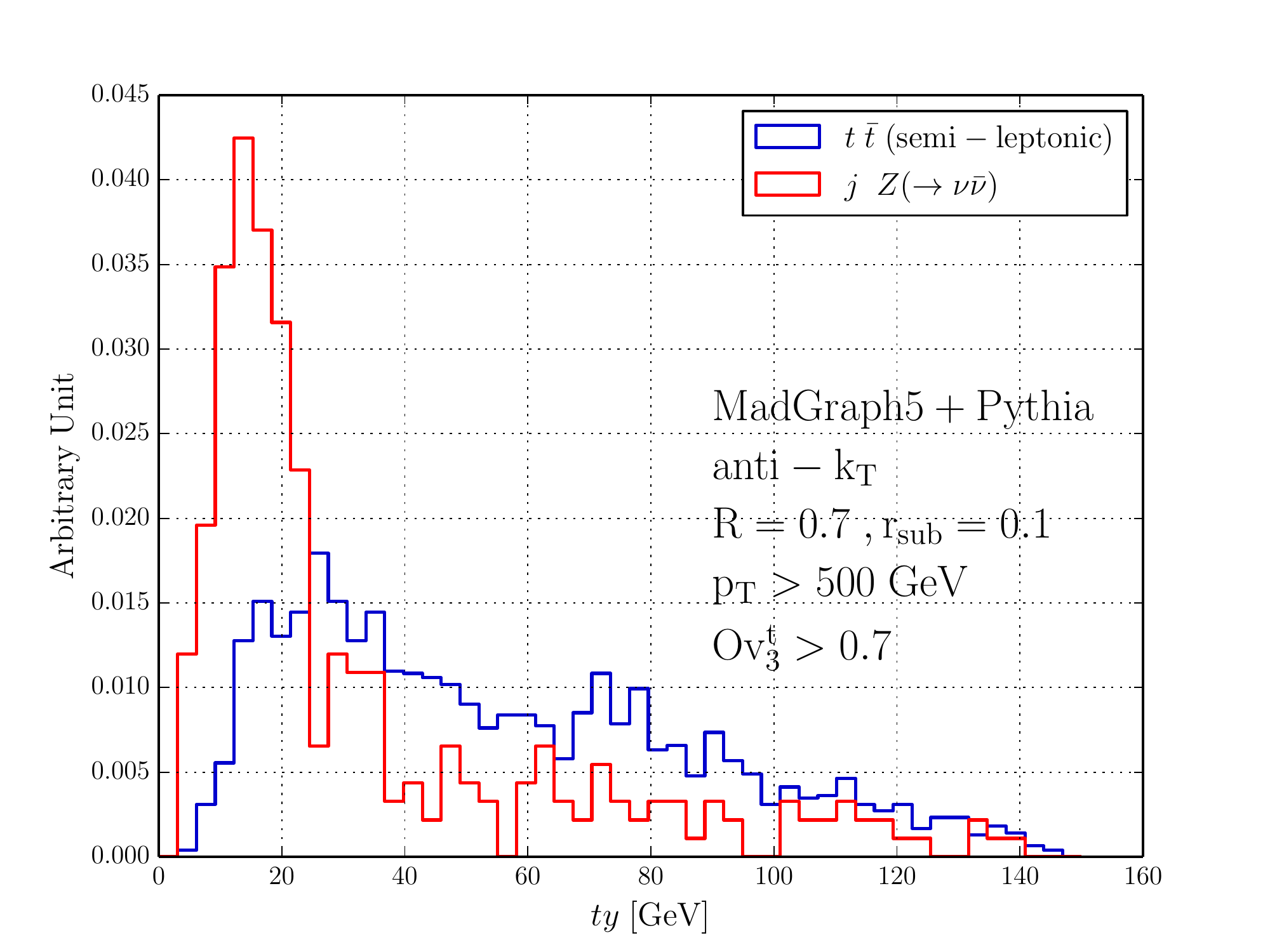} \\[.2cm]
\includegraphics[scale=0.4]{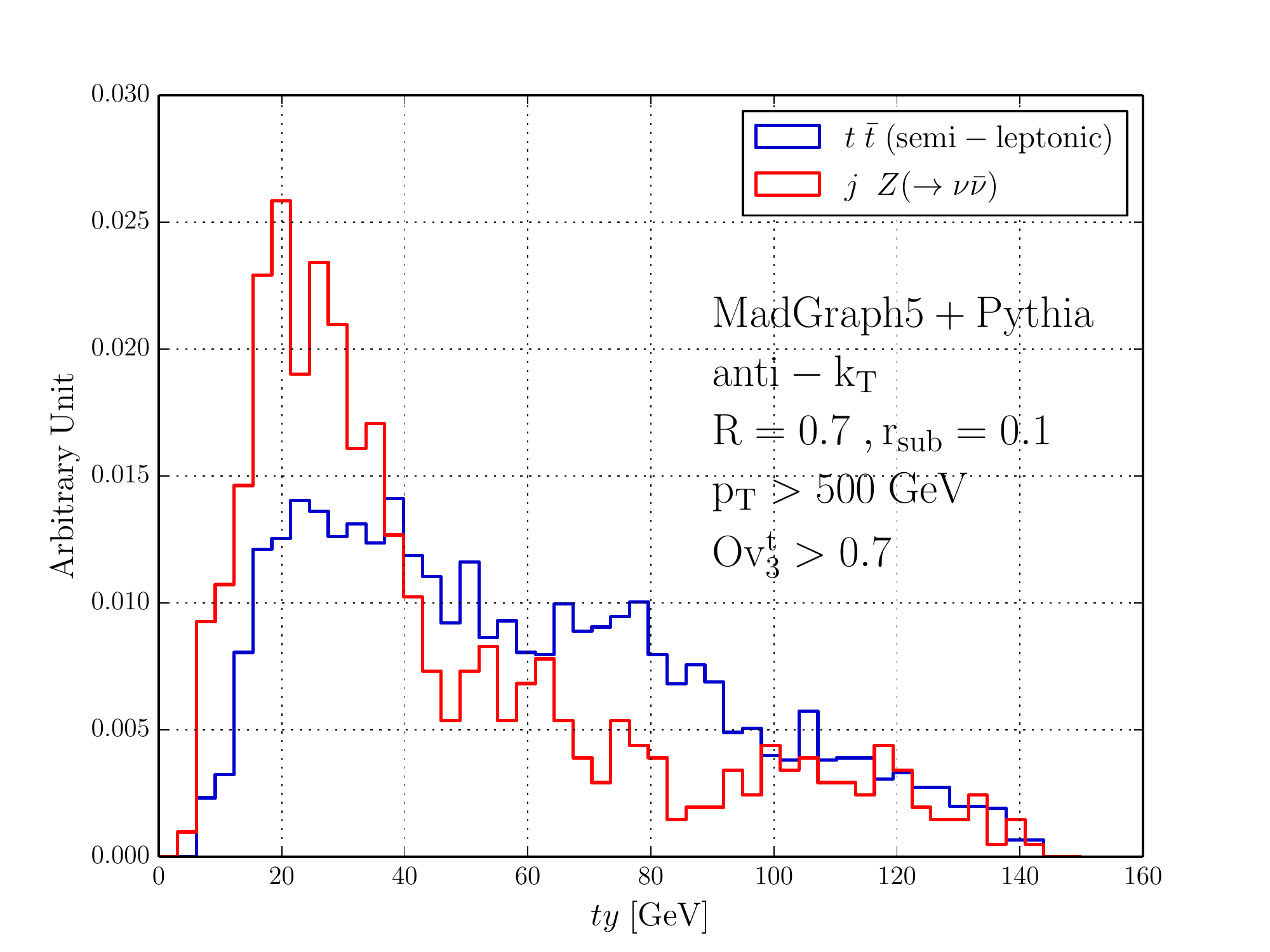}
\caption{$ty$ distributions of the hardest top-tagged $R = 0.7$ fat jet with (top panel) $r_{\rm sub}=0.1$ and (bottom panel) $r_{\rm sub}=0.15$. }
\label{fig:ty}
\end{figure}

Figure \ref{fig:ty} shows $ty$ distributions of the hardest top-tagged fat jet with (top panel) $r_{\rm sub}=0.1$ and (bottom panel) $r_{\rm sub}=0.15$. We observe a stark difference at $r_{\rm sub}=0.1$ level where the generic $ty$ scale of the top-free $j Z(\rightarrow \nu \bar{\nu})$ samples is much lower than the top-containing $t \bar{t}$ signal events. It renders an additional handle to suppress $j Z(\rightarrow \nu \bar{\nu})$ by demanding a $ty$ cut of $20 \sim 30 \GeV$ on top of the prior $Ov$-selection. 

In order to quantify the efficiency and mis-tag rate, let $n_{\rm t}$ and $n_{\rm j}$ be a number of $t \bar{t}$ and $j Z(\rightarrow \nu \bar{\nu})$ events respectively containing at least one $R = 0.7$ fat jet with $p_T > 500 \GeV$ and $|\eta|< 2.5$. Let $n_{\rm t}'$ and $n_{\rm j}'$ be a number of surviving $t \bar{t}$ and $j Z(\rightarrow \nu \bar{\nu})$ events respectively in which the hardest fat jet passes the boosted top selection. It is convenient to define the efficiency and mis-tag rate by
\begin{equation}
	 \rm{Eff} = \frac{n_{\rm t}'}{n_{\rm t}}   \;\;,\;\;  \rm{Mistag} = \frac{n_{\rm j}'}{n_{\rm j}}   \; ,
\end{equation}
where the details of the top selection scheme and corresponding Eff (Mistag) are summarized in Table \ref{tab:EffTable}. We find that the prior $Ov$-selection ($Ov_3^t > 0.7$) combined with $ty > 25 \GeV$ can achieve an efficiency of $\sim 33 \%$ with  mis-tag rate of $\sim 1.5 \%$.
\begin{table}[h]
\begin{center}
\scalebox{1.0}{
\begin{tabular}{|c|c|c|}
\hline
		                                    	& $r_{\rm sub}=0.1$        &    $r_{\rm sub}=0.15$   \\ \hline	
$Ov_3^t > 0.7$                                  & $41 \%$  ($4.0 \%$)      & $64 \%$  ($9.0 \%$) 	       \\ \hline
$Ov_3^t > 0.7$ , $ty > 5 \GeV$         &  $41 \%$  ($4.0 \%$)     & $64 \%$  ($9.0 \%$)                \\ \hline
$Ov_3^t > 0.7$ , $ty > 10 \GeV$      &  $40 \%$  ($3.6 \%$)     & $63 \%$  ($8.6 \%$)                \\ \hline
$Ov_3^t > 0.7$ , $ty > 15 \GeV$      &  $38 \%$  ($2.7 \%$)     & $61 \%$  ($8.1 \%$)                \\ \hline
$Ov_3^t > 0.7$ , $ty > 20 \GeV$      &  $36 \%$  ($2.0 \%$)     & $58 \%$  ($7.0 \%$)                \\ \hline
$Ov_3^t > 0.7$ , $ty > 25 \GeV$      &  $33 \%$  ($1.5 \%$)     & $53 \%$  ($6.1 \%$)                \\
\hline		
\end{tabular}}
\end{center}
\caption{The details of the top selection scheme and corresponding Eff (Mistag).}
\label{tab:EffTable}
\end{table}

A complete analysis aiming to find an optimal efficiency and mis-tag rate is left for future work. In this paper, we find it useful to eliminate the most dominant semi-leptonic $t\bar{t}$ + jets background in the semi-leptonic channel. The direct influence of the $ty$ cut into practice will be discussed in Section \ref{sec:semi}. In the fully-hadronic and SSDL channels, however, dominant backgrounds turn out to be top-rich processes such as $t\bar{t}$ + jets or $t\bar{t}t\bar{t}$, and therefore we will not apply it into these analyses.

Finally, with regard to our special treatment to mitigate the ISR effects, we combine jet trimming~\cite{Krohn:2009th} with TOM. The jet trimming technique reclusters a fat jet and creates a sub-jet of size $r'_{\rm sub}$ inside, but remove those which fall into $p^i_T / p^{\rm fj}_T  < f_{\rm cut}$ where $p^i_T$ and $p^{\rm fj}_T$ stand for $p_T$ of the $i^{th}$ sub-jet and a fat jet respectively. 
As a consequence, it typically reduces a fat tail in the fat jet invariant mass distribution, and renders a cleaner environment for TOM to undertake the process of the top-jet identification. In this analysis, therefore, all fat jets are subject to the trimming-process with the optimized cut parameters of $r'_{\rm sub} = 0.25$  and  $f_{\rm cut} = 0.05$.

\subsection{$b$-tagging}
\label{sec:b-tagging}

Multiple $b$-tagging plays a central role since we require two $b$ jets from the boosted top decays and one or two additional $b$ jets from the spectator tops. Due to its large impact on the resulting sensitivity, a careful assessment on the $b$-tagging efficiency and associated jet-faking rate is required.

In our semi-realistic $b$-tagging procedure, we assign a $b$-tag to each $r=0.4$ jet if there is a parton level $b$ or $c$ quark within $\Delta r=0.4$ from the jet axis, and we assume a $b$-tagging efficiency of
\begin{equation}
	\epsilon_b = 0.70, \,\,\,\, \epsilon_c = 0.20, \,\,\,\,\, \epsilon_j = 0.01\,, 
\end{equation}
where $\epsilon_{b, c, j}$ are the efficiencies that a $b$, $c$ or a light jet will be tagged as a $b$-jet. 
We note that in recent ATLAS analysis (Ref. \cite{ATLAS13TeV}), the following b-tagging efficiencies are used
$\epsilon_b = 0.77$, $\epsilon_c = 0.22$, and $\epsilon_j = 0.0079$, which are slightly better than what we have used in our analysis.

For a fat jet to be $b$-tagged, we require that a $b$-tagged $r=0.4$ jet lands within $\Delta R = R$ from the fat jet axis, where $R$ is the size of a fat jet. We take into account that more than one $b$-jet might land inside the fat jet, whereby we reweigh at least $1b$-tagging efficiencies of a fat jet depending on the $b$-tagging scheme described in Table \ref{tab:btag}.
\begin{table}[htb]
\begin{center}
\scalebox{0.85}{
\begin{tabular}{|c|c|c|}
\hline
	$b$-tagged score	                 &  Efficiency  (at least 1 $b$-tag)                                                                                                                               & value         \\ \hline
 	 0 (jet: u,d,s,g)				& 	 	$\epsilon_j$				                                                                                                                &  0.01           \\  \hline
	1 (1c)					& 	 	$\epsilon_c$					                                                                                                        &  0.20	       \\  \hline
	2 (2c)					& 	 	$2\, \epsilon_c (1-\epsilon_c)+ {\epsilon_c}^2$		                                                                               &  0.36      \\  \hline
	3 (1b)					& 	 	$\epsilon_b$					                                                                                                         &  0.70	     \\  \hline
	4 (1b+1c)					& 	 	$\epsilon_b (1-\epsilon_c) + \epsilon_c (1-\epsilon_b)+\epsilon_b \epsilon_c$	                                     & 	0.76	     \\  \hline
	\multirow{2}{*}{ 5 (1b+2c)}		& 	 	$\epsilon_b (1-\epsilon_c)^2 + 2 (1-\epsilon_b) (1-\epsilon_c) \epsilon_c$ 		                                      & \multirow{2}{*}{0.81}   \\  
							&		$ +  2 \epsilon_b \epsilon_c (1- \epsilon_c) + \epsilon_c^2 (1-\epsilon_b) +\epsilon_b {\epsilon_c}^2$	    &	                                \\ \hline
	6 (2b)					& 	 	$2 \epsilon_b (1-\epsilon_b)+ {\epsilon_b}^2$			                                                                       &   0.91		   \\  \hline
	7 (2b+1c)					& 	 	$1 - (1-\epsilon_c) (1-\epsilon_b)^2$				                                                                                &   0.93		   \\  \hline
	8  (2b+2c)                     		& 	 	$1 - (1-\epsilon_c)^2 (1-\epsilon_b)^2$                                   			                                              &  0.94	   \\  \hline
	9  (3b)					& 	 	$1 - (1-\epsilon_b)^3$						                                                                                 &  0.97                         \\  \hline					 
\end{tabular}\par}
\caption{Efficiency that a top-tagged fat jet will be $b$-tagged assuming that it contains a specific number of light, $c$ or $b$ jets within $\Delta R = R$ from the jet axis, where $R$ is a size of a fat jet. $\epsilon_j$, $\epsilon_c$ and $\epsilon_b$ are $b$-tagging efficiencies for light, $c$ and $b$ jets respectively. We neglect the possibilities beyond three proper $b$-tagged jets within a fat jet.} \label{tab:btag} 
\end{center}
\end{table}

\section{Searches For A Top-Philic Resonance at the LHC14}\label{sec:results}

\subsection{Fully-hadronic Channel}
\label{sec:had}
The fully-hadronic channel derives benefit from a large branching ratio of $\sim 20 \%$, but receives enormous contamination from the QCD background which is orders of magnitude larger than the signal. Using boosted hadronic top-taggers in conjunction with a multiple b-tagging, however, it is possible to reduce the QCD background to a manageable level. What remains to be most difficult is to suppress the $t\bar{t}$ + jets process which contains two proper hadronic tops with a sizable cross section. It further necessitates an introduction of additional handles such as jet-multiplicity and $M_J$ as in Eq. (\ref{eq:MJ}) to improve the sensitivity of the channel.

The dominant SM backgrounds are irreducible $t\bar{t}$ + jets with up to two additional jets (including b jets) and $t\bar{t} t\bar{t}$. Subdominant backgrounds include the QCD processes where we include multi-jet\footnote{i.e. up to four light-flavour jets}, $b\bar{b}$ + jets and $b\bar{b} b\bar{b}$ in our simulation. We also consider $Z_{b \bar{b}}$ + jets with up to two additional jets (including b jets) when a $Z$ boson decays into $b \bar{b}$. The single top quark process $t \bar{b}$ + jets with up to two additional jets gives a negligible contribution.

We generate a signal and all backgrounds with the pre-selection cuts described in section \ref{sec:simulation} requiring $H_T > 850 \GeV$ to improve the statistics. Table~\ref{tab:TotalBackGroundsHad} summarizes the background cross sections including a conservative NLO K-factor of 2. 
\begin{table}[h]
\begin{center}
\scalebox{1.0}{
\begin{tabular}{|c|c|c|}
\hline
		Channels								 &	Backgrounds              & $\sigma (H_T > 850\GeV) [{\rm fb}]$ \\ \hline	
\multirow{5}{*}{Fully-hadronic}                                              &$t\bar{t} t\bar{t}$              &	$  3.1 $                          	              \\
                                                                                              &$t\bar{t}$ + jets                &	$  2.6    \times 10^4 $	              \\
											 &$t \bar{b}$ + jets              &    $ 2.8     \times 10^3 $	              \\
											 & QCD                               &     $ 4.2   \times 10^6 $	                      \\
											 &$Z_{b \bar{b}}$ + jets       &	$  3.4    \times 10^3 $	              \\
\hline		
\end{tabular}}
\end{center}
\caption{The simulated cross sections of SM backgrounds (including a conservative estimate of NLO K-factor of 2 after preselection cuts described in section \ref{sec:simulation}).}
\label{tab:TotalBackGroundsHad}
\end{table}

All events are subject to pass $Basic \; Cuts$ of requiring at least two fat jets ($R=0.7$) with $p_T^{\rm fj} > 500 \GeV$ and $|\eta_{{\rm fj}} |< 2.5$ (Table~\ref{tab:BasicCutshad} for summary), which are then trimmed subsequently according to the rule described in Section \ref{sec:TOM}.
The specific ditop selection (Table \ref{tab:Ditophad}) begins with the overlap analysis applied to all trimmed-fat jets. We demand at least two top jets (\ie~trimmed-fat jets satisfying $Ov$-selection criterion, $Ov_{3}^t > 0.6$), and identify the first two hardest tops as the candidates from a resonance decay.
\begin{table}[h]
\begin{center}
\setlength{\tabcolsep}{1em}
{\renewcommand{\arraystretch}{1.5}
\begin{tabular}{c|c}
							& Fully-hadronic                                          	    	  \\ \cline{1-2}
\multirow{2}{*}{ \textbf{Basic Cuts} }	& $N_{\rm fj}  \geq 2$ ($R=0.7$) ,                                 \\ 
		 					& $p_T^{\rm fj} > 500 \GeV$, $|\eta_{\rm fj} |< 2.5$ 
\end{tabular}}\par
\caption{Summary of Basic Cuts for the fully-hadronic channel. ``fj" stands for the fat jet.} \label{tab:BasicCutshad} 
\end{center}
\end{table}
\begin{table}[h]
\begin{center}
\setlength{\tabcolsep}{1em}
{\renewcommand{\arraystretch}{2.0}
\begin{tabular}{c|c}
							        & Fully-hadronic                                          	    	  \\ \cline{1-2}
 \textbf{Ditop Selection}                             & $N_{t} \geq 2$, (for $t$: $Ov_3^t > 0.6$)                                  \\
\end{tabular}}\par
\caption{Ditop Selection for the fully-hadronic channel. ``$Ov$'' selection applies to the trimmed-fat jets ($R=0.7$) in the event, and $N_{t}$ is the number of top-tagged fat jets. } \label{tab:Ditophad} 
\end{center}
\end{table}

\begin{figure}
\includegraphics[scale=0.39]{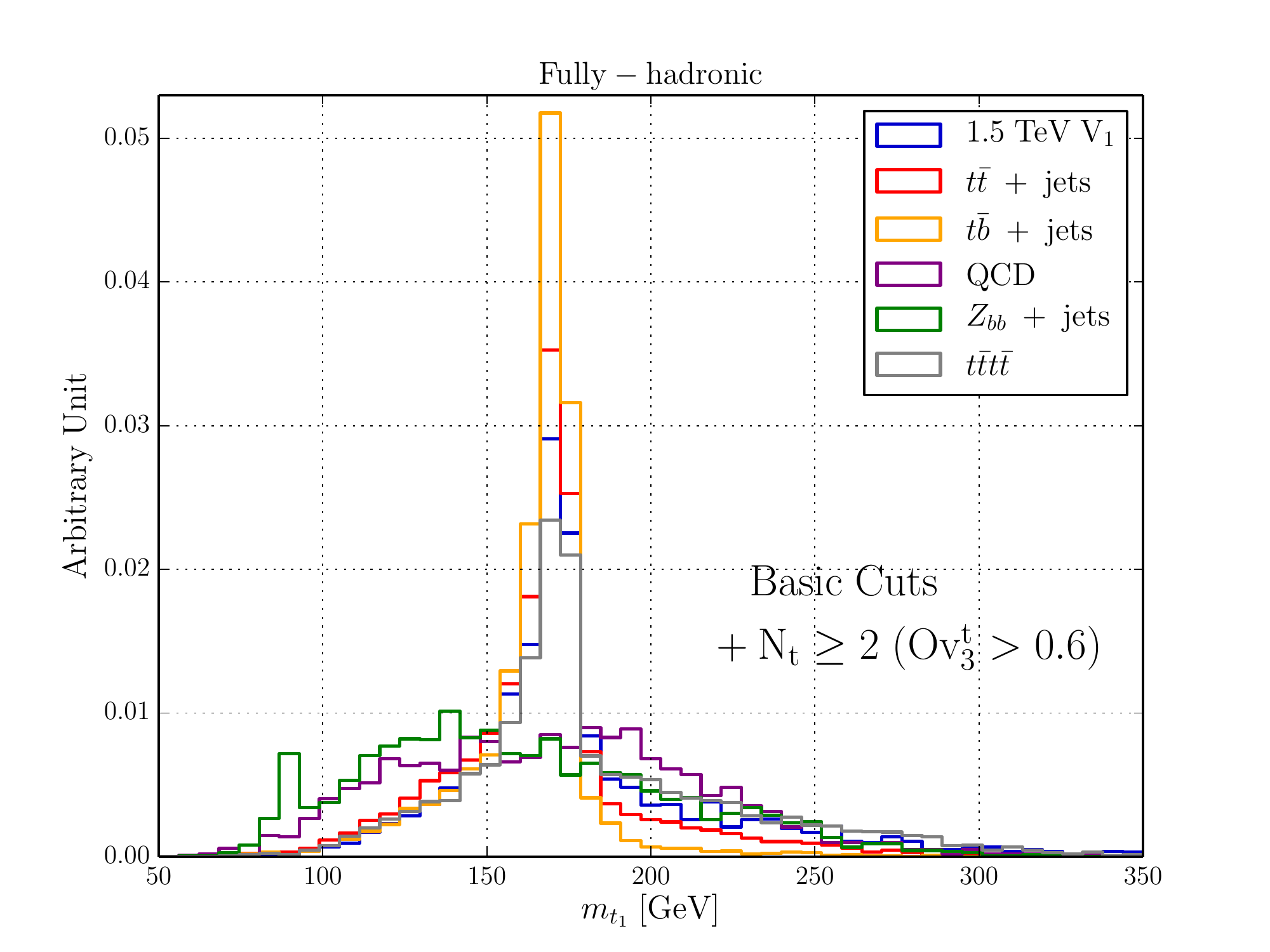} \\
\includegraphics[scale=0.39]{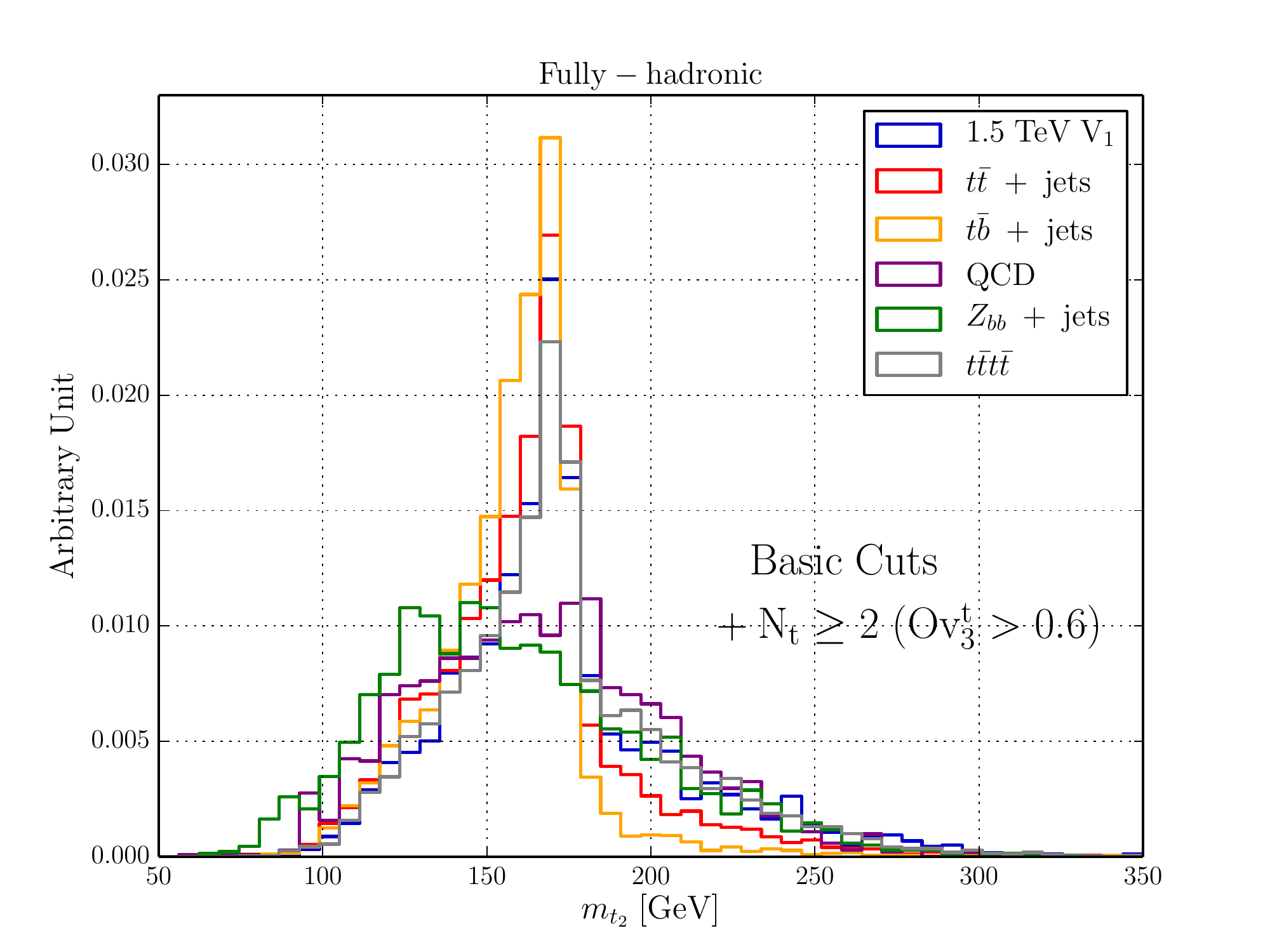} \\
\includegraphics[scale=0.39]{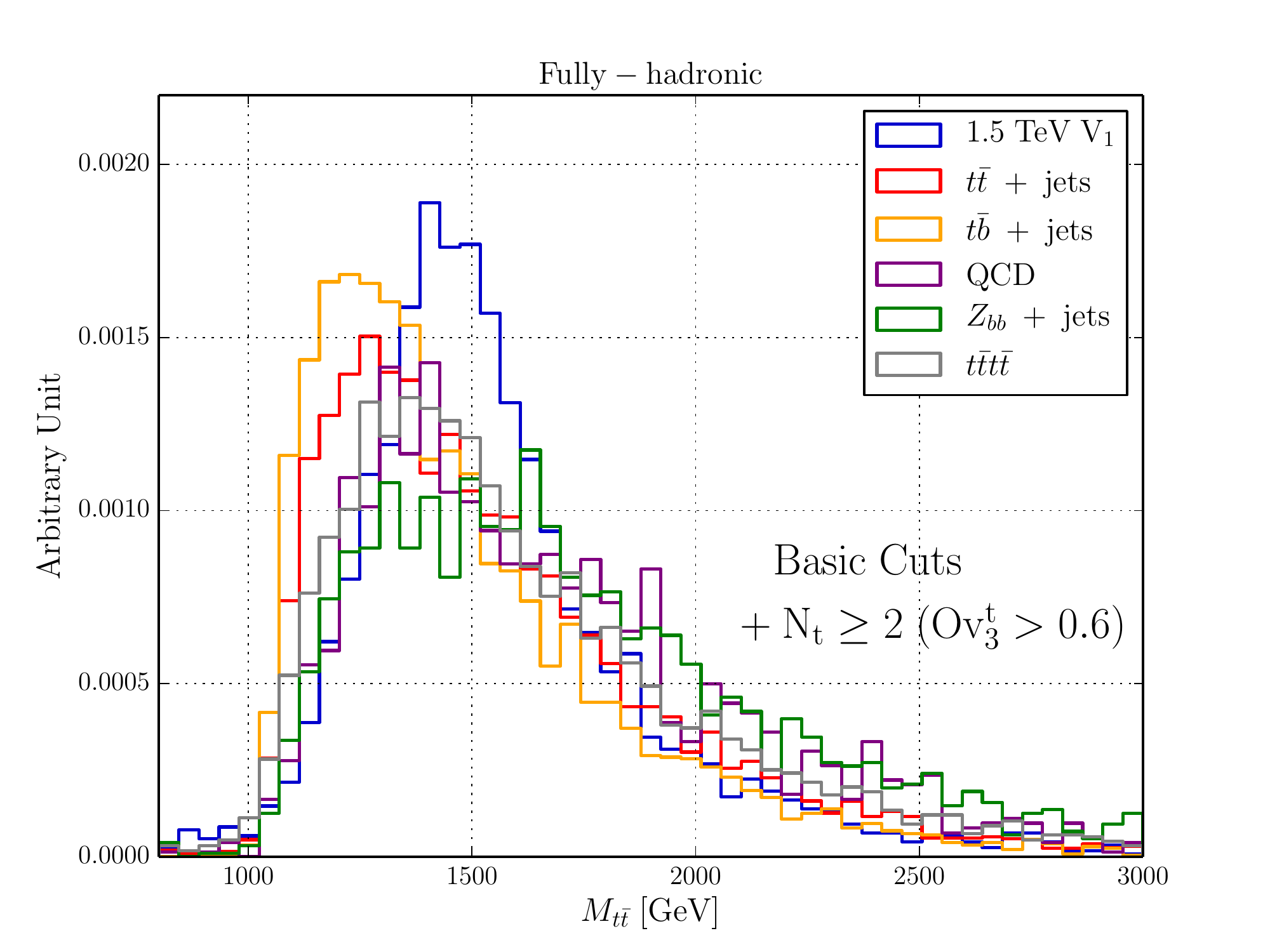}
\caption{Invariant mass distributions of the (top panel) first and (middle panel) second hardest top jets after the Basic Cuts and the Ditop Selection. The $1.5 \TeV$ $V_1$ resonance is then reconstructed using the boosted ditop system in the bottom panel.}
\label{fig:m_top}
\end{figure}

Figure \ref{fig:m_top} (top and middle panels) shows invariant mass distributions of the first two hardest top jets after Basic Cuts and the ditop selection. The top-containing signal and SM backgrounds peak at the physical top mass, while the QCD background is spread out over the wide mass range. The $1.5 \TeV$ $V_1$ resonance is then reconstructed using the boosted ditop system shown in the bottom panel.

\begin{figure}[htb]
\includegraphics[scale=0.39]{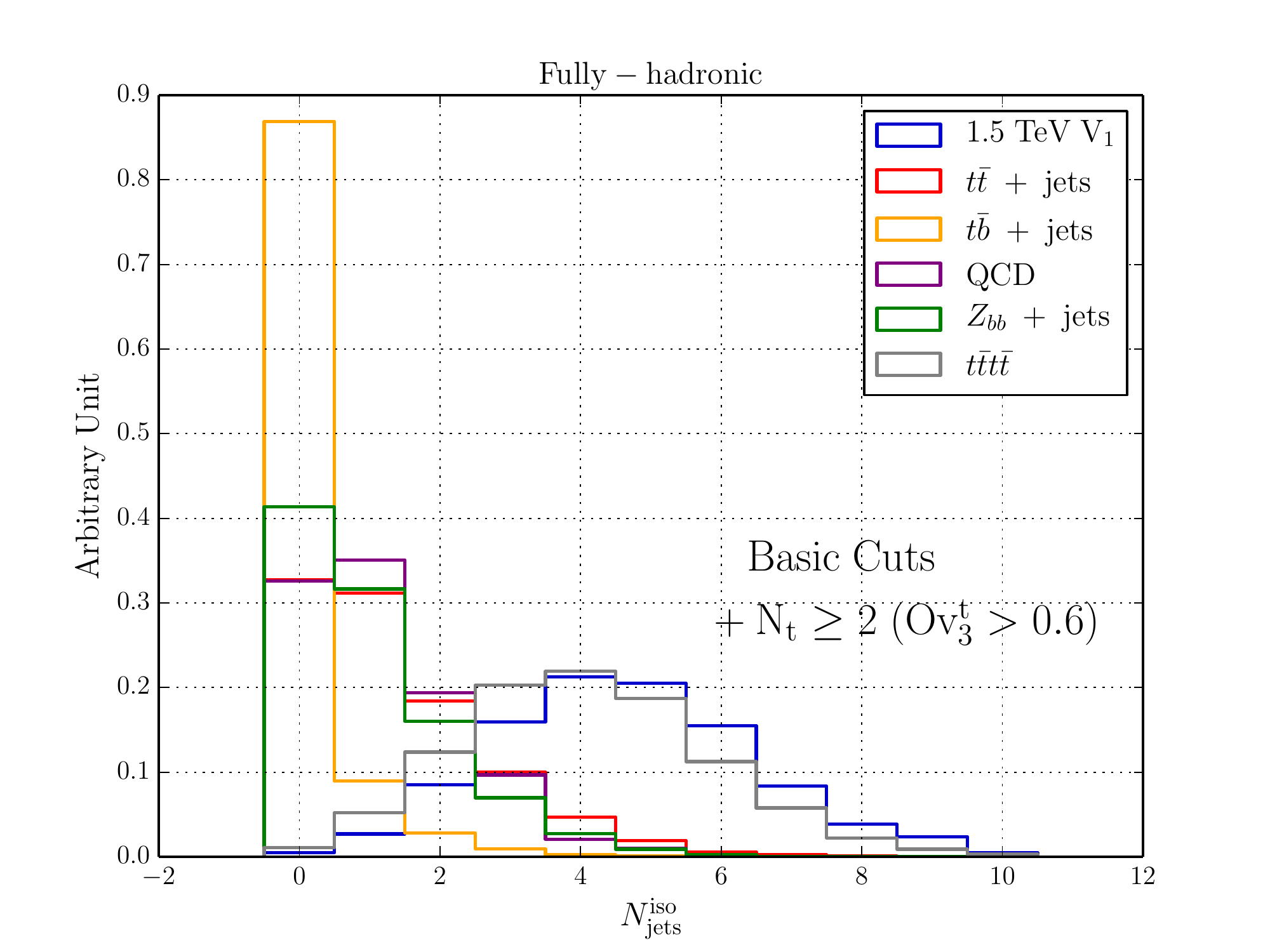} \\
\includegraphics[scale=0.39]{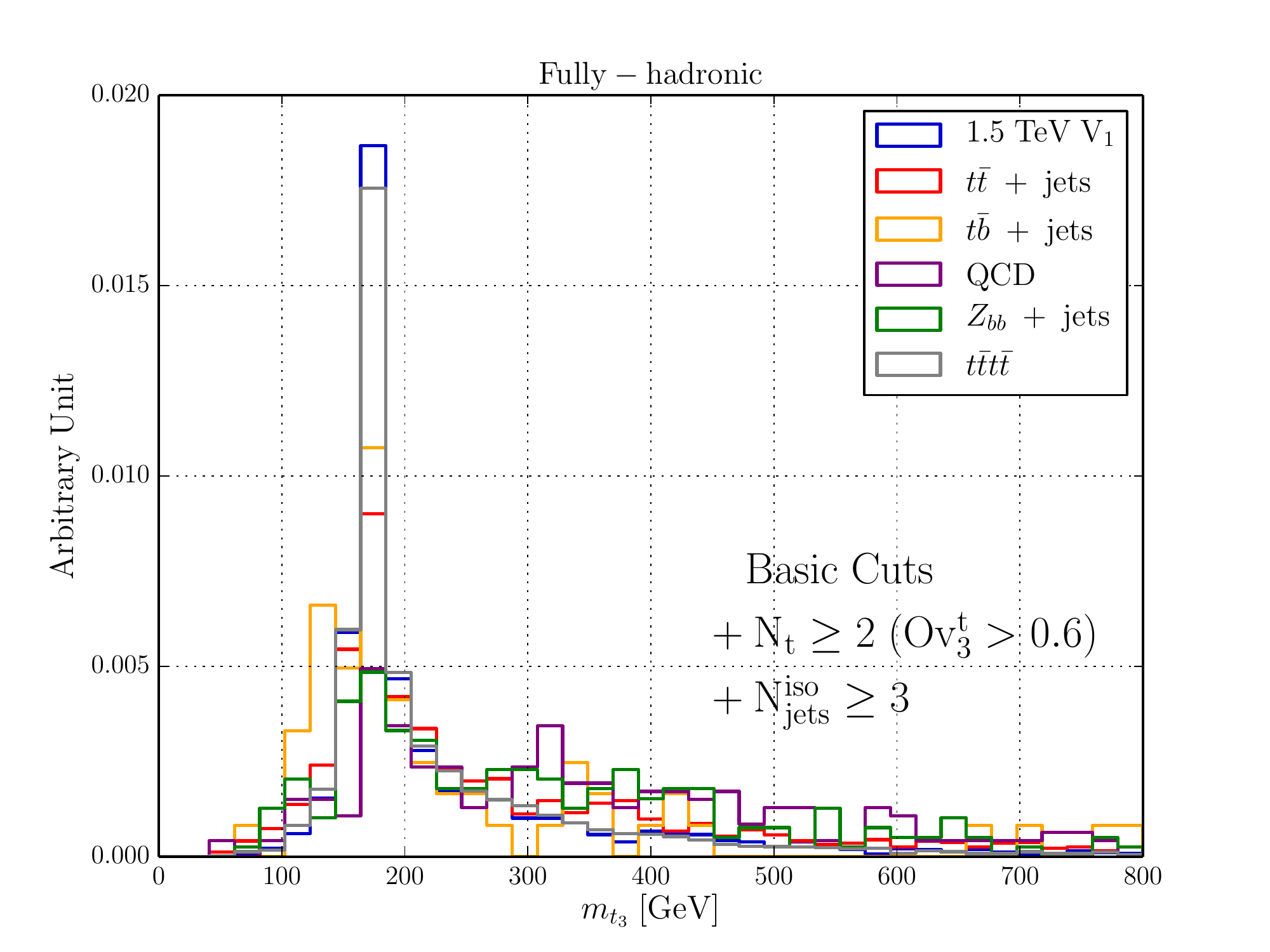} \\
\includegraphics[scale=0.39]{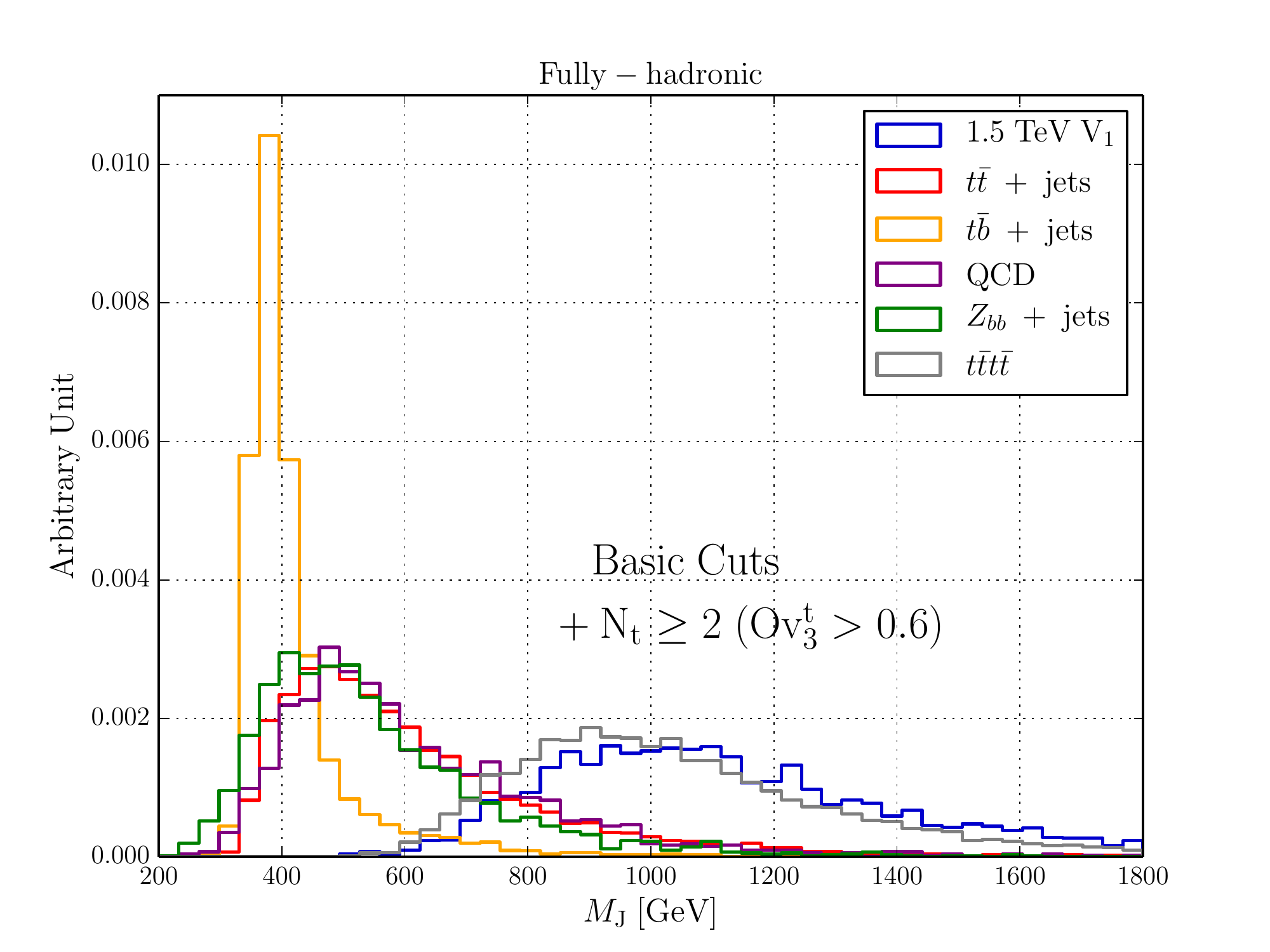} 
\caption{Various distributions of (top panel) a number of isolated jets after Basic Cuts and ditop selection, (middle panel) an invariant mass of three isolated jets which minimizes $\chi^2$ defined in Eq.\ref{eq:chi2}, and (bottom panel) $M_J$ scalar sum of the masses of large radius ($R=1.5$) jets.}
\label{fig:Iso}
\end{figure}

The complexity of the signal delivers additional handles for reducing the backgrounds. Typically a number of isolated $r = 0.4$ jets with $p_T^{\rm j} > 25 \GeV$ and $|\eta_{{\rm j}} |< 2.5$ that are isolated from the top-tagged fat jets (i.e. $\Delta R_{\rm j, t_{1,2}} > 1.1$ ) is limited in SM backgrounds, hence showing a sharp contrast with the signal distribution in Figure \ref{fig:Iso} (top panel). This enables us to disentangle the substantial amount of the backgrounds from the signal by demanding $N^{\rm iso}_{\rm jets}\geq 4$.

The high-multiplicity final states provide a further way to remove the backgrounds. We consider the scalar sum of the masses of large radius ($R=1.5$) jets \cite{CMS:2015uqt,Hook:2012fd,Cohen:2012yc,Hedri:2013pvl}
\begin{equation}
M_J = \sum\limits_{J_i=\rm{large \; R \; jets }} m(J_i) \; .
\label{eq:MJ}
\end{equation}
Typically, a jet mass generated by a parton shower receives a suppression factor of $\alpha_s$, whereas a jet tends to acquire a higher mass when it is formed from partons through the decay of heavy objects. Figure \ref{fig:Iso} (bottom panel) demonstrates an $M_J$ distribution of signal events that is well-separated from SM backgrounds. We can achieve a high background rejection power by demanding $M_J > 900 \GeV$.

On the possibility of reconstructing additional spectator tops, we can use jets clustered with a cone size of $r = 0.4$ which resemble partons from the hadronically-decaying non-boosted spectator tops. Since not all isolated jets fall into the central region (see Figure \ref{fig:parton}), we can reconstruct only one spectator top using three properly-selected isolated jets. These three jets are selected such that they minimize the value of $\chi^2$ among all possible combinations, where $\chi^2$ is defined by
\begin{equation}
\chi^2 = \frac{(m_{\rm{j j j}} - m_t)^2}{\Gamma^2_t} + \frac{(m_{\rm{j j}} - m_W)^2}{\Gamma^2_W},
\label{eq:chi2}
\end{equation}
with $m_t = 172 \GeV$, $m_W = 80 \GeV$, $\Gamma_t = 1.5 \GeV$ and $\Gamma_W = 2.1 \GeV$.

\begin{figure*}[ht]
\begin{center}
\setlength{\tabcolsep}{0em}
{\renewcommand{\arraystretch}{1}
\begin{tabular}{cc}
\hspace{-20pt} \includegraphics[scale=0.39]{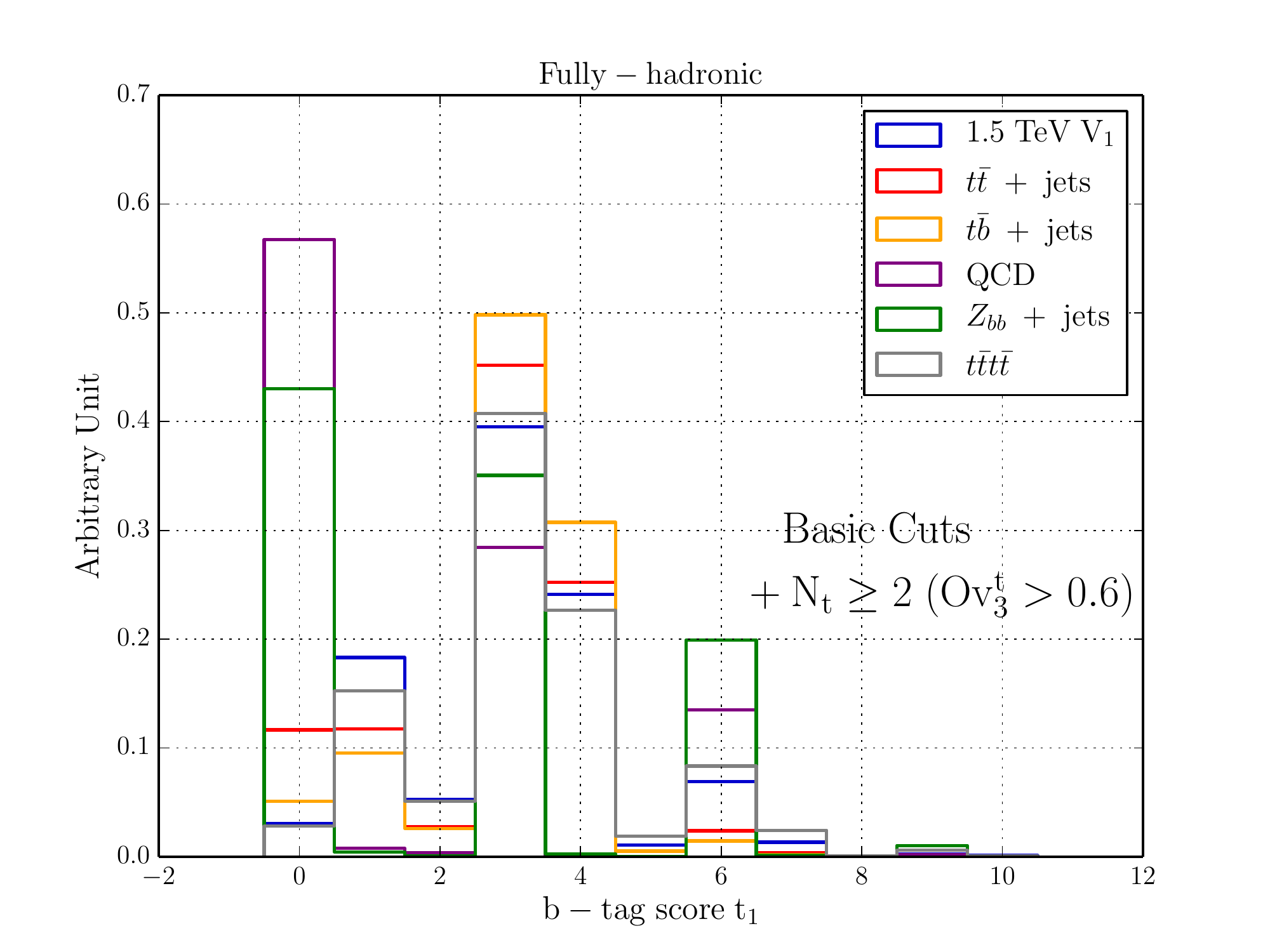} &
\hspace{20pt} \includegraphics[scale=0.39]{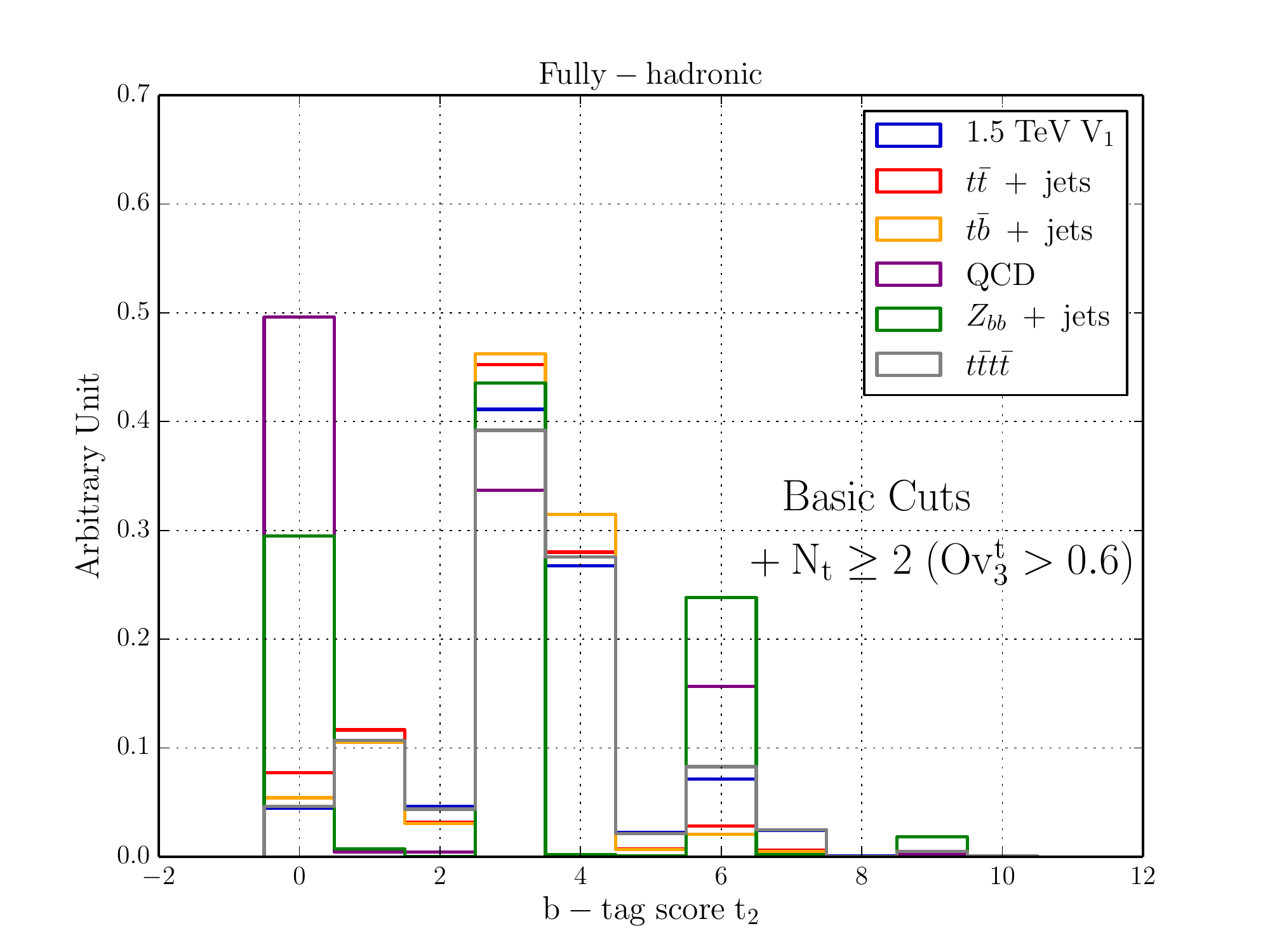}  \\
\hspace{-20pt} \includegraphics[scale=0.39]{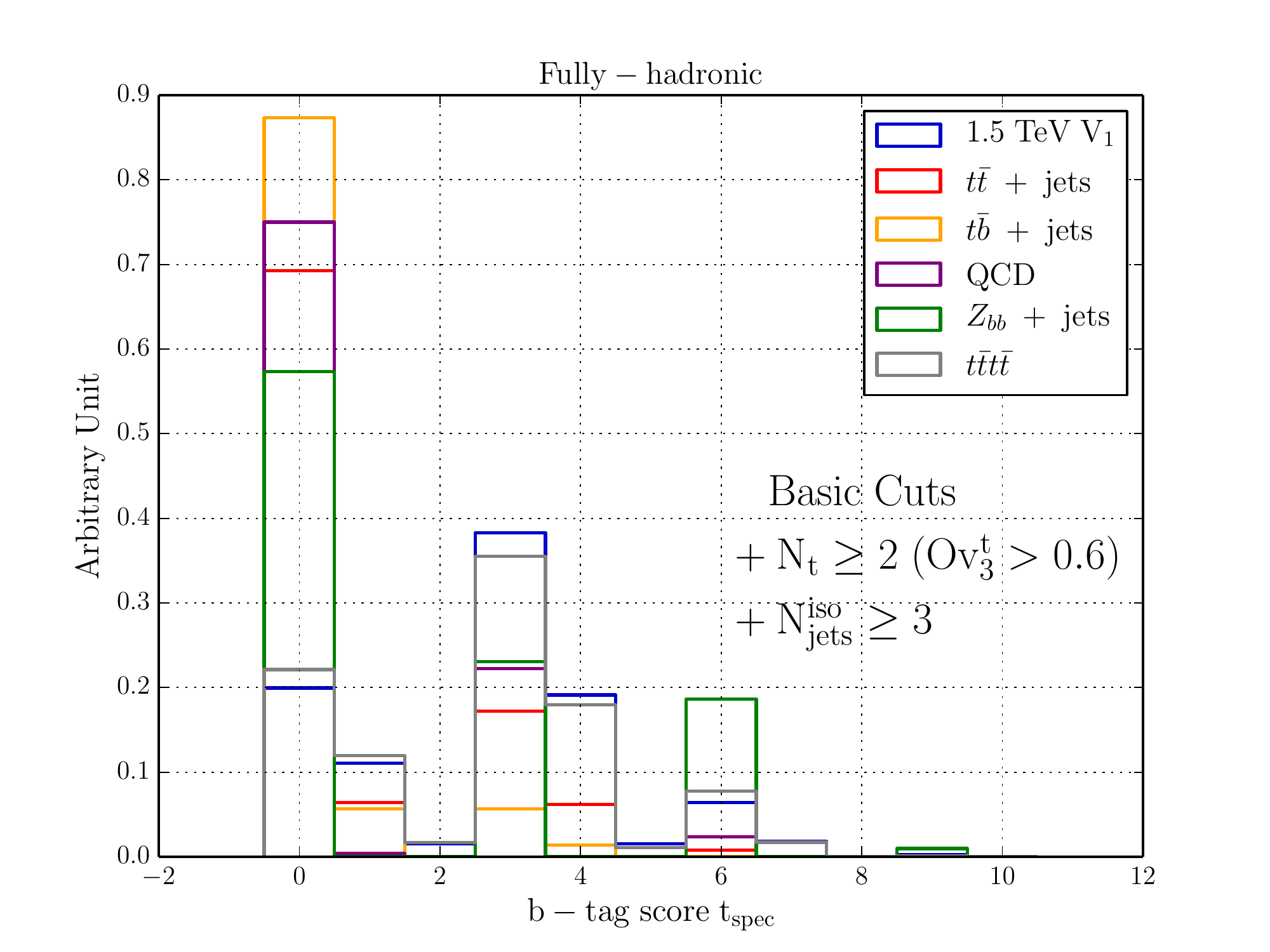} &
\hspace{20pt} \includegraphics[scale=0.39]{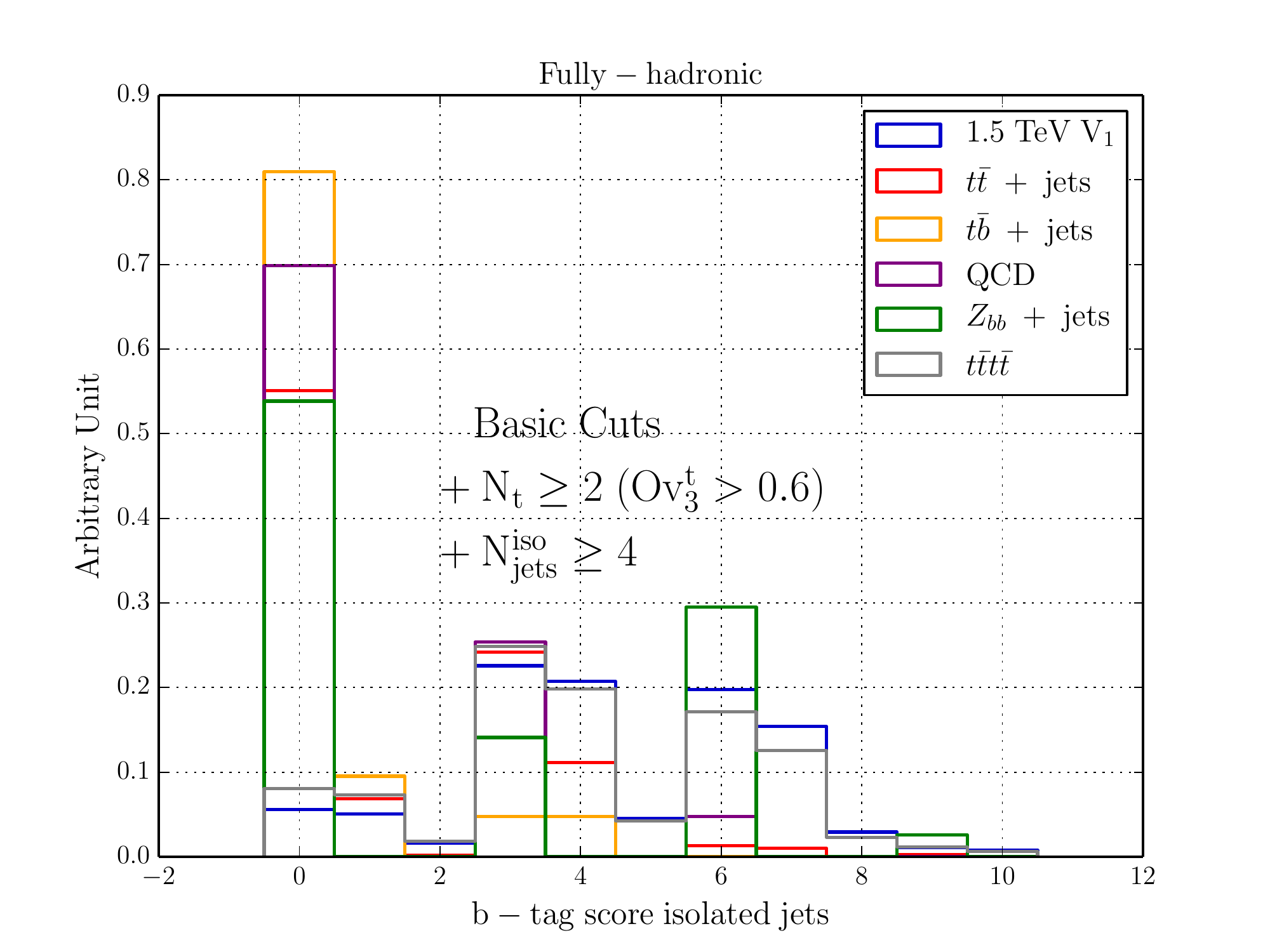}  
\end{tabular}
}
\end{center}
\caption{$b$-tag scores of (upper-left panel) the first and (upper-right panel) the second hardest top jets,  (lower-left panel) the reconstructed spectator top (lower-right panel) isolated $r = 0.4$ jets from the first two top jets (i.e. $\Delta R_{\rm j, t_{1,2}} > 1.1$ ).}
\label{fig:bTag}
\end{figure*}

Figure \ref{fig:Iso} (middle panel) shows the invariant mass distribution of the reconstructed spectator top using three selected jets. We have a sharp peak at $172 \GeV$ with a fat tail in the signal events, and similar patterns are observed in the top-containing backgrounds. If we look at $b$-tag scores of the spectator top in Figure \ref{fig:bTag} (lower-left panel), a substantially large amount of $b$-jets are captured in the signal events, while $70\%$ of $t\bar{t}$ + jets fail to contain a $b$-jet in it. This gives a positive implication that the additional reconstruction of the spectator top combined with $b$-tagging delivers high background rejection power. The situation gets far better, however, when we exploit a multiple $b-$tag on the isolated jets in Figure \ref{fig:bTag} (lower-right panel). Our search strategy, therefore, is targeted for the final states of $t\bar{t}_{\rm{boost}}$ + $1b$ or $2b$ + jets, without reconstructing the additional top.
\begin{table*}[ht]
\setlength{\tabcolsep}{2.5mm}
\renewcommand{\arraystretch}{1.2}
\scalebox{1.0}{\begin{tabular}{c|cccccc|c}
                                                            Fully-hadronic     & Signal   [fb]                         & $t\bar{t}t\bar{t}$   [fb]     & $t\bar{t}$ + jets   [fb]& $t \bar{b}$ + jets   [fb] & QCD   [fb]
                                                                                        & $Z_{b \bar{b}}$ + jets   [fb] & $S/\sqrt{B}$                     \\ \hline \hline
                                                                Preselection    &  \bf{0.67}                             & $3.1$                             & $2.6 \times 10^4$                      & $2.8  \times 10^3 $ & $4.2  \times 10^6 $
                                                                                        & $3.4  \times 10^3 $            &$0.018$                               \\ 
                                                                   Basic Cuts    &   0.29                                 &  $0.26$                             & $2.3  \times 10^3 $                    & $420$                 & $6.5  \times 10^5 $
                                                                                        & $680$                                 & $0.020 $                             \\ 
                                                            Ditop Selection    &   0.17                                 & $0.16$                           & $790 $                                         & $150 $                    & $6.0  \times 10^3 $
                                                                                        & $14$                                   & $ 0.11 $                            \\       
                                    $N^{\rm iso}_{\rm jets}\geq 4$    &    0.13                                & $0.095$                          & 60                                                & $0.59$                     & 200
                                                                                        & $0.60 $                                 & $ 0.43 $                            \\ 
                                    $M_J > 900 \GeV$                      &   0.11                                 & $0.073$                           & 32                                               & $0.23$                     & 89
                                                                                        & $0.28 $                               & $ 0.53 $                              \\  \hline
                                                                       $3b$-tag   &  $0.029$                             & $0.019$                          & $0.35$                                        & $1.1 \times 10^{-4}$ & $1.7 \times 10^{-3}$
                                                                                        & $1.2  \times 10^{-4}$          &  $  \bf{2.6} $                       \\  \hline
                                                                       $4b$-tag   &  $0.010$                              &  $6.0  \times 10^{-3}$      & $0.016$                                     & $3.0 \times 10^{-6}$ & $3.7 \times 10^{-5}$
                                                                                        & $1.2 \times 10^{-5}$            &  $  \bf{3.7} $                        \\ 
\hline
\end{tabular}}
\caption{Effects of our selection strategies in the fully-hadronic channel for the illustrative benchmark
  parameters of \mbox{$M_{V_1} = 1.5 \TeV$} and \mbox{$c_t=2.0$}. We show the
  resulting background and signal cross sections in fb after each of the selection
  steps, together with the related significance that has been calculated for a
  luminosity of 3000~fb$^{-1}$.}
\label{tab:HadCutflow}
\end{table*}

We proceed to show the cutflow Table~\ref{tab:HadCutflow}. For the purpose of illustration, we present a benchmark parameter point of \mbox{$M_{V_1} = 1.5 \TeV$} and \mbox{$c_t=2.0$}. We show the resulting backgrounds and signal cross sections in fb after each of the selection steps, together with the related significance that has been calculated for a luminosity of 3000~fb$^{-1}$.

Table \ref{tab:HadCutflow} shows that the boosted ditop selection can efficiently suppress the background channels which do not contain a top quark (QCD and $Z_{b \bar{b}}$ + jets), where we find an overall improvement in $S/\sqrt{B}$ by a factor of $\sim 6$ at a $60\%$ signal efficiency relative to Basic Cuts. The combined cuts on the $N^{\rm iso}_{\rm jets}$ and $M_J$ are able to suppress low-multiplicity SM backgrounds (even including $t\bar{t}$ + jets) delivering a remarkable improvement in $S/\sqrt{B}$ by a factor of $\sim 5$. Finally, at least $1b$-tag on both of the boosted top jets is applied in addition to at least $1b$-tag ($2b$-tag) on the isolated jets. 
Putting them together (abbreviated to $3b$-tag ($4b$-tag)), we find that the best expectation comes from $4b$-tag with the drastic improvement of a factor of $\sim 7$ in $S/\sqrt{B}$. In the end, we observe that $S/\sqrt{B} \sim 3.7$ is achievable at an overall signal efficiency of $1.5 \%$ for a given luminosity of 3000~fb$^{-1}$.


\subsection{Semi-leptonic  Channel }
\label{sec:semi}
\begin{figure*}[t]
\begin{center}
\setlength{\tabcolsep}{0em}
{\renewcommand{\arraystretch}{1}
\begin{tabular}{cc}
\hspace{-20pt} \includegraphics[scale=0.39]{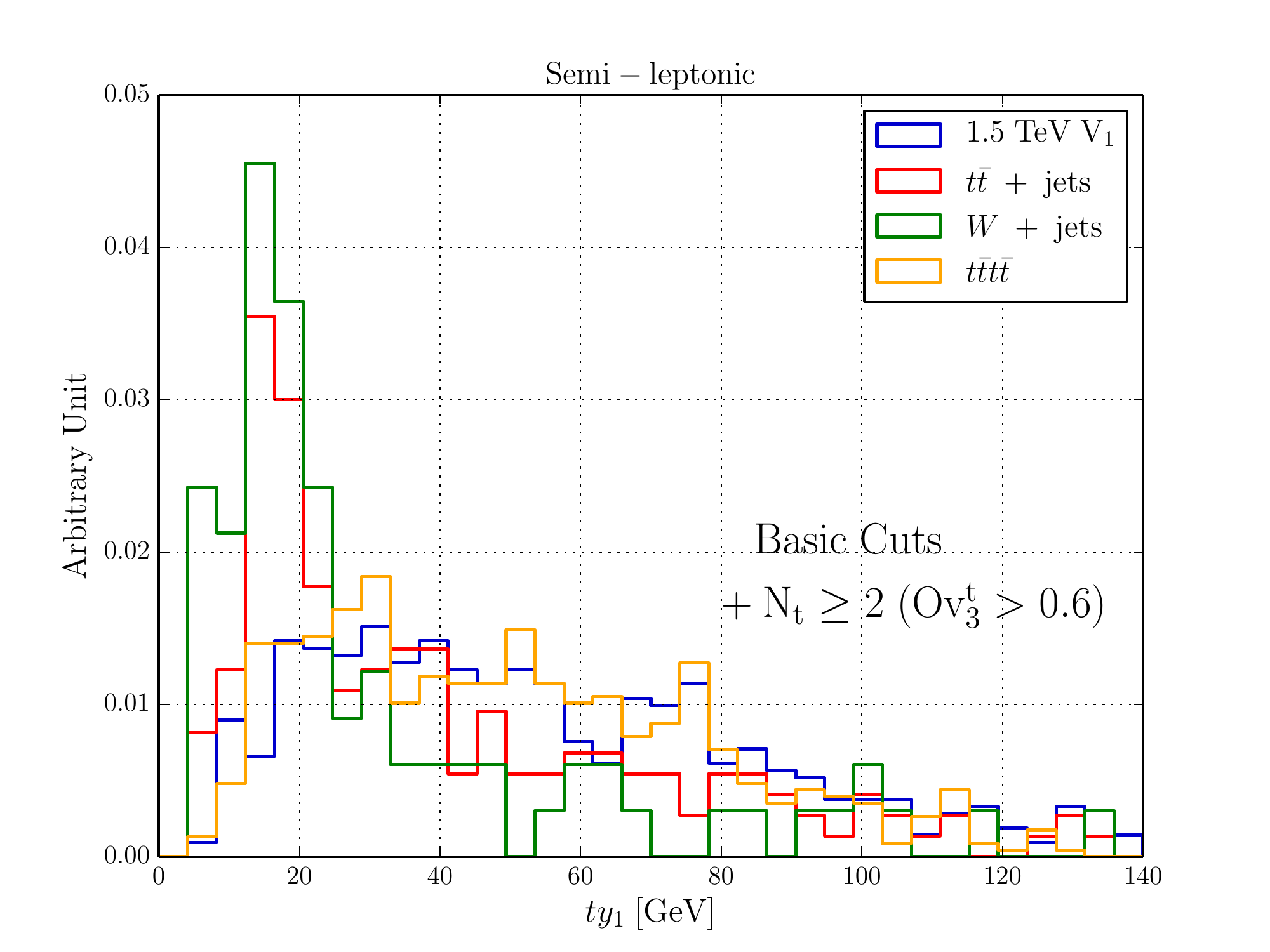} &
\hspace{20pt} \includegraphics[scale=0.39]{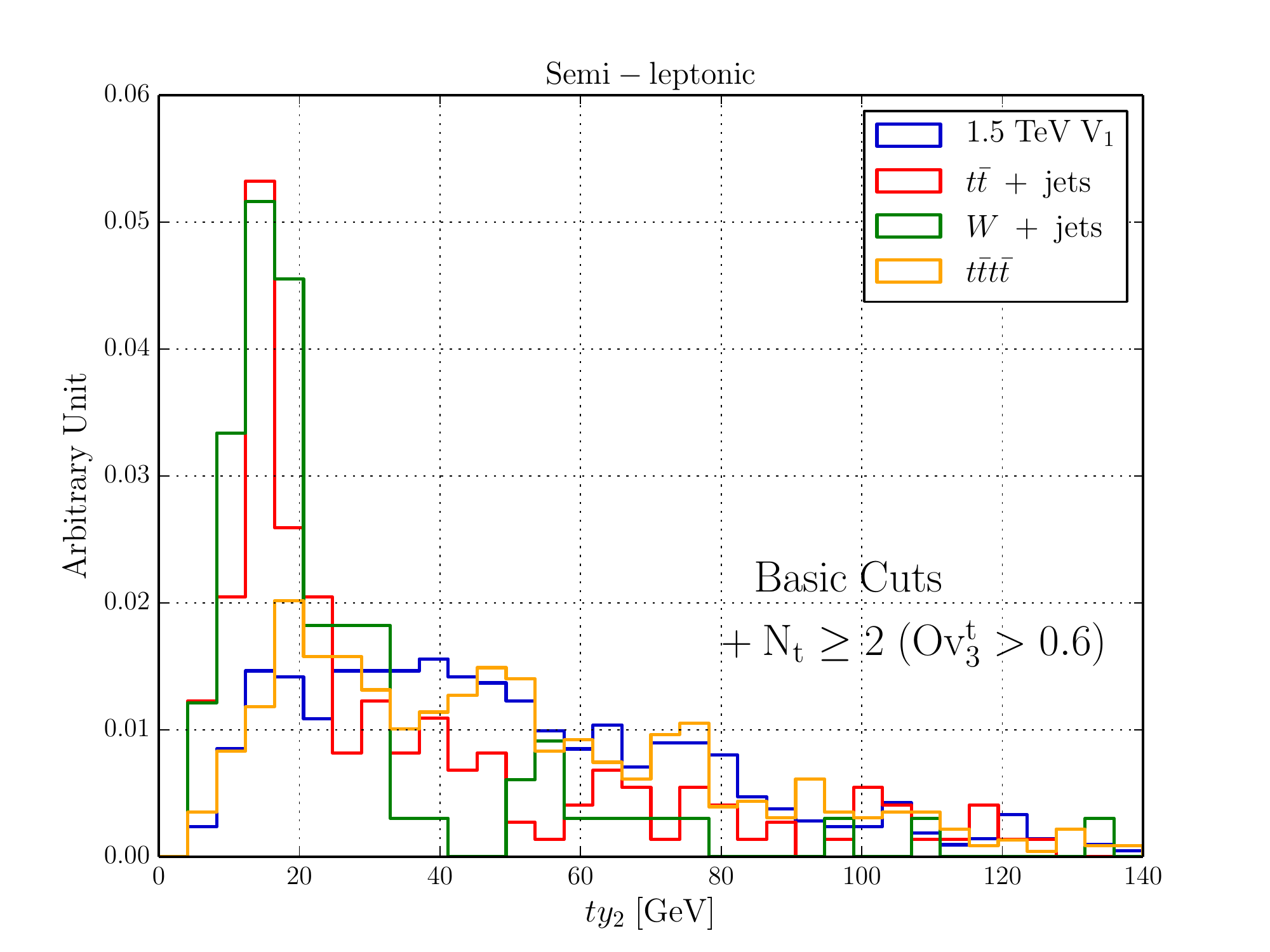}  \\
\hspace{-20pt} \includegraphics[scale=0.39]{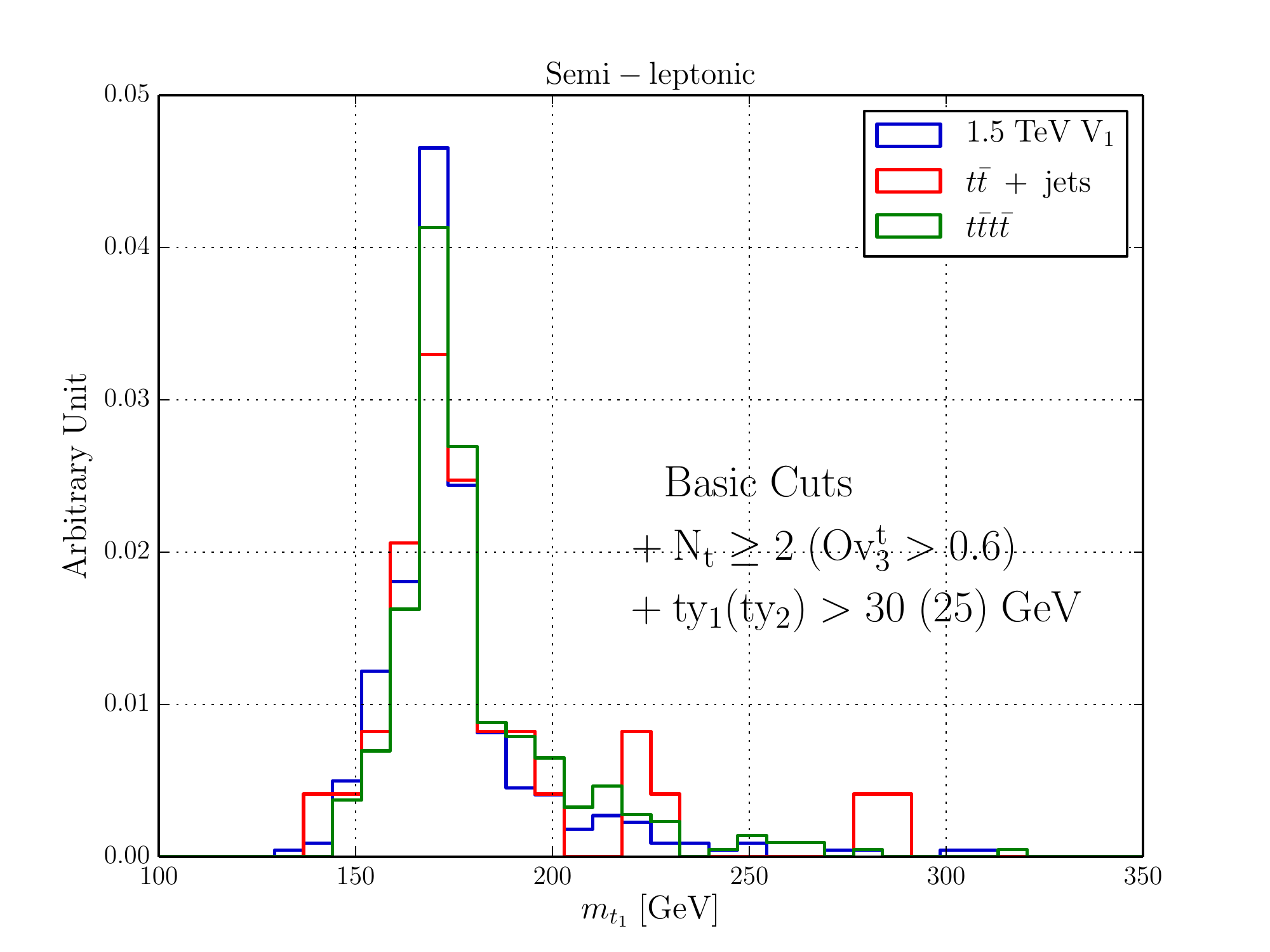} &
\hspace{20pt} \includegraphics[scale=0.39]{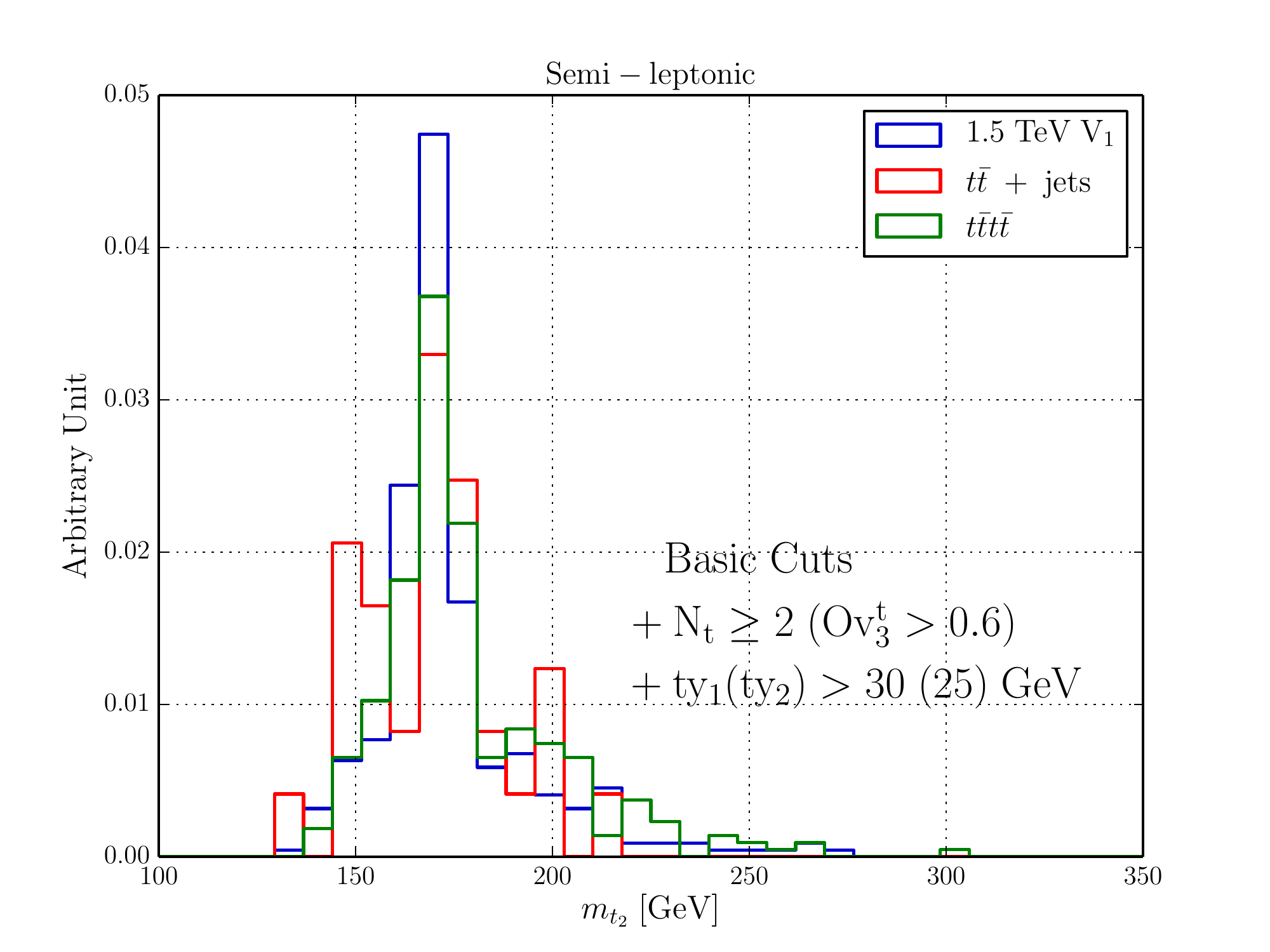}  \\
\end{tabular}
}
\end{center}
\caption{The panels in the first row show $ty$ distributions of the (left) first and (right) second hardest top jets, and the panels in the second row represent the respective invariant mass distributions after Basic Cuts and the Ditop Selection.}
\label{fig:SemiPlot}
\end{figure*}

\begin{table}[t]
\begin{center}
\scalebox{0.93}{
\begin{tabular}{|c|c|c|c|}
\hline
		Signal Channel			&	Backgrounds 						& $\sigma (H_T > 700\GeV) [{\rm fb}]$ \\ \hline
\multirow{3}{*}{Semi-leptonic}            & $W(\rightarrow l \nu)$ + jets 	                          &  $ 1.5  \times 10^4 $	                     \\
                                                           &$t\bar{t} t\bar{t}$                                                &	4.2                          		            \\
							&$t\bar{t}$(semi-leptonic) + jets                          &	$ 1.5    \times 10^4 $		            \\ 
\hline		
\end{tabular}}
\end{center}
\caption{The simulated cross sections of SM backgrounds (including a conservative  NLO K-factor of 2 after preselection cuts described in section~\ref{sec:simulation}).}
\label{tab:SemiBackGround}
\end{table}

\begin{table}[t]
\begin{center}
\setlength{\tabcolsep}{1em}
{\renewcommand{\arraystretch}{1.5}
\begin{tabular}{c|c}

							 &  Semi-leptonic	   			                                                    \\ \cline{1-2}
\multirow{3}{*}{\textbf{Basic Cuts} }	 &  $N_{\rm fj}  \geq 2$ ($R=0.7$),	$N^{\rm iso}_{\rm lepton} = 1$ ,   \\ 
		 					 &  $p_T^{\rm fj} > 500 \GeV$, $|\eta_{\rm fj} |< 2.5$ .                            \\  
							 &  $ \met > 15 \GeV$  \\   \cline{1-2}
%
\multirow{2}{*}{ \textbf{Ditop Selection} }	& $N_{t} \geq 2$, (for $t$: $Ov_3^t > 0.6$)  ,           \\ 
		 					          & $ty_1 \;(ty_2) > 30 \;(25) \GeV$                    \\ 
\end{tabular}}\par
\caption{(top) Summary of Basic Cuts (top) and Ditop Selection (bottom).  
``fj" stands for the fat jet with $|\eta_{\rm fj}| < 2.5$ and $p_T > 500 \GeV$ and $N^{\rm iso}_{\rm lepton}$ represents the number of isolated leptons with mini-ISO $> 0.7$, $p_T^{\ell} > 25\GeV$ and $|\eta_{\ell} |< 2.5$.
``$Ov$'' selection applies to the fat jets ($R=0.7$) in the event, and $N_{t}$ is the number of tagged top fat jets. 
} \label{tab:CutSemi}    
\end{center}
\end{table}

In the previous section, we managed to remove a substantial amount of the QCD background, except for the resilient $t\bar{t}$ (\text{hadronic}) + jets. This leads us to the semi-leptonic channel (cf. Figure \ref{fig:Diagram}) since $t\bar{t}$ (semi-leptonic) + jets now contains a single hadronic top (hardly expected to pass the ditop selection). On top of that, requiring an isolated lepton with mini-ISO $> 0.7$~\cite{Kaplan:2008ie} and $p_T^{\ell} > 25 \GeV$ safely removes any room for the gigantic QCD background, rendering relatively clean final states compared to the fully-hadronic channel (throughout the paper, we refer to ``leptons'' as muons and electrons only).

The main irreducible SM background is the semi-leptonic $t\bar{t}$ + jets with up to two additional jets (including $b$ jets), and $t\bar{t} t\bar{t}$ where we exclusively make one top decay leptonically and other three tops hadronically. $W (\rightarrow \ell \nu)$ + jets constitutes a subdominant background where we included up to 3 extra light-flavour jets (including b jets). The contribution from the single top production $t (\rightarrow \ell \nu b)$ + $\bar{b}$ + jets turns out to be negligible, so we do not include it here. Table~\ref{tab:SemiBackGround} summarizes the background cross sections including a conservative K-factor of 2.

\begin{table*}[t]
\setlength{\tabcolsep}{2.5mm}
\renewcommand{\arraystretch}{1.2}
\scalebox{1.0}{
\begin{tabular}{c|cccc|c}
                                                                      Semi-leptonic     & Signal   [fb]                 & $W$ + jets   [fb]       &    $t\bar{t} t\bar{t}$   [fb]   &  $t \bar{t}(\text{semi-leptonic})$ + jets   [fb] & $S/\sqrt{B}$      \\ \hline \hline
                                                                        Preselection    &  \bf{0.45}                     & $1.5 \times 10^{4}$ &     $4.2$                         & $1.5 \times 10^{4}$                                         &  $0.14 $          \\ 
                                                                           Basic Cuts    &   0.12                          & $890$                      &    $0.15$                       & 340                                                                   &  $0.19   $        \\ 
                                                                    Ditop Selection    & $0.013$                      & $0.10$                     &    $0.012$                     & $1.3 $                                                               &  $ 0.61 $        \\       
                                           $N^{\rm iso}_{\rm jets}\geq 3$     &  $0.011 $                     & $0.067$                   &    $8.8 \times 10^{-3}$  & $0.13$                                                              &  $1.3$            \\
                                                             $M_J > 650 \GeV$     & $0.011$                       & $0.033$                   &    $8.3 \times 10^{-3}$  & $0.13$                                                              &  $1.4$            \\  \hline
                                                                             $3b$-tag     &  $3.3 \times 10^{-3}$  & $1.1 \times 10^{-7}$ &  $2.5 \times 10^{-3}$   & $3.7 \times 10^{-5}$                                          & $  \bf{3.6} $   \\ 
\hline
\end{tabular}
}\\[.2cm]
\caption{Effects of our selection strategies in the semi-leptonic channel for the illustrative benchmark
  parameters: \mbox{$M_{V_1} = 1.5 \TeV$} and \mbox{$c_t=2.0$}. We show the
  resulting background and signal cross sections in fb after each of the selection
  steps, together with the related significance that has been calculated for a
  luminosity of 3000~fb$^{-1}$.}
\label{tab:SemiCutflow}
\end{table*}

$Basic \; Cuts$ require exactly one isolated lepton in the event (mini-ISO $> 0.7$  with $p_T^{\ell} > 25 \GeV$ and $|\eta_{\ell} |< 2.5$) and $ \met > 15 \GeV$. We require at least two fat jets with $p_T^{\rm fj} > 500 \GeV$ and $|\eta_{\rm fj} | < 2.5$.
The specific ditop selection (Table \ref{tab:CutSemi}) takes the same approach in the fully-hadronic channel, except that we additionally require $ty_1 \;(ty_2) > 30 \;(25) \GeV$. In Figure \ref{fig:SemiPlot}, $ty$ distributions of the (upper-left panel) first and (upper-right panel) second hardest top jets indicate that $ty$ cuts significantly reduce $t \bar{t} (\text{semi-leptonic})$ + jets and $W$ + jets, bringing it into the regime where effectively $t\bar{t} t\bar{t}$ is the only background remaining with the resulting signal efficiency of $\sim 90 \%$.

%

As a consequence, Figure \ref{fig:SemiPlot} (bottom panels) shows the invariant mass distributions of the first two hardest top jets after Basic Cuts and the ditop selection, and we observe a cleaner signal peak in contrast to the fully-hadronic channel (cf. Figure \ref{fig:m_top}).

%

%
%
%
%

As demonstrated in the previous section, the high-multiplicity final states allow us to further suppress the backgrounds by demanding cuts of $N^{\rm iso}_{\rm jets}\geq 2$ and  $M_J > 650 \GeV$. Finally, we apply at least $1b$-tag on both of the boosted top jets, and at least $1b$-tag on the isolated jets (abbreviated to $3b$-tag). Since we lost a substantial amount of signal events at the ditop selection already, we are not able to do a $4b$-tag in this case.

The results of the analysis flow are summarized in Table \ref{tab:SemiCutflow} for the same benchmark model point \mbox{$M_{V_1} = 1.5 \TeV$} and \mbox{$c_t=2.0$}. The boosted ditop selection provides significant rejection power on the $W$ + jets and $t \bar{b}$ + jets backgrounds which do not contain a hadronic top jet. As already noted, the ditop selection combined with $ty$ cuts effectively reduce $t \bar{t} ~(\text{semi-leptonic})$ + jets and $W$ + jets delivering an improvement of a factor of $\sim 3$ in $S/\sqrt{B}$. The combined cuts on the $N^{\rm iso}_{\rm jets}$ and $M_J$ further improve $S/\sqrt{B}$ by a factor of $\sim 2$, and finally $3b$-tag leads to the regime where effectively $t\bar{t} t\bar{t}$ is the only background remaining. Overall, we can achieve $S/\sqrt{B} \sim 3.6$ at the cost of signal efficiency $\sim 1 \%$ for a given luminosity of 3000~fb$^{-1}$. The resulting sensitivity is comparable with the one we obtained in the fully hadronic channel (cf. Table \ref{tab:HadCutflow}).
%
%


\subsection{Same-sign Dileptonic (SSDL) Channel }
\label{sec:SameSign}

\begin{table}[h]
\begin{center}
\scalebox{0.95}{
\begin{tabular}{|c|c|c|c|}
\hline
		Signal Channel			&	Backgrounds 			            & $\sigma (H_T > 500\GeV) [{\rm fb}]$ \\ \hline
\multirow{5}{*}{SSDL}                        & $t \bar{t} W^{\pm}$ + jets	            &  $12 $	                                           \\
                                                           & $t \bar{t} Z $ + jets                             &	$8.1$                          		            \\
							& $t \bar{t} W^{\pm} W^{ \pm}$              &	$0.32 $		                                        \\ 
							& $t \bar{t} h $                                        &	$0.84$                       	                     \\ 
							& $t \bar{t} t \bar{t}$                                &	$0.77$                       	                     \\ 
\hline		
\end{tabular}}
\end{center}
\caption{The simulated cross sections of SM backgrounds (including a conservative  NLO K-factor of 2 after preselection cuts described in section~\ref{sec:simulation}).}
\label{tab:SSDLBackGround}
\end{table}

\begin{figure}[h]
\includegraphics[scale=0.39]{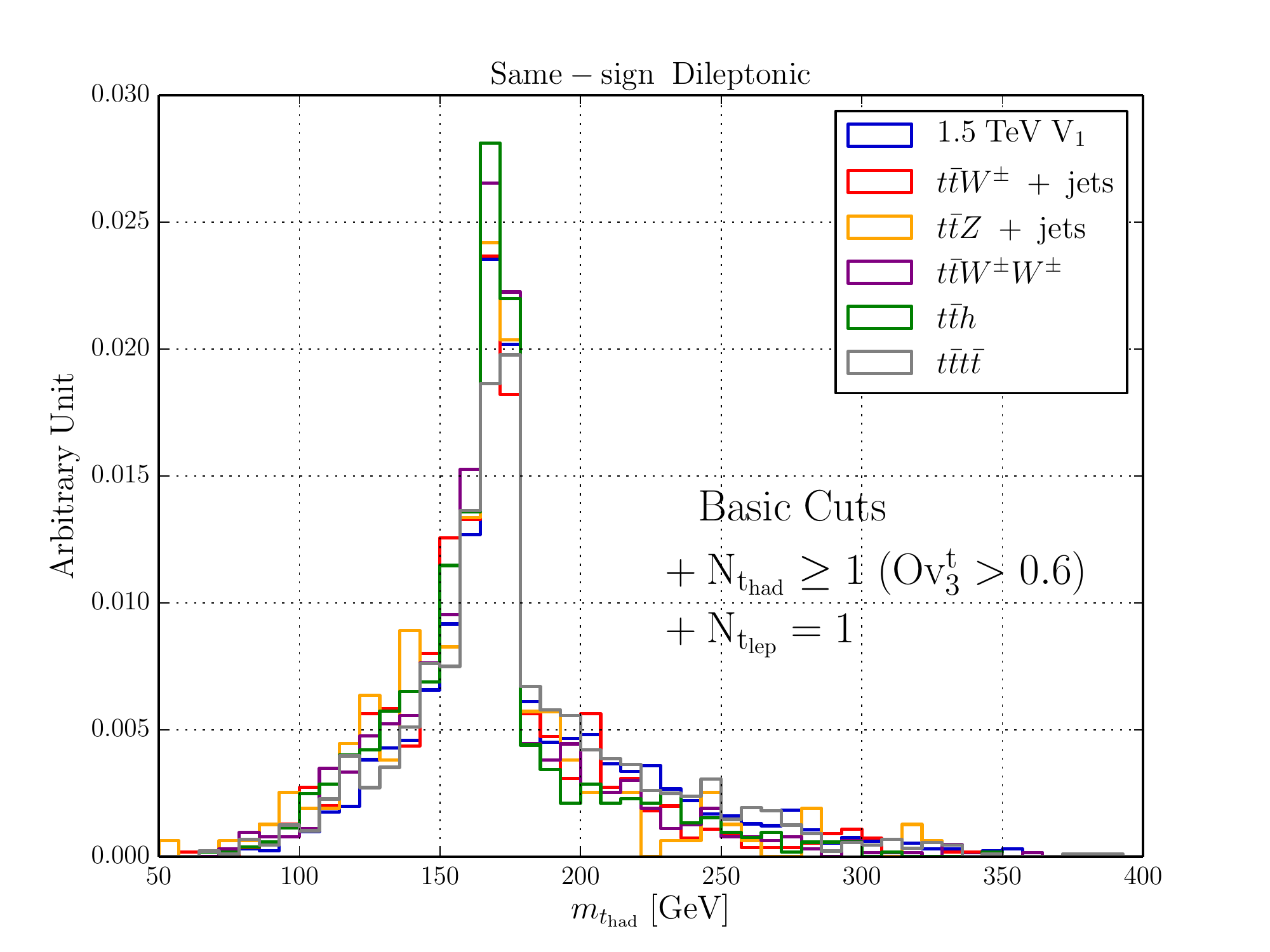}
\includegraphics[scale=0.39]{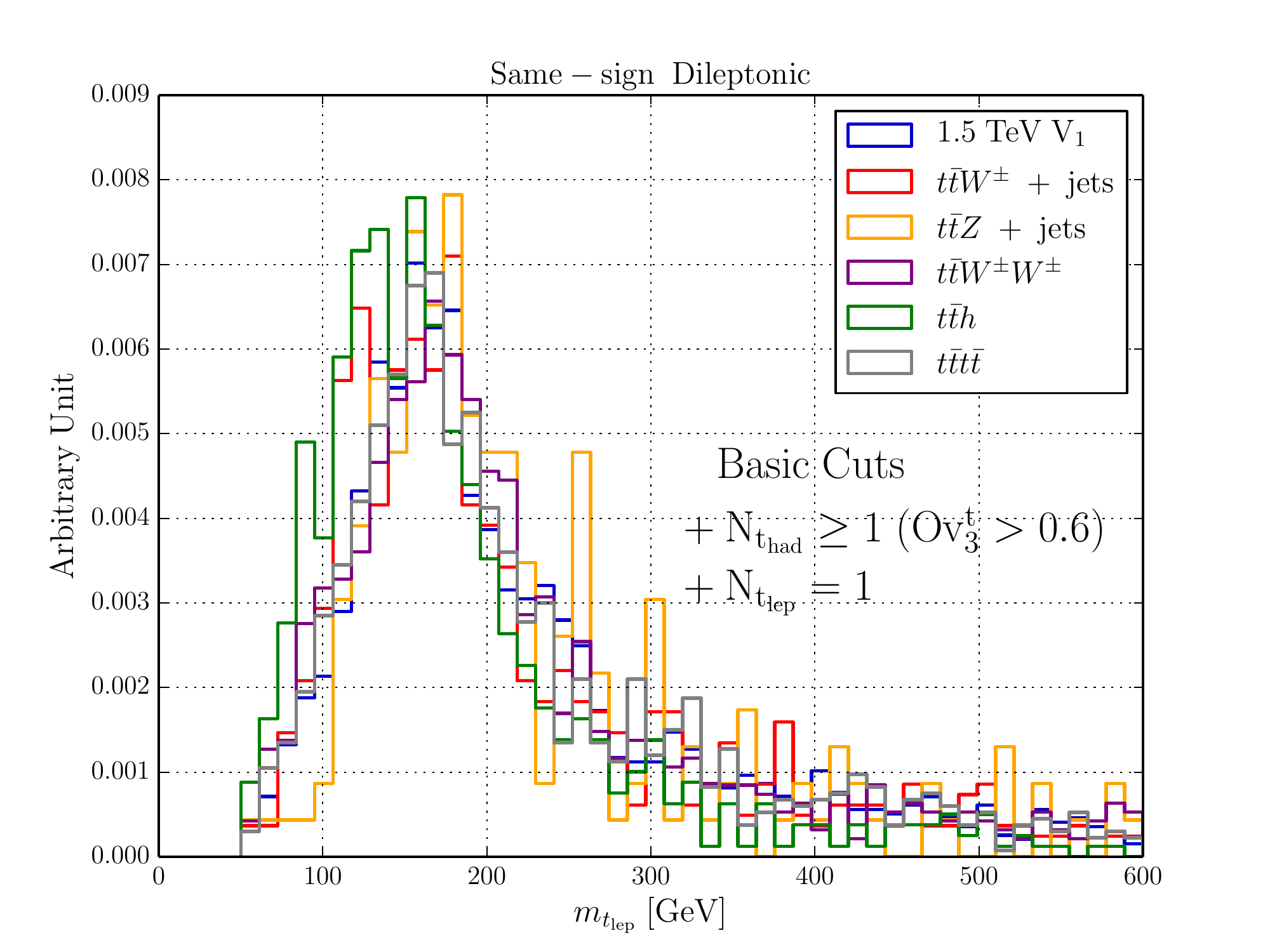}
\includegraphics[scale=0.39]{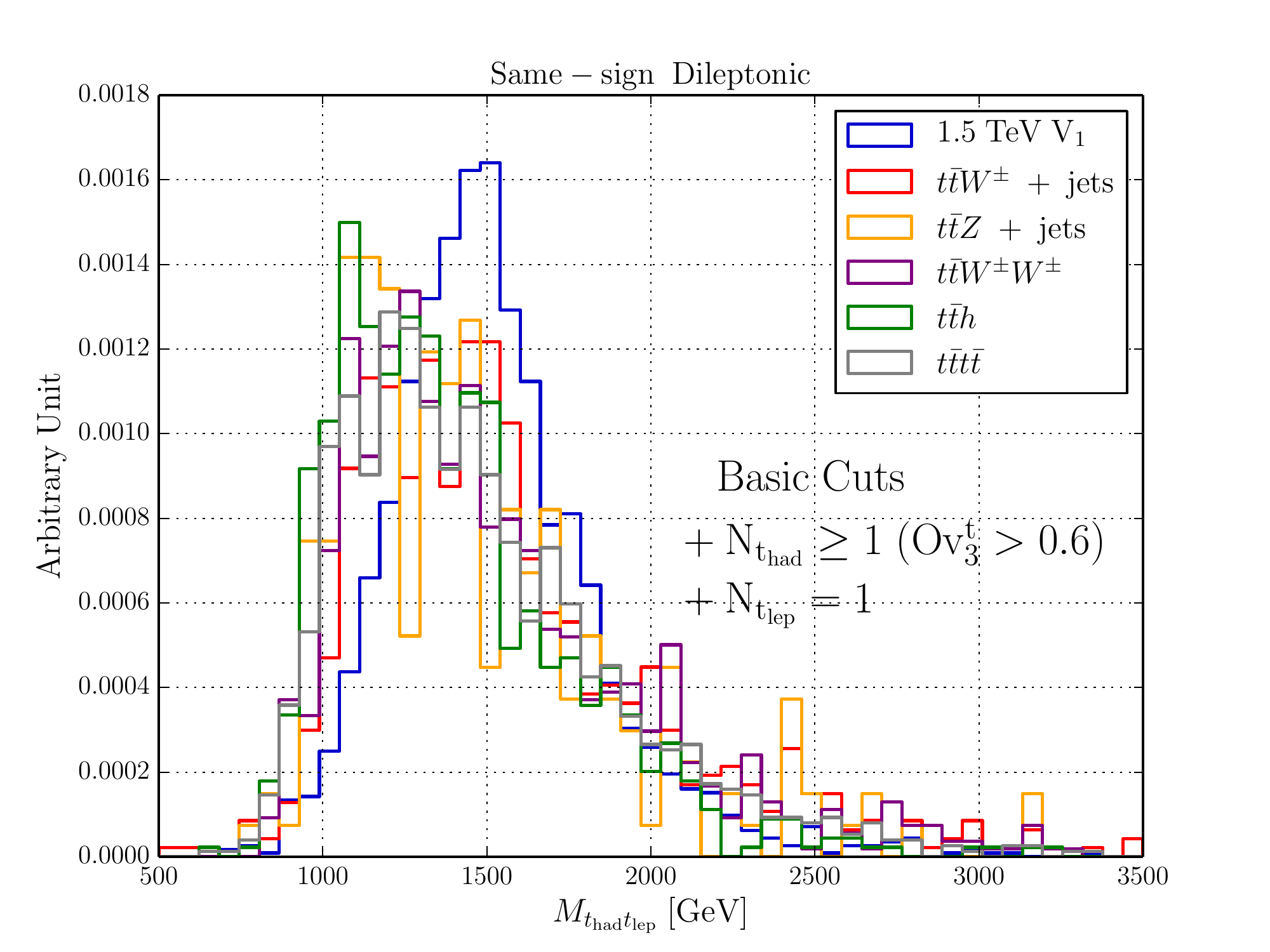}
\caption{Invariant mass distributions of the (top panel) hardest top jet $t_{\rm had}$ and (middle panel) reconstructed leptonic top jet $t_{\rm lep}$ after Basic Cuts and the ditop selection. The $1.5 \TeV$ $V_1$ resonance is reconstructed using the boosted ditop system in the bottom panel.}
\label{fig:SameSign_top}
\end{figure}

Unlike the other channels, the SSDL channel can evade the dominant $t \bar{t} + \; \rm{jets}$ background, at the cost of a small branching ratio $\sim 4.4 \%$. The SSDL channel has been studied in Ref. \cite{Liu:2015hxi} for 14 TeV during the time of the write-up of this paper, with a remarkable resulting significance due to small SM backgrounds. However, on the possibility of fully reconstructing a resonance, it remains less explored so far, and therefore necessitates an additional inquiry on measuring the mass and width of a resonance directly. Conventionally, it is deemed to be non-trivial to fully reconstruct a resonance in the SSDL channel, mainly due to the difficulty in selecting a proper combination of hadronic and leptonic tops.

\begin{table}[h]
\begin{center}
\setlength{\tabcolsep}{1em}
{\renewcommand{\arraystretch}{1.5}
\begin{tabular}{c|c}

							 &  SSDL	   			                                                    \\ \cline{1-2}
\multirow{4}{*}{\textbf{Basic Cuts} }	 &  $N_{\rm fj}  \geq 1$ ($R=0.8$),  \\ 
                                                            &  $N^{\rm iso}_{\rm lepton} = 2 \;(\rm same \; sign)$ ,   \\ 
                                                            &  $N_{j}  \geq 3 $ ,   \\ 
							 &  $ \met > 50 \GeV$ \\ \cline{1-2}
\multirow{2}{*}{ \textbf{Ditop Selection} }	& $N_{t_{\rm had}} \geq 1$, (for $t$: $Ov_3^t > 0.6$)  ,                                 \\ 
							          & $N_{t_{\rm lep}} = 1$      	 
\end{tabular}}\par
\caption{Summary of Basic Cuts (top) and Ditop Selection for the semi-leptonic channel. ``fj" stands for the fat jet with $|\eta_{\rm fj}| < 2.5$ and $p_T > 500 \GeV$ and $N^{\rm iso}_{\rm lepton}$ represents the number of isolated leptons with mini-ISO $> 0.7$, $p_T^{l} > 25\GeV$ and $|\eta_{l} |< 2.5$. ``j" represents the $r = 0.4$ jet with $|\eta_{j} |< 2.5$ and $p_T^{j} > 25 \GeV$. 
``$Ov$'' selection applies to the trimmed-fat jets ($R=0.8$) in the event, and $N_{t_{had}}$ and $N_{t_{lep}}$ are the numbers of tagged hadronic and leptonic tops.} \label{tab:CutsSSDL} 
\end{center}
\end{table}

In the boosted regime, on the other hand, two boosted tops from a heavy resonance decay are characteristically differentiated from two non-boosted spectator tops in terms of $p_T$ scales, resolving the combinatoric issue. Also we can exploit the fact that their decay products are strongly collimated rendering an easy and simple way to reconstruct them. In this section, we demonstrate the capability of reconstructing a resonance using jet-substructure methods as well as a collinear approximation, and then reassess the sensitivity of the SSDL channel.

The main irreducible SM background is the same-sign dileptonic (SSDL) $t \bar{t} t \bar{t}$ even if its production cross section is low. Sub-dominant backgrounds consist of SSDL $t \bar{t} W^{\pm}$ + jets, $t \bar{t} Z$ + jets (up to two additional jets), $t \bar{t} W^{\pm} W^{\pm}$ and $t \bar{t} h~ ( h \rightarrow W^{\pm} W^{* \pm} \rightarrow \ell \nu j j)$. Table \ref{tab:SSDLBackGround} summarizes background constituents of the SSDL channel where we conservatively include a K-factor of 2 to all backgrounds in contrast to Ref.\cite{Liu:2015hxi}.

$Basic \; Cuts$ require exactly two isolated same-sign leptons (mini-ISO $> 0.7$  with $p_T^{\ell} > 25 \GeV$ and $|\eta_{\ell} |< 2.5$) and $ \met > 50 \GeV$. A charge mis-identification probability is not considered in this analysis. We require at least one fat jet with $p_T^{\rm fj} > 500 \GeV$ and $|\eta_{\rm fj} | < 2.5$ where the size of the fat jet is optimized up to $R=0.8$ to increase the ditop-tagging efficiency. In addition, we require at least three central $r=0.4$ jets with $|\eta_{j} |< 2.5$ and $p_T^{j} > 25 \GeV$.
%
%

%
\begin{figure*}[t]
\begin{center}
\setlength{\tabcolsep}{0em}
{\renewcommand{\arraystretch}{1}
\begin{tabular}{cc}
\hspace{-20pt} \includegraphics[scale=0.39]{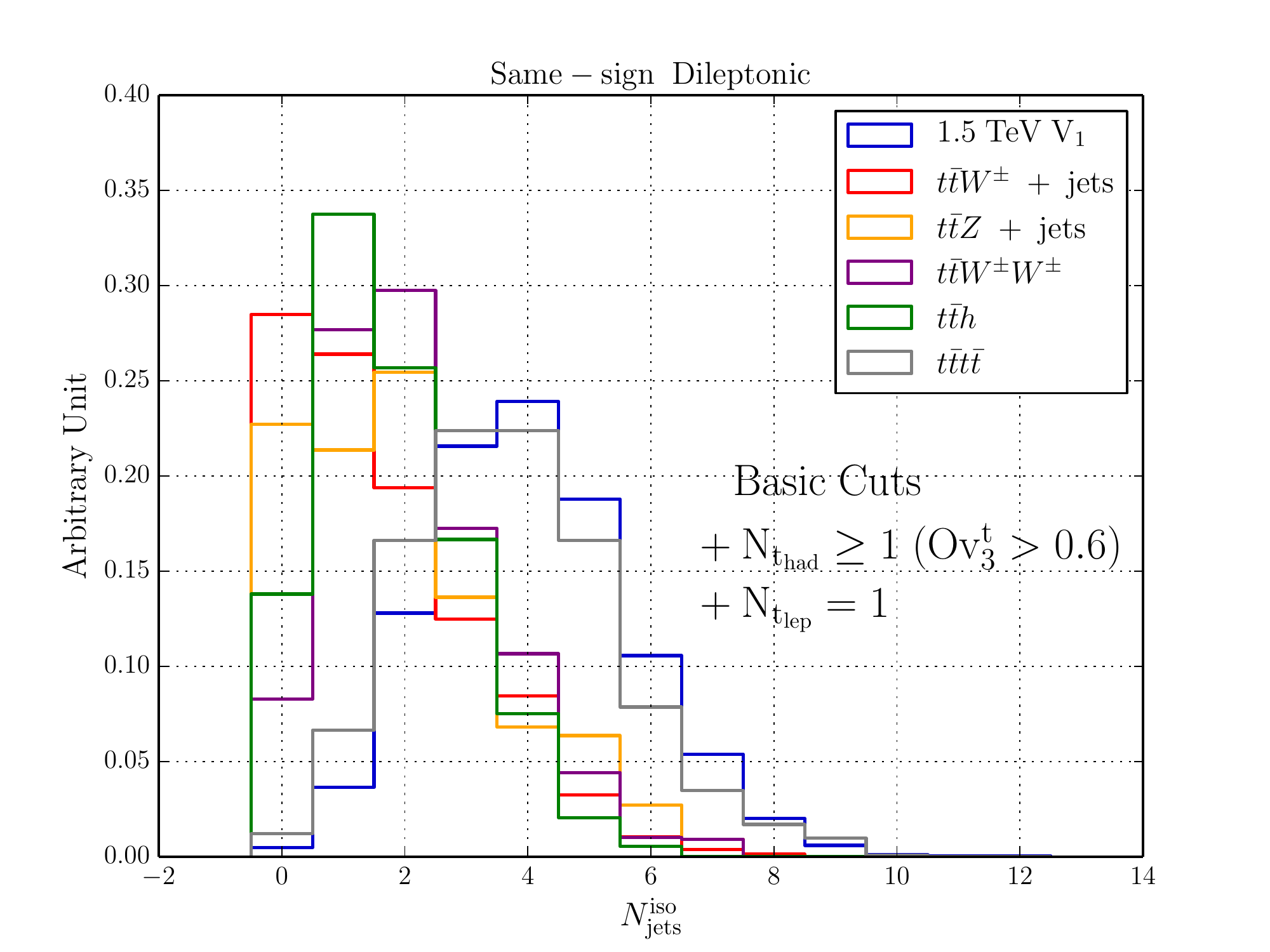} &
\hspace{20pt} \includegraphics[scale=0.39]{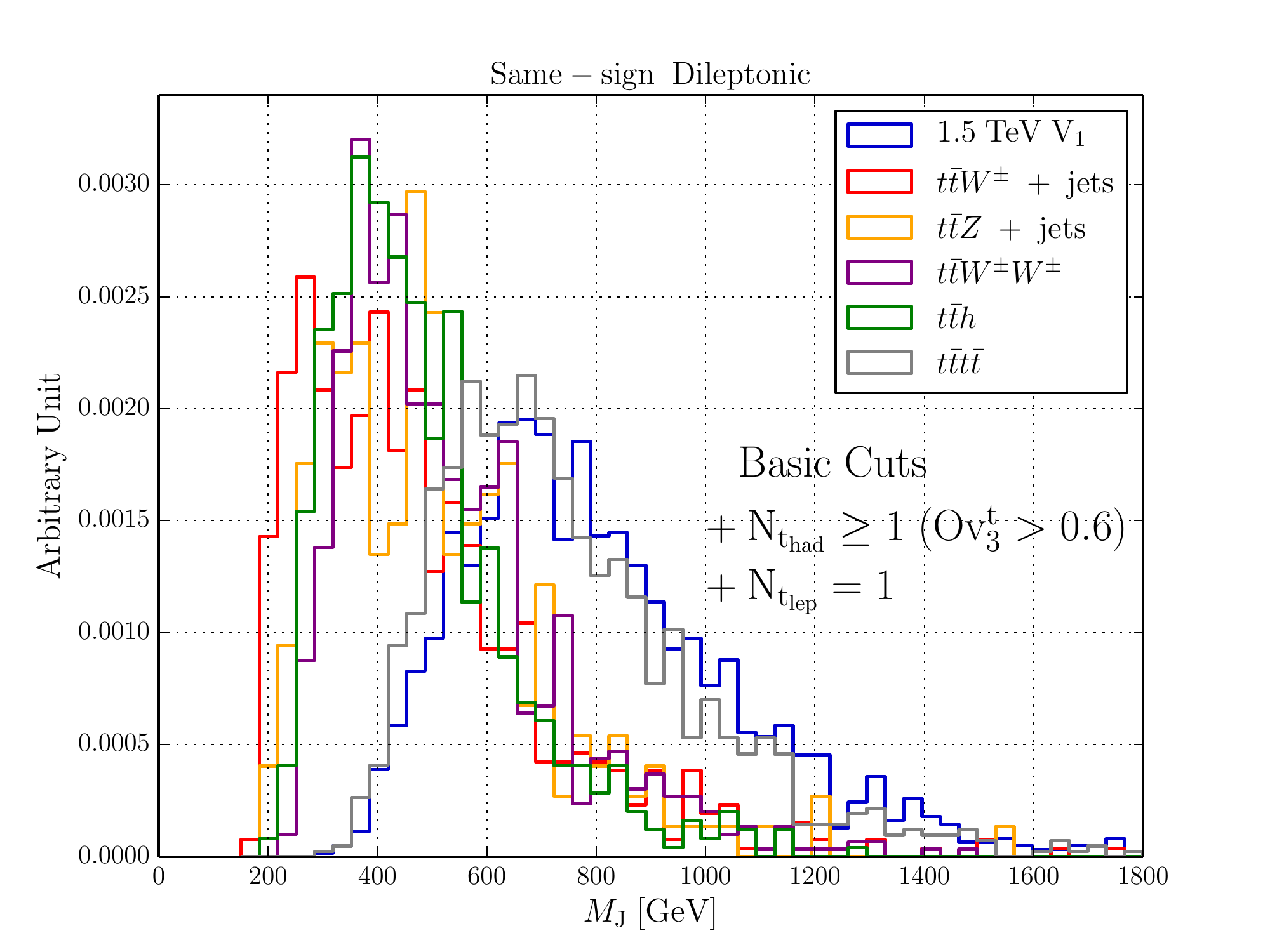}  
\end{tabular}
}
\end{center}
\caption{Distributions of (left panel) a number of jets isolated from $t_{\rm had}$ and $t_{\rm lep}$ by $\Delta R_{\rm j, t_{\rm had, \rm lep}} > 1.2$, and (right panel) $M_J$ scalar sum of the masses of large radius ($R=1.5$) jets. Basic Cuts and the Ditop Selection are applied for the both of cases.}
\label{fig:bTagSSDL}
\end{figure*}

\begin{table*}[t]
\setlength{\tabcolsep}{2.5mm}
\renewcommand{\arraystretch}{1.2}
\scalebox{0.9}{
\begin{tabular}{c|cccccc|c}
                                                                          SSDL     & Signal   [fb]                 & $t \bar{t} W^{\pm}$ + jets   [fb]  &    $t \bar{t} Z $ + jets    &  $t \bar{t} W^{\pm} W^{ \pm}$  [fb] & $t \bar{t} h $   [fb]
                                                                                        & $t \bar{t} t \bar{t}$ [fb] & $S/\sqrt{B}$      \\ \hline \hline
                                                                Preselection    &  \bf{0.15}                     & $12$                                           &     $8.1$                         & $0.32$                                          & 0.84 
                                                                                        &  $0.77$                       & $1.8 $          \\ 
                                                                   Basic Cuts    & $0.051$                        & $0.96 $                                        &    $0.35$                       & $0.027$                                      & 0.039
                                                                                        &  $0.047$                        & $2.3$        \\ 
                                                            Ditop Selection    & $0.028$                       & 0.25                                           &    $0.078$                       & $7.4  \times 10^{-3} $                   & 0.020
                                                                                        & $0.019$                       &   $ 2.5 $        \\       
                                   $N^{\rm iso}_{\rm jets}\geq 3$     &  $0.023$                      & 0.065                                         &    $0.024$                      & $2.5 \times 10^{-3}$                      & $5.5 \times 10^{-3}$
                                                                                         &  $0.014$                      &   $3.9$            \\
                                                      $M_J > 350 \GeV$     & $0.023$                       & 0.063                                        &    $0.023$                      & $2.5 \times 10^{-3}$                       & $5.3 \times 10^{-3}$
                                                                                        &  $0.014$                      &  $3.9$            \\ 
                                              $M_{V_1} > 1100 \GeV$   & $0.022$                       & 0.055                                        &    $0.018$                      & $2.1 \times 10^{-3}$                       & $3.9 \times 10^{-3}$
                                                                                        &  $0.011$                      &  $4.0$            \\ \hline
                                                                     $3b$-tag     & $0.012$                       & $3.458423 \times 10^{-3}$       & $1.1 \times 10^{-3}$      &  $2.13534 \times 10^{-4}$             & $3.4 \times 10^{-4}$
                                                                                        &  $6.0 \times 10^{-3}$   &   $  \bf{6.3} $   \\  \hline
\end{tabular}
}\\[.4cm]

\scalebox{0.9}{
\begin{tabular}{c|cccccc|c}
                                                                                  SSDL     & Signal   [fb]                 & $t \bar{t} W^{\pm}$ + jets   [fb]  &    $t \bar{t} Z $ + jets  & $t \bar{t} W^{\pm} W^{ \pm}$  [fb] & $t \bar{t} h $   [fb]
                                                                                                & $t \bar{t} t \bar{t}$ [fb] & $S/\sqrt{B}$      \\ \hline \hline
                                                                        Preselection    &  \bf{0.15}                     & $12$                                           &     $8.1$                     & $0.32$                                        & 0.84 
                                                                                                &  $0.77$                       & $1.8 $          \\ 
                                                                       Basic Cuts 2    & $0.092$                       & $5.4 $                                        &    $3.2$                       & $0.17$                                       & 0.34
                                                                                                &  $0.39$                      &   $1.6 $        \\ 
                                                          $S_T > 1500 \GeV$     & $0.047$                       & $0.46$                                        &    $0.19$                    & $0.014$                                      & $0.013$
                                                                                                &  $0.037$                     &  $3.0$            \\ \hline
                                                                             $3b$-tag     & $0.024$                      & $0.023$                                      & $9.5 \times 10^{-3}$  &  $1.1 \times 10^{-3}$                  & $9.9 \times 10^{-4}$
                                                                                                &  $0.018$                      &   $  \bf{5.6} $   \\  \hline
\end{tabular}
}\\[.2cm]
\caption{Effects of our selection strategies in the SSDL channel for the illustrative benchmark
  parameters of \mbox{$M_{V_1} = 1.5 \TeV$} and \mbox{$c_t=2.0$}. We show the
  resulting background and signal cross sections in fb after each of the selection
  steps, together with the related significance that has been calculated for a
  luminosity of 3000~fb$^{-1}$.}
\label{tab:SameSign_flow}
\end{table*}

The specific ditop selection (Table \ref{tab:CutsSSDL}) begins with the overlap analysis applied to all trimmed-fat jets with $|\eta_{\rm fj}| < 2.5$ and $p_T > 500 \GeV$. We demand at least one top jet (i.e. fat jet satisfying $Ov$-selection criterion, $Ov_{3}^t > 0.6$), and identify the first hardest top ($t_{\rm had}$) as the candidate from the resonance decay. The boosted topology offers even simpler ways to reconstruct the leptonic top ($t_{\rm lep}$) where its decay products are highly collimated, so that it allows for an efficient use of the simple collinear approximation $\eta_\nu = \eta_\ell, $ where $\nu$ is the neutrino from the decay of a boosted top and its $p_T$ is set to the total missing transverse momentum in the event\footnote{It should be noted that there is another source contributing to the total missing transverse momentum from a leptonic spectator top. We assume that, in the boosted regime, the leading contribution comes from a boosted leptonic top. We have verified that this is a reasonably good approximation.} and $\ell$ is the hardest isolated lepton. 
Then any $r=0.4$ jet with $p_T > 50 \GeV$ (while simultaneously demanding that the jet be located within $\Delta R_{j,\,\ell} < 0.8$ from the lepton) which gives the lowest value of $\chi^2$ is selected, and $\chi^2$ is defined by
\begin{equation}
\chi^2 = \frac{(m_{j l \nu} - m_t)^2}{\Gamma^2_t} \; ,
\label{eq:chi1}
\end{equation}
with $m_t = 172 \GeV$ and $\Gamma_t = 1.5 \GeV$. Combining the hardest lepton, $\met$ and the selected $r=0.4$ jet, we can reconstruct $t_{\rm lep}$ with $p_T > 300 \GeV$. We abbreviate the boosted leptonic top identification to $N_{t_{\rm lep}} = 1$.

Figure \ref{fig:SameSign_top} shows successful invariant mass reconstructions of $t_{\rm had}$ and $t_{\rm lep}$ by the virtue of efficient hadronic top-taggers (TOM) and the collinear approximation. Put them together, it is possible to reconstruct the invariant mass of the $1.5 \TeV$ $V_1$ resonance as in the bottom panel.
\begin{figure*}[t]
\begin{center}
\setlength{\tabcolsep}{0em}
{\renewcommand{\arraystretch}{1}
\begin{tabular}{cc}
\hspace{-20pt} \includegraphics[scale=0.53]{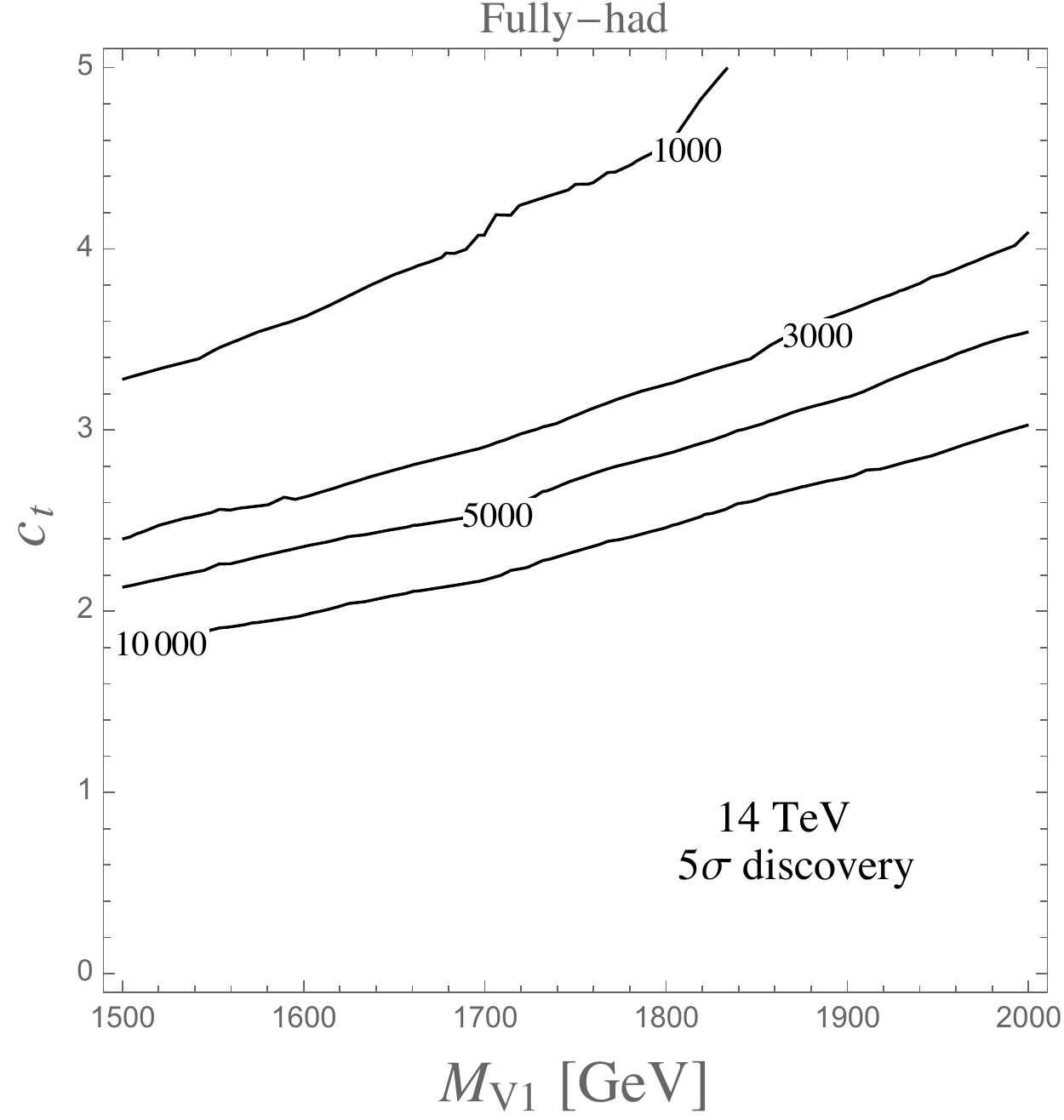} &
\hspace{30pt} \includegraphics[scale=0.53]{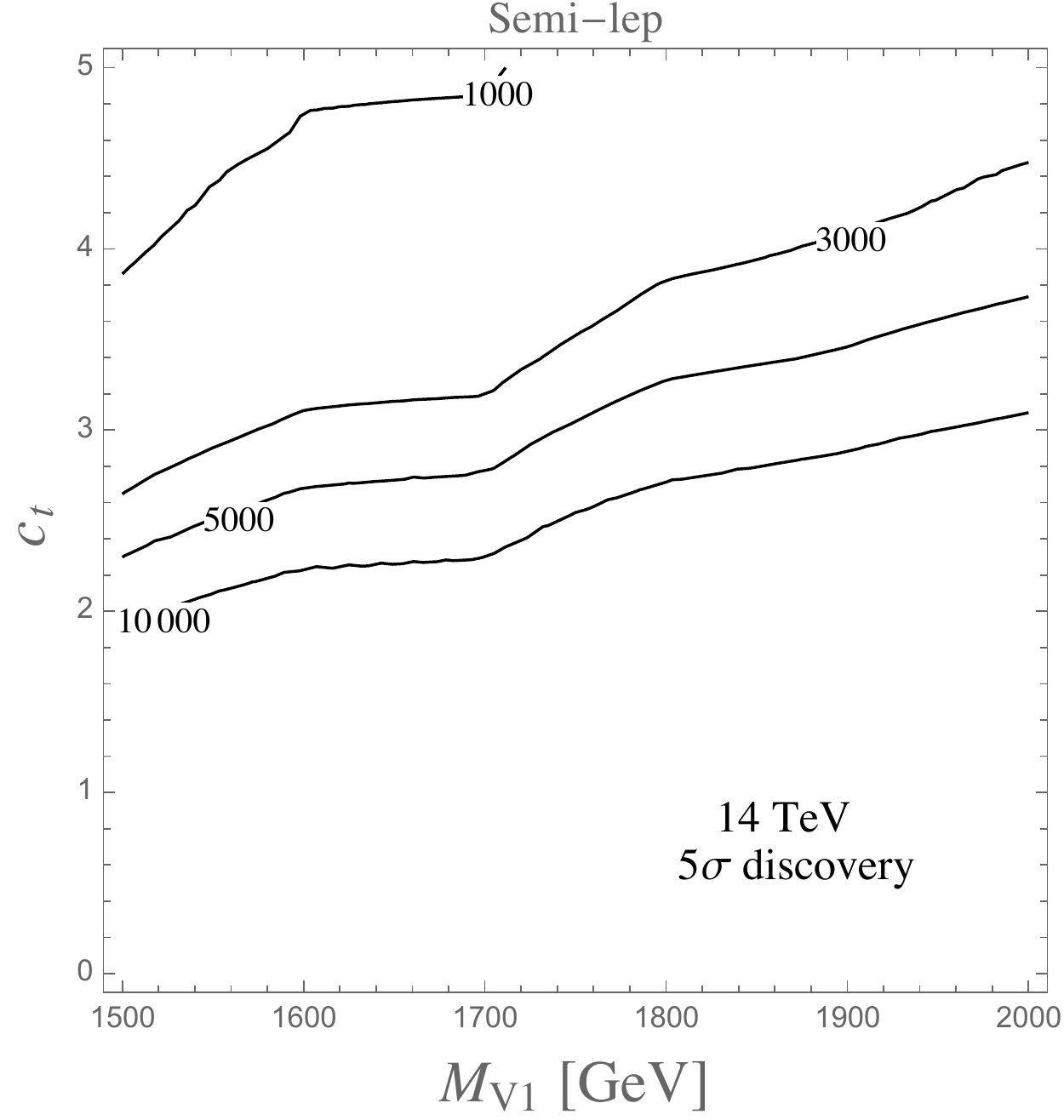}  \\ [.4cm]
\hspace{-20pt} \includegraphics[scale=0.53]{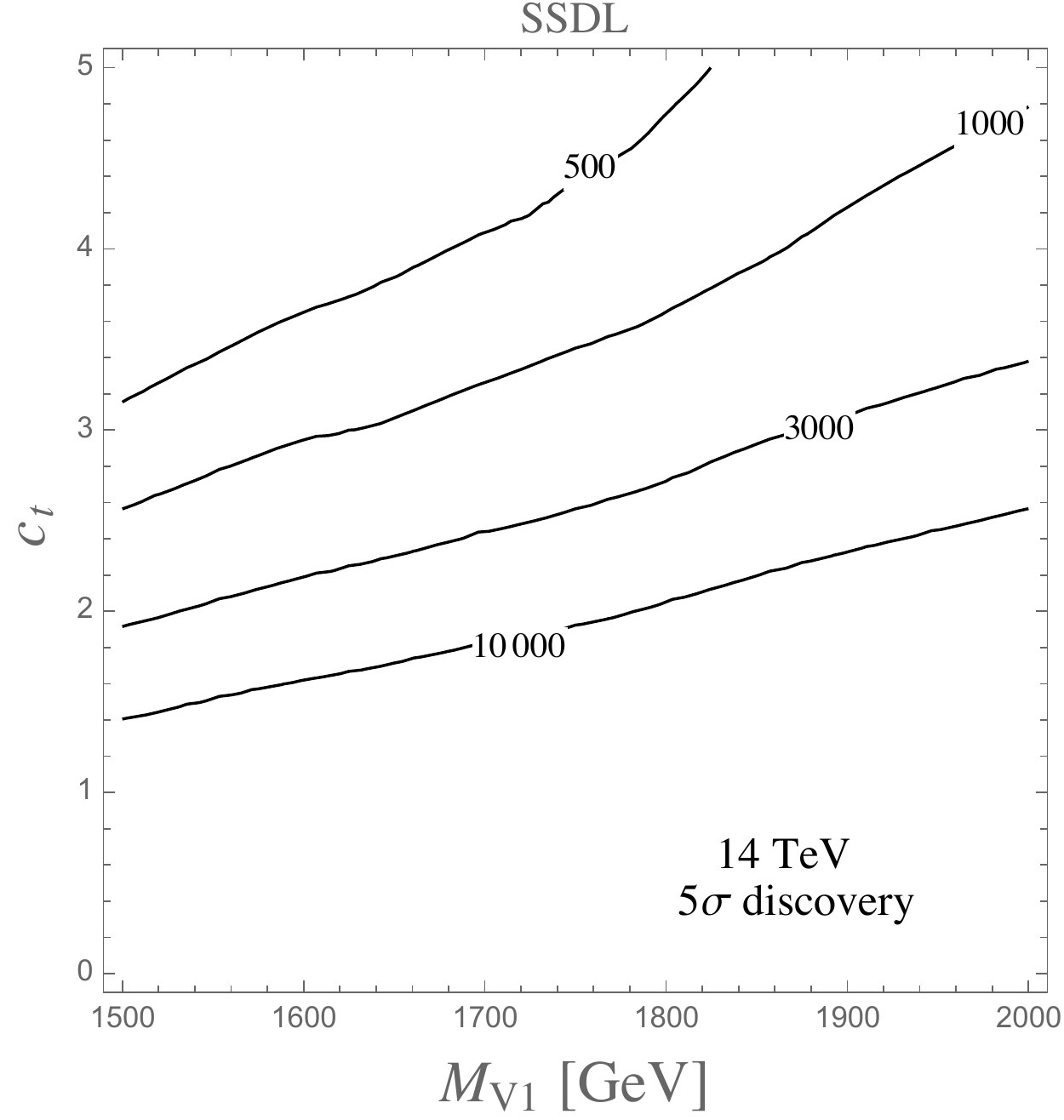} &
\hspace{30pt} \includegraphics[scale=0.53]{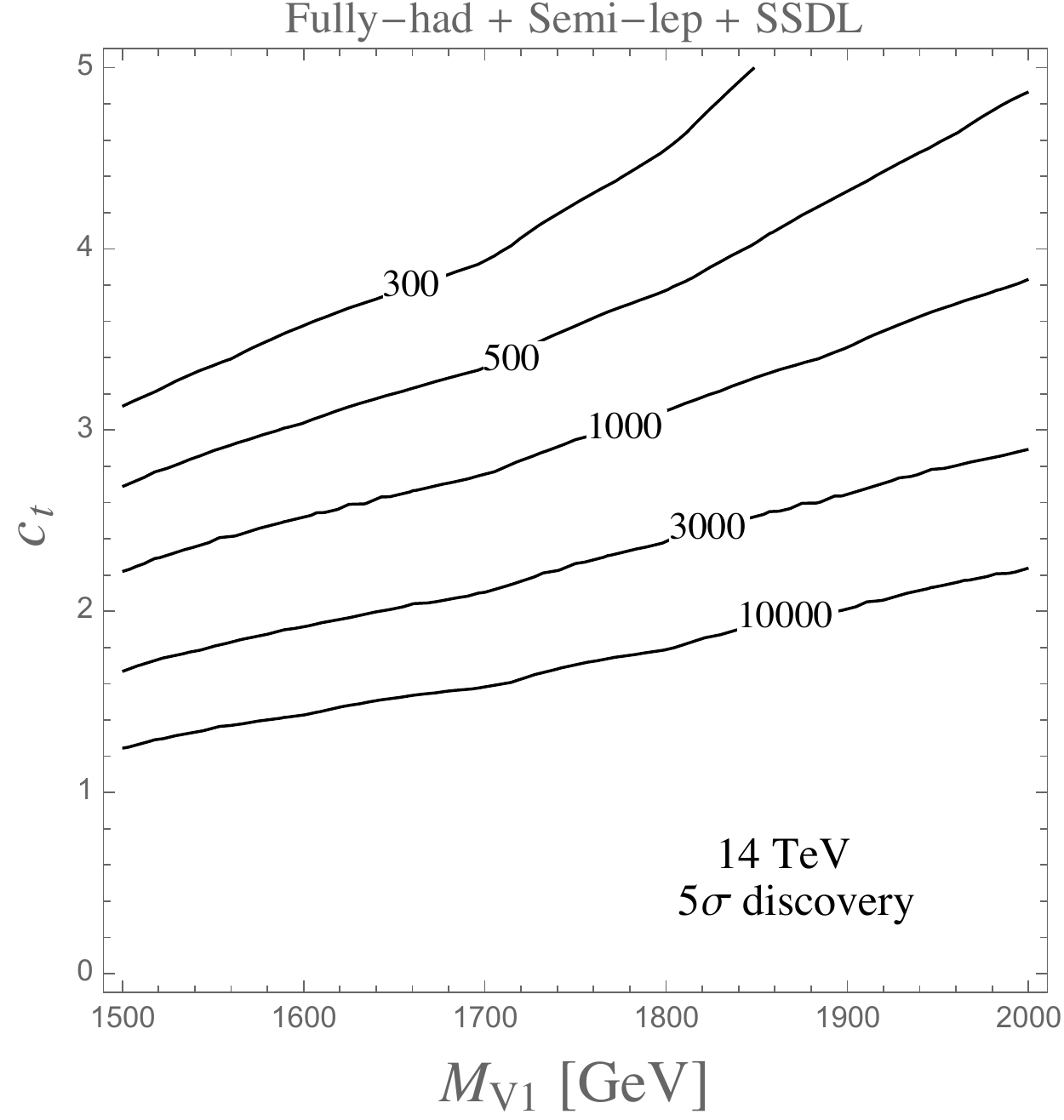}  \\
\end{tabular}
}
\end{center}
\caption{Required luminosities in fb$^{-1}$ of the combined fully-hadronic, semi-leptonic and SSDL channels for
$5\sigma$ discovery for HL LHC run at $\sqrt{s} = 14$ TeV.}
\label{fig:SigMap}
\end{figure*}

Next we require at least three isolated $r=0.4$ jets (isolated from $t_{\rm had}$ and $t_{\rm lep}$ by $\Delta R_{\rm j, t_{\rm had, \rm lep}} > 1.2$). The high-multiplicity final states allow us to further suppress the backgrounds by demanding the cut of $M_J > 650 \GeV$ (see Figure \ref{fig:bTagSSDL}.). Our $b$-tagging strategy requires at least $3b$-tag on the $r=0.4$ jets with $|\eta_{j} |< 2.5$ and $p_T^{j} > 25 \GeV$.

Finally, for a fair comparison, we perform an independent inquiry employing a similar search strategy as in Ref. \cite{Liu:2015hxi}. In this analysis, $Basic \; Cuts \; 2$ requires exactly two isolated same-sign leptons in the event (mini-ISO $> 0.7$  with $p_T^{\ell} > 25 \GeV$ and $|\eta_{\ell} |< 2.5$) and at least three central $r=0.4$ jets with $|\eta_{j} |< 2.5$ and $p_T^{j} > 25 \GeV$. Similarly we define a measure 
\begin{equation}
S_T = \sum\limits_{\text{all} \; j \;, \ell } |p_T| \; ,
\label{eq:chi2}
\end{equation}
and demand $S_T > 1500 \GeV$. At least $3b$-tag is applied to the $r=0.4$ jets with $|\eta_{j} |< 2.5$ and $p_T^{j} > 25 \GeV$.

Table \ref{tab:SameSign_flow} shows cutflow for two very different selection strategies, with (without) the boosted ditop selection, simulated at the benchmark model point of \mbox{$M_{V_1} = 1.5 \TeV$} and \mbox{$c_t=2.0$}. We find that the boosted ditop selection suppresses the top-rich backgrounds at the price of the signal efficiency $\sim 50\%$ marginally improving $S/\sqrt{B}$. Requiring at least three isolated jets and $M_J > 350 \GeV$ further delivers an extra improvement in $S/\sqrt{B}$ by a factor of $\sim 1.6$. Final improvement in $S/\sqrt{B}$ is driven by at least $3b$-tag on the $r=0.4$ jets by a factor of $\sim 1.6$, and the resulting sensitivity after all cuts reaches up to $S/\sqrt{B} = 6.3$ given a luminosity of 3000~fb$^{-1}$. That is slightly higher than $S/\sqrt{B} = 5.6$ in the search without the boosted technology, and twice as high as any other channels in this analysis.

As a consequence, the boosted technique leads to a higher sensitivity and better background management with the capability of resonance reconstruction in the SSDL channel. In the next section, we will further proceed to combine the significances of all three channels to estimate the discovery potential of the $V_1$ resonance.


\subsection{Combining Multiple Channels}
\label{sec:Combo}

\begin{figure*}[t]
\begin{center}
\setlength{\tabcolsep}{0em}
{\renewcommand{\arraystretch}{1}
\begin{tabular}{cc}
\hspace{-20pt} \includegraphics[scale=0.53]{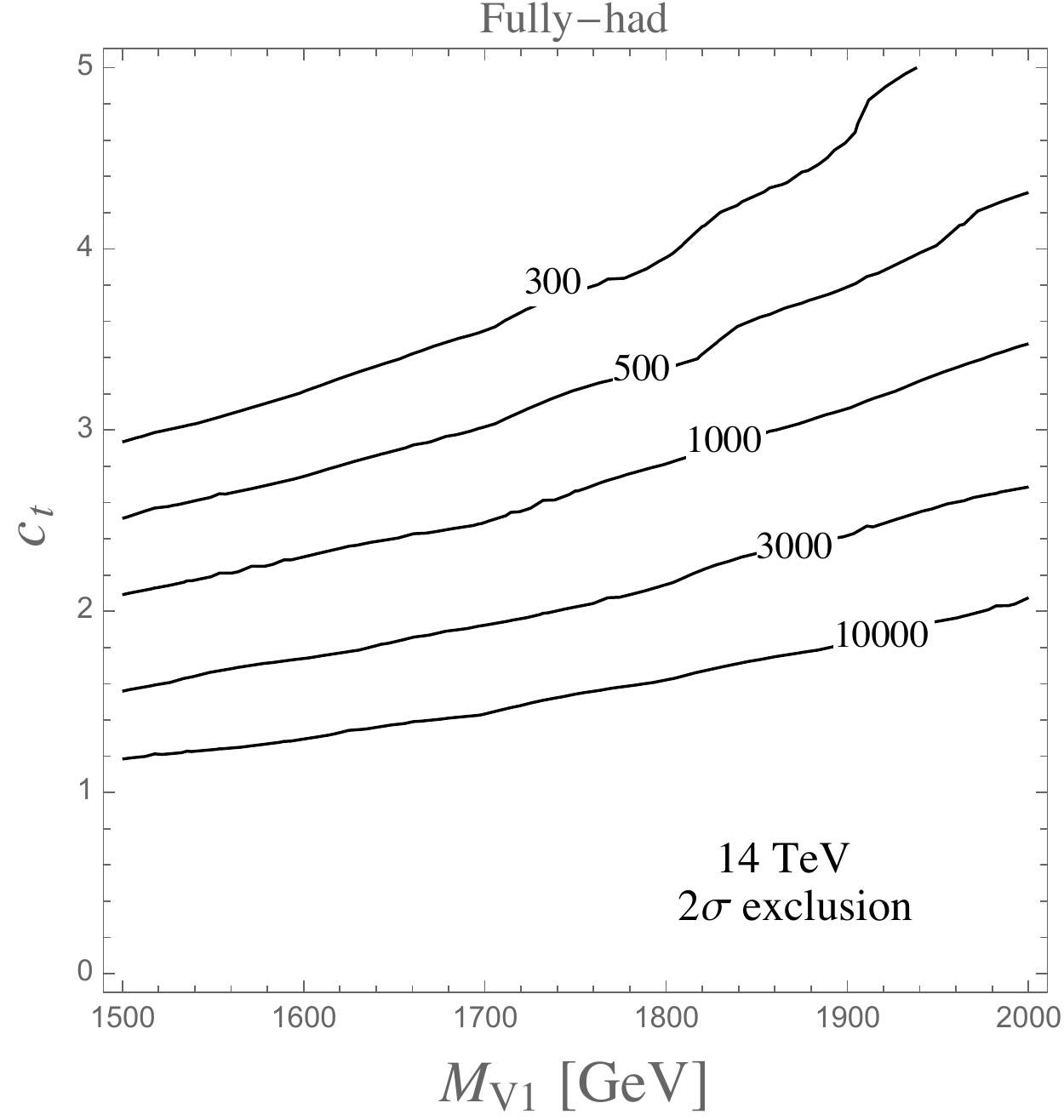} &
\hspace{30pt} \includegraphics[scale=0.53]{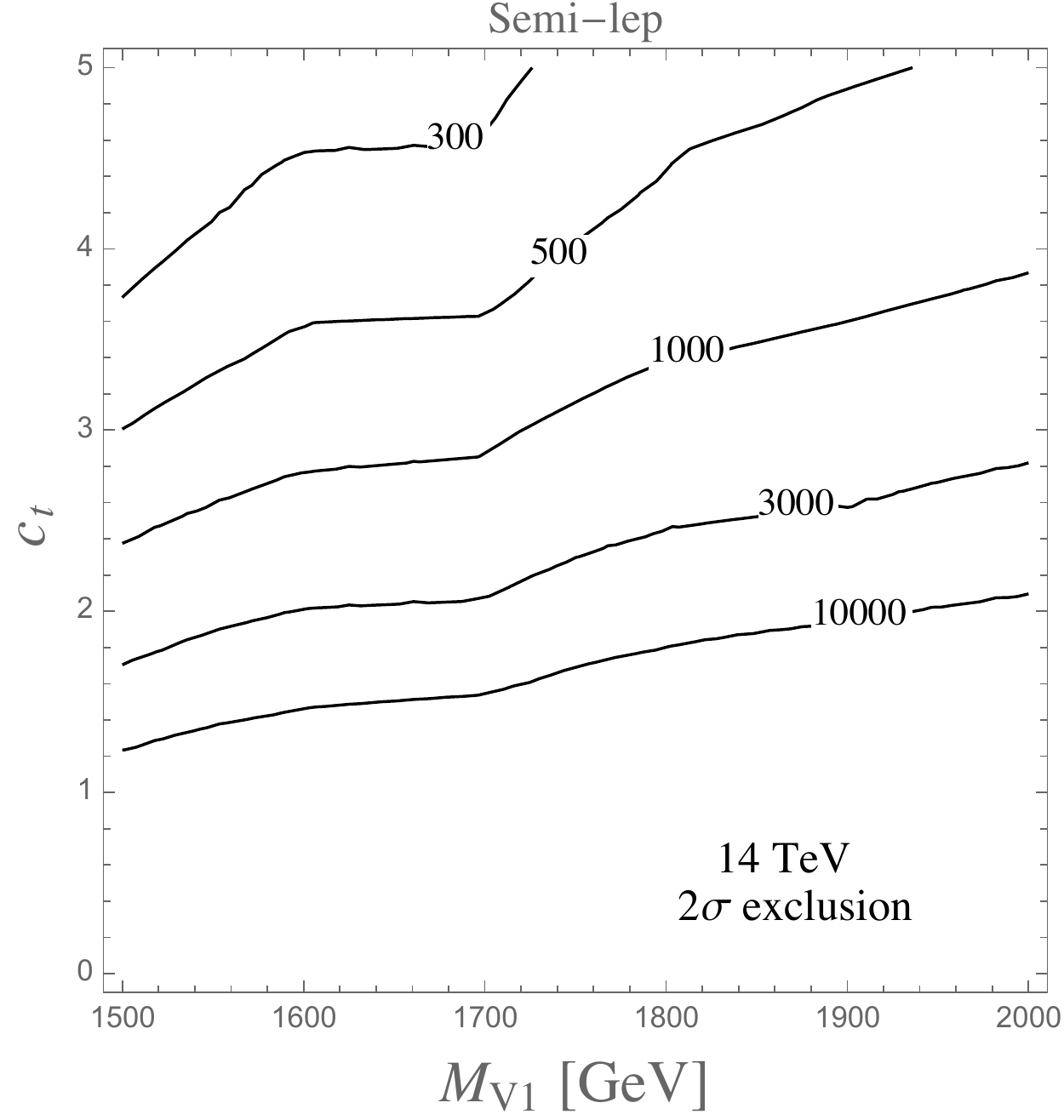}  \\ [.4cm]
\hspace{-20pt} \includegraphics[scale=0.53]{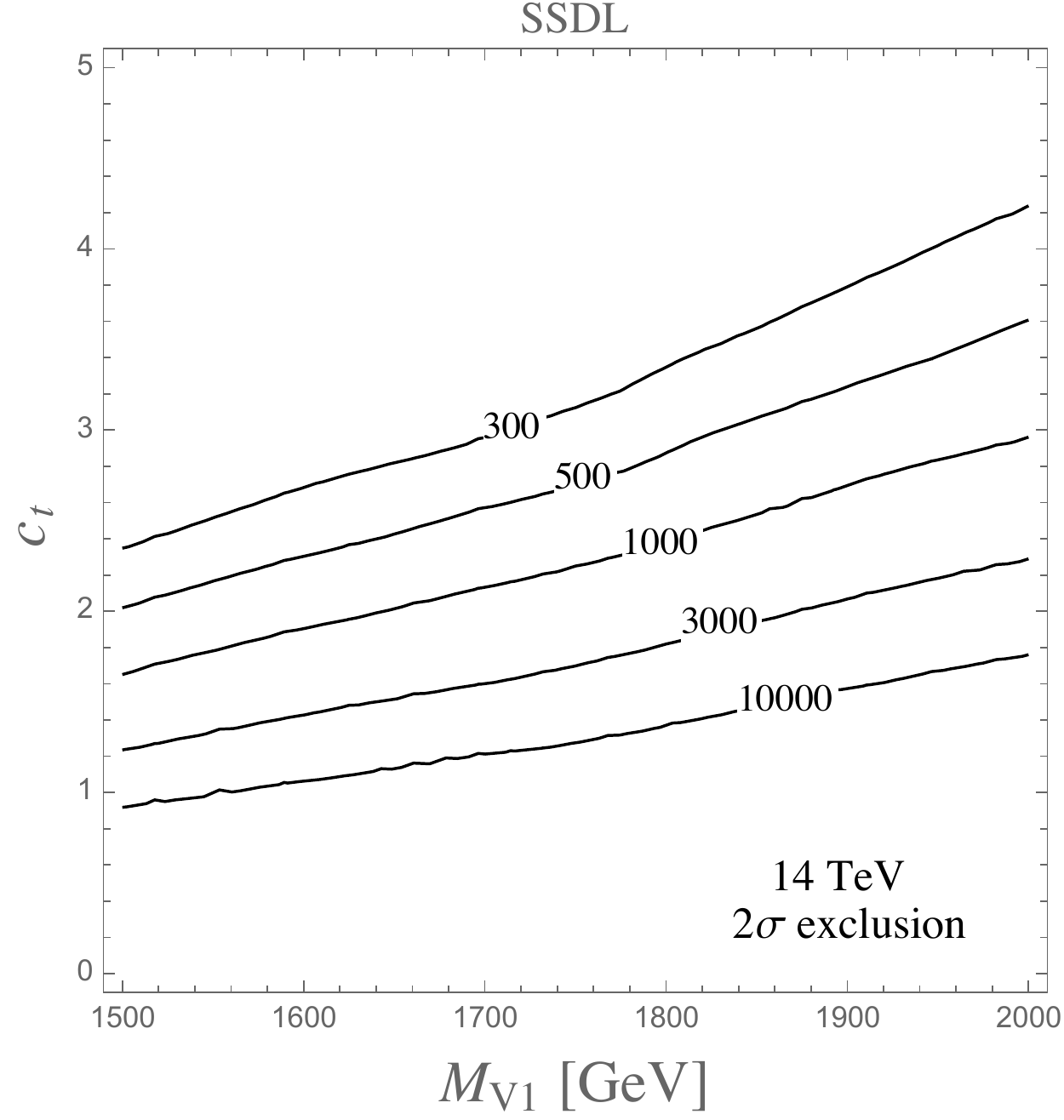} &
\hspace{30pt} \includegraphics[scale=0.53]{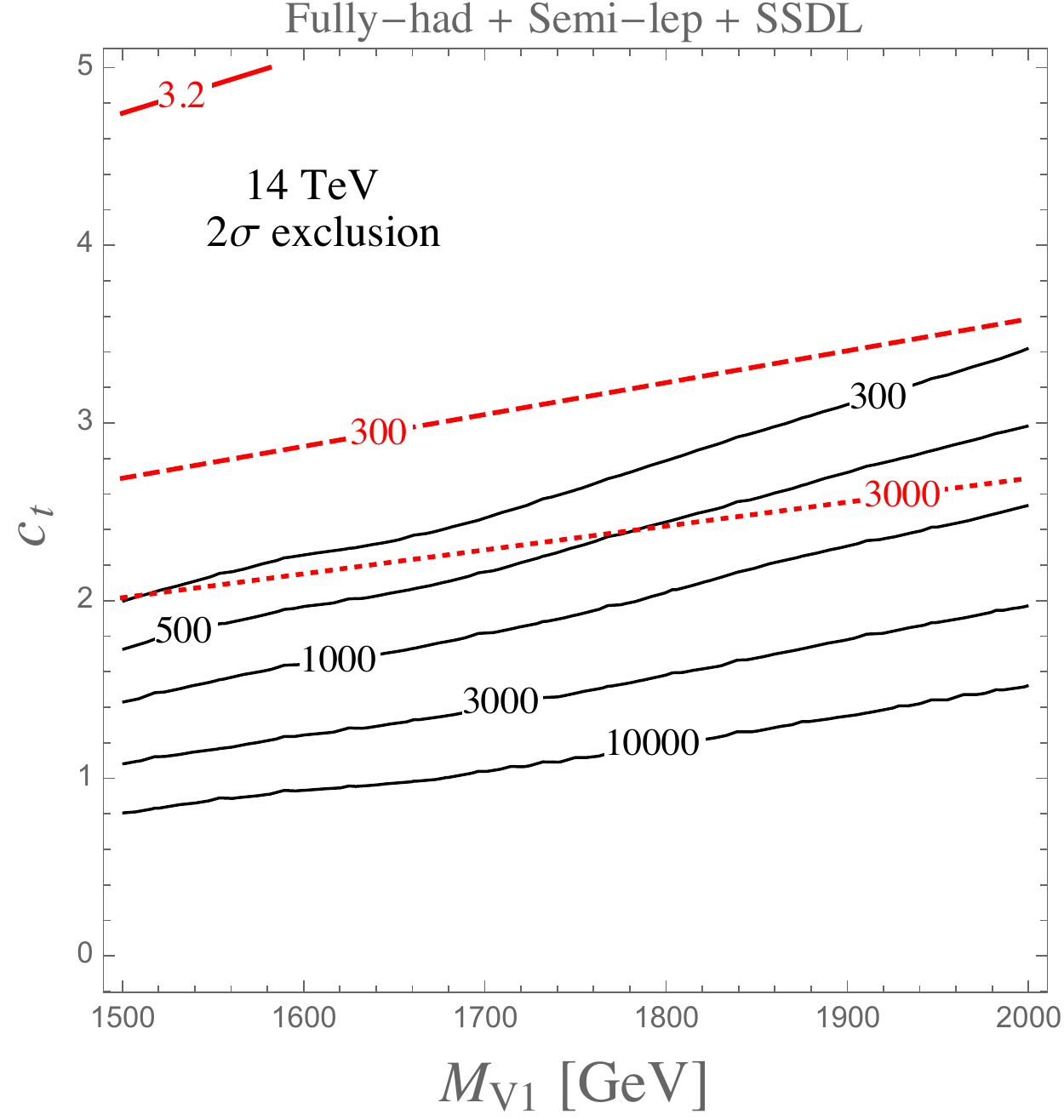}  \\
\end{tabular}
}
\end{center}
\caption{Required luminosities in fb$^{-1}$ of the combined fully-hadronic, semi-leptonic and SSDL channels for (left) $2\sigma$ exclusion for HL LHC at $\sqrt{s} = 14$ TeV. We also show current (red, solid for 3.2 fb$^{-1}$) and projected (dashed for 300 fb$^{-1}$ and dotted for 3000 fb$^{-1}$) bounds with the combined results.}
\label{fig:ExcMap}
\end{figure*}

The discovery potential of the $V_1$ resonance can be further improved by combining all three channels. In this section, we combine the results from the fully-hadronic, semi-leptonic and SSDL channels (each of which has a disjoint final state).

In order to estimate the discovery reach, we define the significance $\sigma_{\rm dis}$ as a likelihood ratio~\cite{Cowan:2010js},
\beq
	\sigma_{\rm dis} \equiv \sqrt{-2\, {\rm ln}\left( \frac{L(B | \mu S + B)}{L(\mu S + B| \mu S + B)} \right) }\,,
\eeq
where $S$ and $B$ are the expected number of signal and background events respectively, and $\mu$ denotes a signal modifier parameter relevant for reflecting correlated systematic uncertainties when combining different searches. 
Assuming all three channels are statistically independent, we use a combined likelihood given by the product of individual likelihoods
\beq
	L(x |n) = \prod_{i=1}^N \frac{x_j^{n_j}}{n_j\!} e^{-x_j} \, ,
\eeq
where $i$ runs over the fully-hadronic, semi-leptonic and SSDL channels.
Since correlated systematic uncertainties in three very different searches are unavailable for us,
we simply take the signal modifier parameter $\mu = 1$, and for a discovery we demand
\beq
	\sigma_{\rm dis} \geq 5.
\eeq

An exclusion limit, on the other hand, is estimated by using the likelihood ratio
\beq
	\sigma_{\rm exc} \equiv \sqrt{-2\, {\rm ln}\left( \frac{L(\mu S + B|B)}{L(B|B)} \right) }\,,
\eeq
with a signal strength parameter $\mu = 1$. The 2$\sigma$ exclusion bound is obtained by
\beq
	\sigma_{\rm exc} \geq 2.
\eeq

Figure \ref{fig:SigMap} summarizes the required luminosities in fb$^{-1}$ of the (upper-left panel) fully-hadronic, (upper-right panel) semi-leptonic, (lower-left panel) SSDL and (lower-right panel) combined channels for $5\sigma$ discovery at $\sqrt{s} = 14$ TeV LHC run II. For the very high luminosity of $\sim 3000 \; \rm fb^{-1}$ we can at most probe it down to $c_t \sim 1.6$ and $c_t \sim 2.9$ for $1.5 \TeV$ and $2.0 \TeV$ respectively in the combined channel. Since the significance isolines scale as the cross section (see Figure \ref{fig:4_top14}), it will be challenging to get any sensitivity in the $c_t < 1.0$ territory even during the high-luminosity phase of the LHC.

Finally, Figure \ref{fig:ExcMap} summarizes the required luminosities in fb$^{-1}$ of the (upper-left panel) fully-hadronic, (upper-right panel) semi-leptonic, (lower-left panel) SSDL and (lower-right panel) combined channels for $2\sigma$ exclusion at $\sqrt{s} = 14$ TeV LHC run II. For the very high luminosity of $\sim 3000 \; \rm fb^{-1}$ we can at most exclude it down to $c_t \sim 1.0$ and $c_t \sim 2.0$ for $1.5 \TeV$ and $2.0 \TeV$ respectively in the combined channel.
We also show current (red, solid for 3.2 fb$^{-1}$) and projected (dashed for 300 fb$^{-1}$ and dotted for 3000 fb$^{-1}$) bounds in the lower-right corner (also shown in figure \ref{fig:4_top14}). 
We note that one should be careful when comparing ATLAS results against our results, 
as they looked at the channel with one lepton plus multiple jets, 
while our lepton comes from the decay of one of the non-boosted spectator tops while the two boosted tops decay hadronically.
Also we have used a LO signal cross section, while including an NLO K factor of 2 in all backgrounds. 
B-tagging efficiencies are comparable but are on the slightly conservative side.
Adopting the b-tagging efficiencies used in Ref. \cite{ATLAS13TeV}, we find 30\% improvement in the final significance in the hadronic channel.
In the semi-leptonic and same-sign dilepton channel, improvements were 11\% and 8\%, respectively. 
These are expected improvements as the hadronic channel exploits more of the b-tagging efficiencies compared to the other channels.
Overall, our results clearly show that a dedicated analysis could improve the sensitivity significantly in the combined channel.
With the ATLAS b-tagging efficiency, we find a factor of 1.13 improvement in the final significance.
For example, $\sigma_{\rm dis}$ = 7.088 ($\sigma_{\rm exc}$ = 6.352) becomes 8.062 (7.229) for our benchmark point, $M_{V_1}=1.5$ TeV and $c_t=2$ with 3000 fb$^{-1}$.

\section{Summary and Discussion}\label{sec:conclusions}

With the discovery of Higgs boson at the LHC, the next highest priority is the precision measurement of the Higgs interaction with SM particles and searches for new phenomena beyond the SM. 
In both cases, the top quark plays a central role, making any searches associated with top quarks appealing.
Among many others, a $t\bar t$ resonance is very well-motivated and searched for extensively at the LHC.
Current bounds on the resonance mass lie in the TeV range, depending on models.

In this paper we have studied a $t\bar t$ resonance that couples primarily to top-quarks (top-philic), and very weakly to the rest of the SM particles. For concrete discussion, we have considered a case with a color singlet vector resonance, $V_{1}$. 
In a simplified model, we have investigated the discovery potential of such a top-philic resonance in the four-top final state.
In the large mass region ($M_{V_{1}}  \geq 1.5 ~\rm TeV$ in our study), two tops from the decay of $V_1$ are boosted and we have exploited the TOM with a new IR-safe template observable, template $y$-cut, to reconstruct the boosted top-quarks and reduce the dominant backgrounds. 
We combined jet trimming with TOM for the first time to remove soft radiation thus further lowering the background rates.

In our analysis we considered three different channels: fully hadronic decay of all four top-quarks, semi-leptonic decay of the non-boosted tops with hadronic decays of boosted tops, and same-sign dileptonic decay. We found that the SSDL channel provides the best sensitivity, even though it suffers from small branching fractions, and the boosted-top tagging improved the significance up to 10\%-20\% in the same sign dilepton channel.
For example, we obtained $S/\sqrt{B} = 5.6$ for $M_{V_{1}} =1.5 ~\rm TeV$ and $ c_{t} = 2.0 $ without boosted techniques and an improvement to $S/\sqrt{B} = 6.3$ with boosted techniques at the 14 TeV HL-LHC. 
We found that the fully-hadronic and semi-leptonic channels show comparable significances. 
This is due to the $ty$ cut that we used to reduce the mis-tag rate of QCD jets, which shows a remarkable performance with the background reduction in the semi-leptonic channel.

After combining all three channels we showed that the 14 TeV LHC with 300 fb$^{-1}$ can exclude such a top-philic resonance 
up to a coupling strength $c_t \approx 2$ for a resonance mass of 1.5 TeV and $c_t \approx 3.4$ for 2 TeV.
The HL-LHC with 3000 fb$^{-1}$ pushes down to $c_t \approx 1$ for 1.5 TeV and $c_t \approx 2$ for 2 TeV, respectively.
Roughly our combined results show about 60\% (50\%) reduction in the required luminosity for 2$\sigma$ exclusion (5$\sigma$ discovery), compared to the SSDL channel alone.
The boosted top-tagging not only improves the sensitivity but also helps in reconstructing the mass of the top-philic resonance. 

Finally it is interesting to note that a light top-philic resonance ($M_{V_1} \lesssim 300$ GeV) can decay into different final states such as $Zh$ ($\sim 60\%$-$80\%$), $b\bar b$ ($\sim 20\%$-$40\%$) or $W^+W^-$ (a few \% of branching fraction) \cite{Cox:2015afa}.
Moreover the top-philic resonance may be a bridge to the dark sector, through which dark matter can annihilate to the $t\bar t$ final state.
In this case, other collider signatures such as $ j + \met$ and $t \bar{t} +  \met  $ will open up.
We show that a top-philic resonance provides a rich phenomenology at the LHC and hence encourage the experimental collaborations to pursue a dedicated study on it.

\bigskip
\emph{Acknowledgements:} 
We would like to thank HTCaaS group of KISTI for providing useful cluster resources during the full course of this project. 
We are grateful to Chul Kim, Jae-Hyeok Yoo and Hwidong Yoo  for discussion and comments, and would like to thank Michael Spannowsky for encouragement toward this study.
JHK is supported by the IBS Center for Theoretical Physics of the Universe (IBS-R018-D1) and Center for Axion and Precision Physics Research (IBS-R017-D1-2016-a00). 
KK is supported by the U.S. DOE under Grant No. DE-FG02-12ER41809.
GM is supported by the Fermilab Graduate Student Research Program in Theoretical Physics and in part by the National Research Foundation of South Africa, Grant No. 88614. 
SL and JHK (in part) have been supported by the National Research Foundation of Korea grant funded by the Korea government (NRF- 2015R1A2A1A15052408). We also acknowledge the Korea Future Collider Study Group (KFCSG) for motivating us to proceed with this work.


\bibliography{draft}
\end{document}